\documentclass[rmp,aps,twocolumn,showpacs,nofootinbib]{revtex4}
\usepackage{graphicx} 
\usepackage{epsfig}
\usepackage{dcolumn}  
\usepackage{bm}       

\def\ii{{\rm i}}  \def\ee{{\rm e}}  \def\mb{{\bf m}}
\def\gb{{\bf g}}  \def\Rb{{\bf R}}  \def\pb{{\bf p}}  \def\Qb{{\bf Q}}
  \def\kb{{\bf k}}  \def\Eb{{\bf E}}  \def\rb{{\bf r}}
\def\xx{\hat{\bf x}}   \def\Hb{{\bf H}}
\def\yy{\hat{\bf y}}   \def\ksp{{k_\parallel^{\rm SP}}}
\def\zz{\hat{\bf z}}
\def\aE{{\alpha_{\rm E}}}  \def\aM{{\alpha_{\rm M}}}
\def\apE{{\alpha'_{\rm E}}}  \def\apM{{\alpha'_{\rm M}}}

\begin{document}
\title{Colloquium: Light scattering by particle and hole arrays}

\author{F. J. Garc\'{\i}a de Abajo}

\affiliation{Instituto de \'Optica - CSIC, Serrano 121, 28006
Madrid, Spain}

\begin{abstract}
This colloquium analyzes the interaction of light with
two-dimensional periodic arrays of particles and holes. The enhanced
optical transmission observed in the latter and the presence of
surface modes in patterned metal surfaces are thoroughly discussed.
A review of the most significant discoveries in this area is
presented first. A simple tutorial model is then formulated to
capture the essential physics involved in these phenomena, while
allowing analytical derivations that provide deeper insight.
Comparison with more elaborated calculations is offered as well.
Finally, hole arrays in plasmon-supporting metals are compared to
perforated perfect conductors, thus
assessing the role of plasmons in these types of structures through
analytical considerations.
\end{abstract}
\date{\today}
\maketitle \tableofcontents


\section{Introduction}
\label{introduction}

The scattering of waves in periodic media plays a central role in
areas of physics as diverse as low-energy electron diffraction
\cite{P1974} or atomic-beam scattering from crystal surfaces
\cite{FR98}. Valence electrons in solids, sound in certain ordered
constructions \cite{MSS95}, or light in photonic crystals
\cite{JVF97,C03} undergo diffraction that under certain conditions
can limit their propagation in frequency regions known as band gaps
\cite{AM1976}. Among these examples, the scattering of
electromagnetic waves is particularly important because it allows
obtaining structural and spectroscopic information over a
fantastically wide range of lengths, going from atomic dimensions in
x-ray scattering \cite{HGD93} to macroscopic distances in radio and
microwaves. Actually, Maxwell's equations are written in first-order
derivatives with respect to spatial coordinates, so that light
scattering in the absence of nonlinear effects is solely controlled
by the shape and permittivity of diffracting objects with distances
measured in units of the wavelength, and therefore, the same
phenomena are encountered over entirely different length scales.

We can classify the performance of periodic structures in three
distinct categories according to the ratio of the period $a$ to the
wavelength $\lambda$. For $\lambda\gg a$, an effective homogeneous
medium description is possible. This is in fact what happens in most
naturally-occurring substances when $a$ has atomic dimensions. But
also in certain artificially textured materials (metamaterials),
which allow achieving exotic behavior like magnetic response at
visible frequencies \cite{GGG05} and media with negative refraction
index \cite{YPF04}, without neglecting the exciting possibility of
using nanoparticles as building blocks to tailor on-demand optical
properties \cite{L06}. The opposite limit ($\lambda\ll a$) is
generally well accounted for by classical rays, although keeping
track of phases proves to be crucial near points of light
accumulation, like in the self-imaging of gratings described by
Talbot nearly two centuries ago \cite{T1836,paper119}. Nevertheless,
it is the intermediate regime, when $\lambda$ is comparable to $a$,
in which diffraction shows up in full display. We find examples of
this in both three-dimensional (3D) photonic crystals, which offer a
promising route to fully controlling light propagation over
distances comparable to the wavelength \cite{JVF97,C03}, and
two-dimensional (2D) crystals, in which an impressive degree of
optical confinement has been accomplished \cite{AAS03}.

In this colloquium, we shall focus on light scattering by planar
structures of particles or holes, which have become a current
subject of intense research driven to some extent by advances in
nano-patterning techniques. 
Our main purpose is to explain the phenomena observed within this
context in a tutorial but nevertheless comprehensive fashion. We
shall first review experimental and theoretical developments in
Sec.\ \ref{overview}. Then, we shall formulate in Sec.\
\ref{tutorialapproach} a simple powerful model that deals with the
response of particle and hole arrays on a common footing, leading to
analytical expressions that capture the main physical aspects of
these systems. Finally, metals with plasmons will be discussed, and
the main differences with respect to plasmon-free perfect conductors
elucidated, in Sec.\ \ref{realmetals}. We shall use Gaussian units,
unless otherwise stated.

The beginning of the last century witnessed important developments
in diffraction of light in gratings after Wood's observation of
anomalous reflection bands \cite{W1902,W1912,W1935} and their
subsequent interpretation \cite{R1907,F1936,F1941}. Two types of
anomalies were identified, one of them occurring when a diffracted
beam becomes grazing to the plane of the grating, the Rayleigh
condition \cite{R1907}, giving rise to a sharp bright band, and the
other one showing up to the red of the former as an extended feature
containing two neighboring dark and bright bands \cite{F1936,F1941}.

The century concluded with another significant discovery
\cite{ELG98}: periodic arrays of subwavelength holes drilled in thin
metallic films can transmit much more light per hole at certain
frequencies than what was previously expected for single openings,
based upon Bethe's prediction of a severe cutoff in transmission as
$(b/\lambda)^4$ for large $\lambda$ compared to the hole radius $b$
\cite{B1944}. Previous knowledge gathered by electrical engineers in
the microwave domain \cite{U1967,C1971,M1980} had already exploited
the use of periodically-drilled surfaces as frequency-selective
filters and discussed the occurrence of 100\% transmission at
wavelengths slightly above the period. However, the hole sizes that
were considered in that context lied in the region of sizeable
transmission for single holes. The more recently discovered
extraordinary transmission phenomenon was however observed for
narrower holes (relative to the wavelength), the transmission of
which exceeded orders of magnitude what was expected from the sum of
their individual transmissions \cite{ELG98}. For square arrays under
normal incidence, a transmission minimum occurred at a wavelength
close to the period $a$, coinciding with the Rayleigh condition
\cite{R1907}, and a transmission maximum showed up at longer
wavelength, thus revealing its connection to Wood's anomalies
\cite{GTG98,SVV03}. However, the explanation of the effect is still
a subject of debate, as some authors understand that it originates
mainly in the interaction of the apertures with surface plasmons
\cite{GTG98,PNE00,MGL01,W01,SGZ01,BMD04}, whereas other authors make
emphasis in dynamical light diffraction \cite{T99,T02,SVV03,LT04}.
While the latter works well to understand the observed extraordinary
optical transmission in drilled plasmon-free perfect conductors
\cite{MCC1988,GSH03,CN04,MH04}, supporters of the surface-plasmon
interpretation argue that the enhanced transmission relies in this
case on plasmon-like lattice-surface-bound modes sustained by
patterned perfect-conductor surfaces \cite{PMG04}. Actually,
evidence of such modes had been observed before in periodically
perforated metallic screens for wavelengths several times larger
than the period \cite{UT1972}. We shall see below how these are in
fact complementary views of the same phenomenon and how diffraction
in particle arrays contains already the essential features that can
be translated to understand the phenomenology of hole arrays. But
let us first summarize experimental and theoretical findings in this
area.

\section{Overview of Existing Results}
\label{overview}

A huge amount of literature has been accumulated on transmission
through periodic structures, and it is an interesting exercise to
reexamine it in connection to recent developments.

\subsection{Single holes}
\label{singleholesexp}

Bethe's predicted cutoff in the transmission of a single hole in a
perfect-conductor thin screen as $(b/\lambda)^4$ is the
leading-order term of the expansion of the transmission cross
section in powers of $b/\lambda$ \cite{B1944}. Subsequent
higher-order analytical corrections \cite{B1954,CSS05}, and
eventually rigorous numerical calculations \cite{R1987,paper069},
demonstrated that the cross section lies below the hole area up to a
radius $b\approx 0.2\lambda$. These results have found experimental
corroboration down to the NIR regime \cite{OK95}, with new localized
plasmon resonances showing up at shorter wavelengths
\cite{DLY04,paper118}.

Two different mechanisms have been however suggested to achieve
enhanced transmission in a single hole: filling it with a material
of high permittivity \cite{paper069,GMP05,WL06}, thus creating a
partially-bound cavity mode that couples resonantly to incident
light (see Sec.\ \ref{interplaybetween}); and decorating the
aperture with periodic corrugations \cite{LDD02} in much the same
way as highly-directional antennas are capable of focusing
electromagnetic radiation on a central dipole element by means
of concentric, periodically-spaced metallic rings \cite{J1977}.

\subsection{Optical transmission through hole arrays}
\label{extraordinaryoptical}

The intensity of light passing through holes is boosted at certain
wavelengths when we arrange them periodically. Pioneering
calculations and microwave experiments showed already zero
reflection in thin films perforated by periodic arrays of small
apertures of radius $b\approx 0.36\lambda$ \cite{C1971}. Further
seminal experiments focused on the relation between hole arrays in
thin metal screens and their complementary screens \cite{U1967},
putting Babinet's principle to a test in the far-infrared region.
This was followed by numerous applied studies of hole arrays
(regarded as frequency-selective surfaces) in the engineering
community, including filters for solar energy collection and
elements to enhance antennae performance
\cite{M1980,M1980_2,CLM1987,MCC1988}.

The work of \textcite{ELG98} demonstrated in the optical domain
extraordinary light transmission, which for the first time occurred
for openings of radius below the cutoff of the first propagating
mode in a circular waveguide, $b<0.29\lambda$. Since then, this
phenomenon has been consistently observed for a varied list of
metallic materials \cite{PDL06}, over a wide range of wavelengths
[e.g., for microwaves \cite{GSH03,CN04}, to which metals respond as
nearly perfect conductors, in the infrared \cite{SWT06}, and in the
VUV, using a good conductor in this regime like Al \cite{ESD07}],
and with different types of array symmetries, including the recent
demonstration of the effect in 2D quasi-crystal arrangements
\cite{STL06,SPF2006,PGE06,MAN07}.

\begin{figure}
\includegraphics[width=80mm,angle=0,clip]{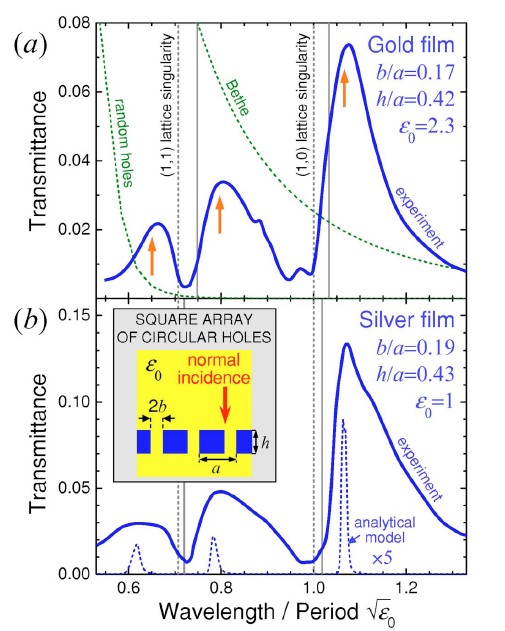}
\caption{\label{experiment} (Color in online edition) Extraordinary
optical transmission in hole arrays. The measured transmittance
(solid curves) is shown for apertures drilled in gold (a) and silver
(b) films, taken from \onlinecite{KTK01}, and \onlinecite{MGL01},
respectively. The silver film is self-standing in air, while the
gold is deposited in quartz and immersed in an index-matching
liquid. The lattice constant is $a=600$ nm in Ag and $a=750$ nm in
Au. The transmittance of the perforated gold goes well above that
predicted for non-interacting apertures in a perfect-conductor film
or by Bethe's formula for a thin screen (dashed curves). Rayleigh's
condition for the (1,0) and (1,1) beams becoming grazing are
indicated by vertical dashed lines. Analytical results are shown as
arrows in (a) and as a dashed curve in (b) (see Sec.\
\ref{differencesinthelattice}). The transmittance is presented vs
the wavelength in the dielectric environment of the metal,
normalized to the lattice constant.}
\end{figure}

Two examples of enhanced transmission, taken from
\onlinecite{KTK01}, and \onlinecite{MGL01}, are shown in Fig.\
\ref{experiment}. The transmission is several times larger in the
infrared peak than the prediction of Bethe for non-interacting holes
in a thin screen, and four orders of magnitude larger than what is
expected for non-interaction apertures in a perfect-conductor film
of the same thickness (dashed curves).

Light transmission through hole arrays has been examined
theoretically for over four decades
\cite{EC1962,C1971,M1980,DMW1989}, although a detailed account of
extraordinary optical transmission in real metals had to wait until
the new century began \cite{PNE00,MGL01,W01,SGZ01,SVV03} and the
advance in computation power allowed predictive capacity
\cite{KES04,CGS05}.

The influence of various geometrical and environmental factors has
been extensively studied. In particular, the role of hole shape has
been shown to yield nontrivial effects
\cite{KES04,GBM04,ESZ04,KSZ05,VKE05}, such as larger enhancement and
red shift of the transmission peaks with respect to the Rayleigh
condition for light polarized along the short axis of elongated
apertures. Finite arrays have been found to exhibit interesting
shifts in the transmission maxima as well, depending on the number
of apertures \cite{LT04,BGM04}. More exotic shapes like annular
holes have been also simulated \cite{RM1988,BV02} and measured
\cite{FZM05}, with the additional appeal that annular waveguides
support always one guided mode at least \cite{J99}.

The transmission is exponentially attenuated with hole depth because
it is mediated by evanescent modes of the apertures regarded as
narrow subwavelength waveguides. However, strong signatures of
interaction between both metal interfaces have been reported
\cite{DLB02}, as well as high sensitivity to dielectric environment,
so that maximum transmission is achieved when the permittivity is
the same on the two sides of the film \cite{KTK01}.

Extraordinary optical transmission has expanded to embrace a wide
range of phenomena \cite{GE07}, like the interaction of hole arrays
with molecules for potential applications in biosensing \cite{DKE06}
and all-optical switching \cite{SZS02,JGH05,DRK06}, and the
demonstration of the quantum nature of plasmons through photon
entanglement preservation after traversing a hole array
\cite{AVW02}.

\subsection{Particles}
\label{particles}

The field of light scattering by small particles has a rich research
tradition \cite{V1981,BH98} that is being continued by hot topics such
as for example novel near-field effects in the coupling of metallic
nanoparticle arrays \cite{KDW99} and strong inter-particle
interactions in dimers \cite{ASN04,NOP04,paper114}. 
Here, we shall single out just two recent exciting developments in
line with the rest of our discussion. The first one refers to
coupled metallic nanoparticle arrays. These particles can sustain
localized plasmon excitations that hop across neighbors. It has been
suggested \cite{QLK98}, and later confirmed by experiment
\cite{MBA01,MKA03}, that this phenomenon can be utilized to transmit
light energy along chains of subwavelength particles, thus providing
some basic constituents for future plasmonic devices.

In a different development, the scattering spectra from 1D and 2D
arrays of metallic nanoparticles were predicted to exhibit very
narrow plasmon lineshapes produced by dynamical scattering
\cite{ZJS04,ZS04}. Experiments performed on lithographically
patterned particle arrays confirmed this effect and achieved
reasonable control over spectral lineshapes \cite{HZS05}. We shall
discuss this further in Sec.\ \ref{narrowingplasmon}.

\section{Tutorial Approach}
\label{tutorialapproach}

A tutorial model will be presented next that becomes exact in the
limit of narrow holes or small particles in perfect-conductor films.
This model will describe the basic physics involved both in
extraordinary light transmission and in lattice surface modes of
structured metals, but it leads to simple analytical expressions
that permit understanding these phenomena in a fundamental way and
making several challenging predictions.

\subsection{Basic relations}
\label{basicrelations}
\begin{figure}
\includegraphics[width=80mm,angle=0,clip]{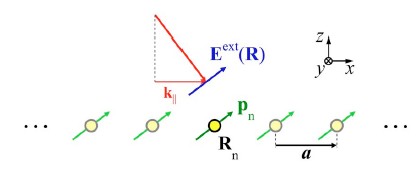}
\caption{\label{particle-array} (Color in online edition)
Two-dimensional array of small identical particles illuminated by a
light plane wave. $\kb_\parallel$ is the momentum component parallel
to the array. The particle at position $\Rb_n$ displays a dipole
$\pb_n$.}
\end{figure}

We shall start with some basic analytical relations for the
scattering of an external light plane wave on a periodic array of
identical particles that are small compared to both the wavelength
and their separation (see Fig.\ \ref{particle-array}). Within
linear, non-magnetic response, the particle at position $\Rb_n$ can
be assumed to respond with an induced dipole
$\pb_n=\aE\,\Eb(\Rb_n)$, determined by its electric polarizability
tensor $\aE$ and the self-consistent field acting on it,
$\Eb(\Rb_n)$. This dipole induces an electric field at point $\rb$
that can be written $\mathcal{G}^0(\rb-\Rb_n)\pb_n$ in terms of the
dipole-dipole interaction tensor,
\begin{eqnarray}
\mathcal{G}^0(\rb)=(k^2+\nabla\nabla)\frac{\ee^{\ii k r}}{r},
\label{Gr}
\end{eqnarray}
where $k$ is the light momentum in free space.\footnote{More
explicitly, $\mathcal{G}^0(\rb)\pb=[\exp(\ii k
r)/r^3]\,\big\{\left[(kr)^2+\ii kr-1\right]\,\pb\,-\left[(kr)^2+3\ii
kr-3\right]\,(\rb\cdot\pb)\,\rb/r^2\big\}$.} Now, the
self-consistent dipole of our particle is found to be
\begin{eqnarray}
\pb_n=\aE \left[\Eb^{\rm ext}(\Rb_n) +\sum_{n'\neq n}
\mathcal{G}^0(\Rb_n-\Rb_{n'})\pb_{n'}\right], \label{eq1}
\end{eqnarray}
where $\Eb^{\rm ext}(\Rb_n)=\Eb^{\rm
ext}\exp(i\kb_\parallel\cdot\Rb_n)$ is the external electric field,
which depends upon the site position $\Rb_n$ just through a phase
factor involving components of the incoming wave momentum parallel
to the array, $\kb_\parallel$, as illustrated in Fig.\
\ref{particle-array}, and the second term inside the square brackets
represents the field induced by the rest of the particles. Bloch's
theorem guarantees that the solution of Eq.\ (\ref{eq1}) must have
the form $\pb_n=\pb\exp(i\kb_\parallel\cdot\Rb_n)$. Direct insertion
of this expression into Eq.\ (\ref{eq1}) leads to
\begin{eqnarray}
\pb=\frac{1}{1/\aE-G(\kb_\parallel)} \Eb^{\rm ext} \label{pp}
\end{eqnarray}
and
\begin{eqnarray}
G(\kb_\parallel)=\sum_{n\neq 0} \mathcal{G}^0(\Rb_n)
\ee^{-\ii\kb_\parallel\cdot\Rb_n}, \label{Gk}
\end{eqnarray}
where we have chosen $\Rb_0=0$. Notice that the denominator of Eq.\
(\ref{pp}) separates the properties of the particles ($\aE$) from
those of the lattice [the structure-factor-type of sum
$G(\kb_\parallel)$], in the spirit of the KKR method in solid state
physics \cite{AM1976}. The lattice sum in Eq.\ (\ref{Gk}) can be
converted into rapidly converging sums using Ewald's method
\cite{GZ1980}, and we have used in particular the procedure
elaborated by \textcite{K1968}.

Incidentally, Eqs.\ (\ref{eq1})-(\ref{Gk}) can be also applied to 3D
particle arrays with $\kb_\parallel$ replaced by a 3D crystal
momentum. This type of approach has been shown to lead to robust
band gaps in atomic lattices \cite{VST96}. Furthermore Eq.\
(\ref{eq1}) together with the Clausius-Mossotti formula
\cite{AM1976} constitute the basis of the discrete-dipole
approximation (DDA) method for solving Maxwell's equations in
arbitrary geometries \cite{PP1973,DF94}. It should also be noted
that the present approach can be extended to larger particles
arranged in ordered \cite{SYM98_1,SYM00} or disordered arrays
\cite{paper040} by including higher-order multipoles, and that this
is one of the methods that can be actually applied to deduce
effective optical properties of composite materials \cite{M02}.

It is useful to represent the dipole-dipole interaction in 2D
momentum space in the plane of the array, which we shall take to
coincide with $z=0$. This is readily done by expressing the scalar
interaction at the right end of Eq.\ (\ref{Gr}) as
\begin{eqnarray}
\frac{\ee^{\ii k r}}{r}=\frac{\ii}{2\pi}\int \frac{d^2\Qb}{k_z}
\ee^{\ii(\Qb\cdot\Rb+k_z|z|)}, \nonumber
\end{eqnarray}
where $k_z=\sqrt{k^2-Q^2}$ is the normal momentum and the notation
$\rb=(\Rb,z)$, with $\Rb=(x,y)$, has been adopted. From here and
Eq.\ (\ref{Gr}) one obtains expressions like
\begin{eqnarray}
\mathcal{G}_{xx}^0(\rb)=\frac{\ii}{2\pi}\int \frac{d^2\Qb}{k_z}
(k^2-Q_x^2) \ee^{\ii(\Qb\cdot\Rb+k_z|z|)} \label{Gxxmomentum}
\end{eqnarray}
for the components of the interaction tensor, here specified for the
$xx$ directions. This allows us to recast Eq.\ (\ref{Gk}) into a sum over
2D reciprocal lattice vectors $\gb$, using the relation
\begin{eqnarray}
  \sum_n\exp(\ii\Qb\cdot\Rb_n)=\frac{(2\pi)^2}{A}\sum_\gb\delta(\Qb-\gb),
  \label{Fourier}
\end{eqnarray}
where $A$ is the area of the lattice unit cell. For example, the
$G_{xx}$ component under normal incidence ($k_\parallel=0$) becomes
\begin{eqnarray}
G_{xx}(0)&=&\lim_{z\rightarrow 0}\Big[\frac{2\pi\ii}{A}\sum_{\gb}
\frac{1}{k_z^g}(k^2-g_x^2)\ee^{\ii k_z^g|z|} \label{GG}
\\ &-&\frac{\ii}{2\pi} \int \frac{d^2\Qb}{k_z}
(k^2-Q_x^2)\ee^{\ii k_z|z|}\Big], \nonumber
\end{eqnarray}
where $k_z^g=\sqrt{k^2-g^2}$, and the integral represents the
subtraction of the $n=0$ term in the sum of (\ref{Gk}). This
expression is important to elucidate some properties of the lattice
sums, as we shall show below.

\subsubsection{Reflection and absorption in particle arrays}
\label{reflectivityofdiffractionless}

The scattered field is given by a Rayleigh expansion similar to the one in Eq.\
(\ref{GG}) \cite{paper106}, with each vector $\gb$ labeling one
reflected and one transmitted beam of parallel momentum
$\kb_\parallel+\gb$ \cite{R1907}. In the far field in particular,
the zero-order ($\gb=0$) reflection and transmission coefficients
under normal incidence reduce to
\begin{eqnarray}
r=\frac{2\pi\ii k/A}{1/\aE-G_{xx}(0)} \label{r2}
\end{eqnarray}
and
\begin{eqnarray}
t=1+r, \label{t2}
\end{eqnarray}
where the first term in the right-hand side of Eq.\ (\ref{t2})
represents the unscattered beam, and the numerator of (\ref{r2}) is
the far-field amplitude produced by a lattice of unit dipoles.

Interestingly, the absorbance of the array is given by
$1-|1+r|^2-|r|^2$ [see Eq.\ (\ref{t2})], which when regarded as a
function of the complex variable $r$, has a maximum of 50\%
coinciding with $r=-1/2$ and $t=1/2$. This condition is easily
attainable near a lattice singularity (see Sec.\
\ref{latticeresonances}), using for instance weakly dissipative
spherical particles. Similar results have been predicted for narrow
cylinder arrays \cite{LAG06}, in which 100\% absorption is possible
in one of the polarization components for the right choice of
parameters.

A particularly simple situation is encountered when the wavelength
is larger than the lattice spacing, so that all diffracted beams
other than the zero-order beam are evanescent
($|\kb_\parallel+\gb|>k$). Then, upon inspection of Eq.\ (\ref{GG}),
one finds the useful relation
\begin{eqnarray}
\Im\{G_{xx}(0)\}=2\pi k/A-2k^3/3, \;\;\;\;\; k<g_1, \label{OT2}
\end{eqnarray}
where $g_1$ denotes the period of the reciprocal lattice
($g_1=2\pi/a$ for square arrays). Moreover, if the particles are
non-absorbing, the optical theorem constrains their polarizability
by the condition $\Im\{-1/\aE\}=2k^3/3$ \cite{V1981}. Combining
these expressions, one obtains
\begin{eqnarray}
r=\frac{-1}{1+\frac{\ii A}{2\pi k}\Re\left\{1/\aE-G_{xx}(0)\right\}}
\label{r1}
\end{eqnarray}
for the reflection coefficient of non-dissipative particles under
normal incidence below the diffraction threshold.

The electrostatic approximation provides a reasonable description of
the electric polarizability of small particles, $\alpha_{\rm E}^{\rm
es}$. However, this needs to be amended in order to comply with the
mentioned optical-theorem constrain, for instance via the
prescription $\aE=1/(1/\alpha_{\rm E}^{\rm es}-2\ii k^3/3)$.
Analytical expressions for $\alpha_{\rm E}^{\rm es}$ exist for a
variety of particle shapes, including homogeneous spheres
($\alpha_{\rm E}^{\rm es}=b^3(\epsilon-1)/(\epsilon+2)$, where $b$
is the radius and $\epsilon$ the permittivity) and ellipsoids
\cite{J1945}.

\begin{figure}
\includegraphics[width=80mm,angle=0,clip]{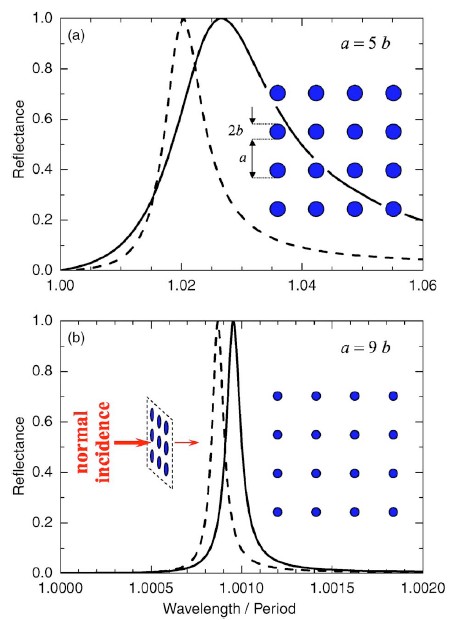}
\caption{\label{particles-T} (Color in online edition) Reflectance
spectra of square arrays of perfectly-conducting thin circular
disks. The wavelength $\lambda$ is normalized to the lattice
constant $a$. The disks radius is $b=a/5$ in (a) and $b=a/9$ in (b).
The light is impinging normal to the array and 100\% reflection is
observed in these two cases at the maximum. Solid curves: full
numerical results. Dashed curves: analytical model for $|r|^2$ [Eq.\
(\ref{r1})].}
\end{figure}

We illustrate the applicability of Eq.\ (\ref{r1}) through an
example consisting of square lattices of perfectly-conducting thin
disks. Fig.\ \ref{particles-T} compares the analytical result of
Eq.\ (\ref{r1}) (dashed curves) with the full solution of Maxwell's
equations obtained by following a layer-KKR multiple-scattering
formalism \cite{SYM98_1,SYM00} to simulate the array together with a
modal expansion solution of the isolated disk similar to the one
available for isolated holes \cite{R1987,paper102}. In the
analytical solution we have used the polarizability of thin metallic
disks as derived from an ellipsoid of vanishing height, $\alpha_{\rm
E}^{\rm es}=4b^3/3\pi$, where $b$ is the radius. The results of the
analytical model describe qualitatively well the presence of zero-
and full-reflection points in the spectra, irrespectively of the
disk size, but we shall discuss this point further in Sec.\
\ref{holearrays}.

\subsubsection{Narrowing lineshapes through dynamical scattering}
\label{narrowingplasmon}

The above formalism can be used to explain the effect of narrowed
plasmon lineshapes in the scattering spectra of 1D and 2D particle
arrays \cite{ZJS04,ZS04,HZS05}. For simplicity, we shall discuss
metallic spherical particles described by the Drude dielectric
function
\begin{eqnarray}
\epsilon(\omega)=1-\frac{\omega_p^2}{\omega(\omega+\ii\eta)},
\label{Drude}
\end{eqnarray}
where $\omega_p$ is the bulk plasma frequency and the plasmon
damping rate is $\approx\eta/2\ll\omega_p$.

Using this expression to obtain the polarizability of a small sphere
of radius $b$ (see Sec.\ \ref{reflectivityofdiffractionless}), we
can recast Eq.\ (\ref{pp}) into a Lorentzian of width
$\approx\eta/2+(\omega_p b^3/2\sqrt{3}) \Im\{G\}$. The natural width
of the isolated particles is now supplemented by a term proportional
to $\Im\{G\}$ [see Eq.\ (\ref{OT2})], which can take negative values
that compensate the $\eta/2$ term to render arbitrarily narrow
collective plasmon resonances for an appropriate choice of array
parameters.

Applying this to a 2D square array under normal incidence with
$\lambda\sim a$, we find that Eq.\ (\ref{OT2}) yields complete
cancelation of the width for $b/a\approx 0.16(\eta/\omega_p)^{1/3}$.
Under such conditions, the narrowing of the width is just limited by
the physical requirement that $|r|^2+|t|^2\le 1$ [see Eqs.\
(\ref{r2}) and (\ref{t2})].

\subsection{Lattice singularities}
\label{latticeresonances}

The interaction among particles in the periodic arrays of Sec.\
\ref{basicrelations} appears to be governed by the lattice sums
$G(\kb_\parallel)$ and is dominated by their singularities, which
originate in accumulation of in-phase scattered fields. Following
similar arguments to previous expositions of this idea
\cite{R1907,F1941}, we just consider a 1D periodic chain of
particles illuminated by an incident plane wave with both
propagation direction and electric field perpendicular to the array,
so that the field induced by a given particle on a distant one
scales with the inverse of their separation, and thus, the
contribution of distant particles to the interaction lattice sum has
the convergence properties of the series $\sum_{n=1}^\infty \ee^{\ii
kan}/n$, which diverges as the wavelength approaches the period $a$
as $-\ln(ka-2\pi)$ \cite{GR1980}. The same is true for 2D arrays.
These singularities in $G(\kb_\parallel)$ are signaled by the
Rayleigh condition of a diffracted beam becoming grazing
\cite{R1907}, as can be seen from Eq.\ (\ref{GG}), where divergent
terms $g\approx k$ (i.e., terms with zero normal momentum $k_z^g$)
dominate the sum.

A remarkable consequence of this analysis is that the array becomes
invisible to the incoming light right at the lattice sum divergence
[$G_{xx}(0)\rightarrow\infty$, so $r\rightarrow 0$, according to
Eq.\ (\ref{r2})], showing 100\% transmission even for absorbing
particles.

Focusing for simplicity on a square array of period $a$, the normal-incidence lattice sum (\ref{GG})
diverges as \cite{paper102}
\begin{eqnarray}
G_{xx}(0)\approx\frac{4\pi^2\sqrt{2}}{a^3}\frac{1}{\sqrt{\lambda/a-1}}-118
\label{Gana}
\end{eqnarray}
for $\lambda\gtrsim a$, where a fitted constant has been subtracted
in order to extend the validity of this expression well beyond the
singularity.

\begin{figure}
\includegraphics[width=80mm,angle=0,clip]{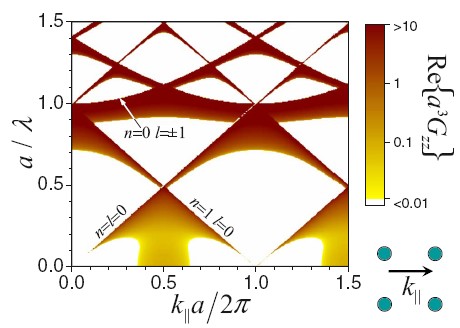}
\caption{\label{lattice-resonances} (Color in online edition)
Lattice sum $G_{zz}(\kb_\parallel)$ [Eq.\ (\ref{Gk})] for a square
lattice of period $a$ as a function of parallel momentum
$k_\parallel$ and wavelength $\lambda$. The direction of
$\kb_\parallel$ is along one of the axes of the lattice.}
\end{figure}

For oblique incidence with $\kb_\parallel$ along one of the lattice unit vectors, proceeding as
in the derivation of Eq.\ (\ref{GG}), one finds that
$G(\kb_\parallel)$ is diagonal and its components diverge as
\begin{eqnarray}
G(\kb_\parallel) \propto \frac{1}{\sqrt{(k_\parallel+2\pi
n/a)^2+(2\pi l/a)^2-k^2}}, \label{div}
\end{eqnarray}
where $n$ and $l$ run over integral numbers (excluding $l=0$ in
$G_{xx}$). This behavior is illustrated in Fig.\
\ref{lattice-resonances}, showing in full display the lattice
singularities exhibited by $\Re\{G_{zz}(\kb_\parallel)\}$.

\subsection{Hole arrays}
\label{holearrays}

\subsubsection{Babinet's principle and hole arrays in thin screens}
\label{thinscreens}

\begin{figure}
\includegraphics[width=80mm,angle=0,clip]{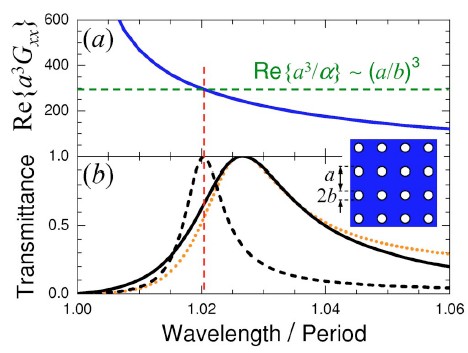}
\caption{\label{Babinet} (Color in online edition) Babinet's
principle applied to disk and hole arrays. The transmittance
(reflectance) of the disk array for light of a given polarization
$\sigma$ (s or p) is identical to the reflectance (transmittance) of
the complementary hole array for orthogonal polarization $\sigma'$
(p or s, respectively).}
\end{figure}

The behavior of hole arrays in perfect-conductor screens can be
directly connected to the properties of the disk arrays considered
in Fig.\ \ref{particles-T}. Indeed, one can invoke the exact Babinet
principle \cite{BW99,J99}, which connects the reflected fields of
the disk array for a given incident polarization with the
transmitted fields of its complementary hole array with orthogonal
polarization, as illustrated in Fig.\ \ref{Babinet} \cite{paper102}.
Therefore, the reflectance spectra shown in Fig.\ \ref{particles-T}
are identical with the transmittance spectra of the complementary
perforated screens.

Focusing again on square arrays and normal incidence, we observe two
characteristic features in the transmittance spectra: (i) the
transmission vanishes when the wavelength $\lambda$ equals the
period $a$, and (ii) a 100\% transmission maximum takes place at a
wavelength slightly above $a$. The origin of these effects can be
traced back to Wood's anomalies in gratings \cite{W1902,W1912,W1935}
and to their interpretation in terms of the following two mechanisms
\cite{F1936,F1941}: (i) accumulation of in-phase scattering events
when the wavelength equals the period (see explanation in Sec.\
\ref{latticeresonances}) and (ii) coupling of the incident light to
a surface resonance. These phenomena persist in hole arrays
perforated in thicker films of non-ideal absorbing metals, for which
the maximum transmission is reduced but still justifies the term
extraordinary optical transmission \cite{ELG98}.

\begin{figure}
\includegraphics[width=80mm,angle=0,clip]{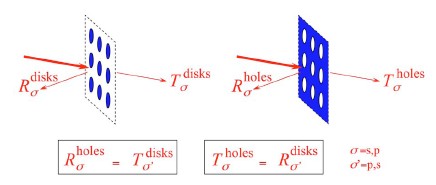}
\caption{\label{maximum} (Color in online edition) Geometrical
construction of the condition of full transmission in a hole array.
(a) Wavelength dependence of the real part of the lattice sum
$G_{xx}$ [Eq.\ (\ref{Gk})] for $k_\parallel=0$. (b) Normal-incidence
transmittance of a hole array complementary of the disk array of
Fig.\ \ref{particles-T}(a) ($b=a/5$): exact calculation (solid
curve), analytical model of Eq.\ (\ref{r1}) (dashed curve), and Fano
profile of Eq.\ (\ref{Fano}) (dotted curve). The transmission
minimum at $\lambda=a$ results from the divergence of $G_{xx}$,
while the transmission maximum (see vertical dashed line) is derived
from the condition that $\Re\{G_{xx}\}$ equals the inverse of the
hole polarizability, according to Eq.\ (\ref{r1}).}
\end{figure}

The analytical simplicity of the transmission coefficient for our
thin-screen hole array, given by the right hand side of Eq.\
(\ref{r1}), allows us to gain deeper insight into the origin of this
phenomenon. The lattice sum $G_{xx}(0)$ was shown to diverge when
$\lambda=a$, as Fig.\ \ref{maximum} illustrates. This leads to
vanishing transmission, which we can interpret in terms of
accumulation of in-phase scattering (see discussion in Sec.\
\ref{latticeresonances}). Furthermore, 100\% transmission is
achieved if the second term in the denominator of Eq.\ (\ref{r1})
becomes zero, a condition that can be rigorously fulfilled for
arbitrarily tiny apertures \cite{paper102}: the smaller the holes,
the larger $1/\aE$, because the polarizability is proportional to
the cube of their radius, but no matter how large this fraction
becomes, there is always one wavelength at which the divergent
lattice sum matches it. This statement is illustrated by geometrical
construction in Fig.\ \ref{maximum}, in which the point of
intersection of the horizontal dotted line and the solid curve
[Fig.\ \ref{maximum}(a)] signals the condition
$\Re\{1/\aE-G_{xx}(0)\}=0$. \footnote{We rely here on the condition
$\Re\{1/\aE\}>0$, which is satisfied by the polarizability of
planar, perfectly-conducting disks. Interestingly, lattice
resonances will be absent in arrays of particles with negative
polarizability, such as metallic nanoparticles under blue-detuned
illumination relative to a nearby plasmon band.} The possibility of
100\% transmission in non-absorbing structures has been pointed out
before \cite{M1980,M1980_2}, and the theory just presented goes
further to show that this is possible for arbitrarily small holes.
Nevertheless, the number of apertures needed to accomplish high
transmission will increase as they become smaller, and at the same
time the transmission resonance will be increasingly narrower and
closer to $\lambda=a$. Therefore, these transmission maxima involve
long-range interaction among holes, dominated by dynamical
diffraction (i.e., multiple-scattering paths). In fact, if only
single-scattering were considered, Eq.\ (\ref{pp}) would become
$\pb=\aE\,(1+\aE G(\kb_\parallel)\aE) \Eb^{\rm ext}$, which wrongly
predicts simultaneous divergence of transmittance and reflectance at
$\lambda=a$.

This collective response in planar periodic arrays can be regarded
as a lattice surface resonance \cite{F1941}, which becomes a true
surface-bound state when evanescent incoming waves are considered,
as we shall see in Sec.\ \ref{surfacemodes}. However, the resonance
is strongly coupled to propagating light for external plane-wave
illumination, a situation described by Fano \cite{F1961} in his
study of a discrete resonance state (our lattice surface-bound mode)
coupled to a continuum (the transmitted light). This type of
approach has been shown to work rather well in theory
\cite{SVV03,CGS05} and in comparison with measured transmission
spectra \cite{GEW03}. Our transmittance calculations should also
respond to Fano profiles of the form \cite{F1961}
\begin{eqnarray}
T=C\,\frac{(q+\varepsilon)^2}{1+\varepsilon^2}, \label{Fano}
\end{eqnarray}
where $\varepsilon$ can be assimilated to the light frequency and
$q$ describes the strength of the coupling to the lattice surface
resonance. Fig.\ \ref{maximum}(b) compares our exact calculation of
the transmittance (solid curve) with a Fano profile corresponding to
parameters $q=-3$ and $C=0.1$ (dotted curve), in which we assume a
linear relationship between $\varepsilon$ and the light frequency,
with $\varepsilon=-0.33$ ($\varepsilon=3$) for $T=1$ ($T=0$). The
agreement is very reasonable, considering that no dependence of the
coupling parameter on wavelength is taken into account. This further
supports an interpretation of extraordinary transmission in terms of
coupling to the lattice surface resonance set up by dynamical
diffraction in the array.

The geometrical construction of Fig.\ \ref{maximum} provides a
visual explanation of transmission in arrays of elongated apertures:
an elongated piece of planar metal (e.g., a rectangle) has larger
electric polarizability along its long-axis direction, and this has
direct consequences for the Babinet-related situation of an
elongated hole with the electric field along the short axis; larger
polarizability involves more red-shifted and broader transmission
maxima [this is so because the point of crossing in Fig.\
\ref{maximum}(a) occurs where $G_{xx}$ is less steep], just as
observed experimentally \cite{KES04,GBM04}.

Incidentally, Eqs.\ (\ref{pp}) and (\ref{Gk}) constitute a good
approximation to describe the extraordinary transmission observed in
2D quasi-crystal hole arrays \cite{STL06,SPF2006,PGE06,MAN07}, in
which the lattice sum $G$ exhibits pronounced, but finite maxima
related to bright spots in the Fourier transform of the hole
distribution. These spots define the reciprocal lattice for periodic
arrays, but have quasi-crystal angular symmetry in quasi-crystals.
In the spirit of Rayleigh's explanation of Wood's anomalies
\cite{R1907}, the cumulative effect of long-distance interaction
among apertures can be claimed to create these reciprocal-space hot
spots, so that the effect of neighboring holes can be overlooked and
an effective homogeneous $\pb$ describes qualitatively well the
extraordinary transmission effect in quasi-crystal arrays
\cite{SPF2006}, as well as the rich Talbot-like structure and
subwavelength light localization observed at distances up to several
wavelengths away from the array \cite{paper119}.

\subsubsection{Single holes in thick films}
\label{singleholes}

Our use of Babinet's principle in the previous section indicates
that, similar to small particles, small holes in perfect conductors
can be assimilated to equivalent induced dipoles, in line with
Bethe's pioneering description of the field scattered by a single
aperture in a thin screen \cite{B1944}, which he regarded as arising
from a magnetic dipole parallel to the screen plus an electric
dipole perpendicular to it.

\begin{figure}
\includegraphics[width=80mm,angle=0,clip]{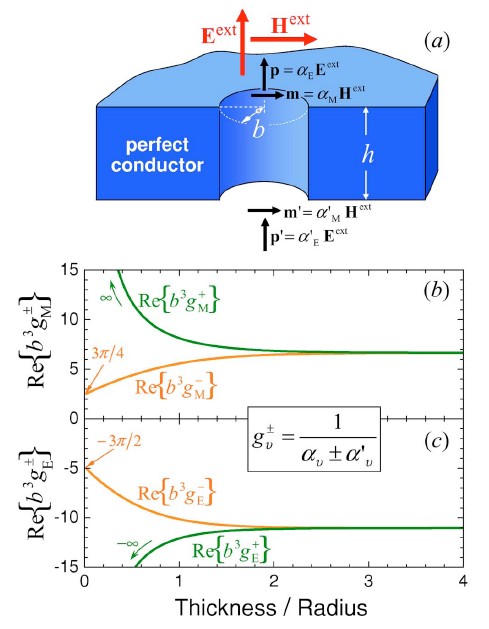}
\caption{\label{thick-hole} (Color in online edition) Response of a
small hole in a perfect-conductor thick film. (a) The field
scattered by a subwavelength aperture in response to external
electric ($E^{\rm ext}$) and magnetic ($H^{\rm ext}$) fields is
equivalent (at large distance compared to the radius $b$) to that of
effective electric ($p$) and magnetic ($m$) dipoles, which allow
defining polarizabilities ($\aE$ and $\aM$, respectively) both on
the same side as the external fields ($\alpha_\nu$) and on the
opposite side ($\alpha'_\nu$). Only the perpendicular component of
the electric field and the parallel component of the magnetic field
can be nonzero at the surfaces of the perfect-conductor film.
(b)-(c) Thickness dependence of the real part of the hole response
functions $g^\pm_\nu$ for $\lambda\gg b$ [the imaginary part
satisfies Eq.\ (\ref{OT})].}
\end{figure}

Narrow holes can still be represented by induced dipoles in thick
screens, as illustrated in Fig.\ \ref{thick-hole}(a). Parallel
electric dipoles and perpendicular magnetic dipoles are forbidden by
the condition that the parallel electric field and the perpendicular
magnetic field vanish at a perfect-conductor surface. This allows
defining electric (E) and magnetic (M) polarizabilities both on the
same side as the applied field ($\alpha_\nu$, with $\nu=$E,M) and on
the opposite side ($\alpha'_\nu$). Furthermore, energy flux
conservation under arbitrary illumination leads to an exact
optical-theorem type of relationship between these polarizabilities
\cite{paper106}: by considering two plane waves incident on either
side of the film and by imposing that the incoming energy flux
equals the outgoing one (because perfect conductors cannot absorb
energy), we obtain the condition
\begin{eqnarray}
  \Im\{g^\pm_\nu\}=\frac{-2 k^3}{3},
\label{OT}
\end{eqnarray}
where we have defined
\begin{eqnarray}
  g^\pm_\nu =\frac{1}{\alpha_\nu\pm\alpha'_\nu} \nonumber
\end{eqnarray}
as hole response functions. The remaining real parts of $g^\pm_\nu$
are obtained numerically from the field scattered by a single hole
\cite{R1987,paper069}. These functions are represented in Fig.\
\ref{thick-hole}(b)-(c) within the electrostatic limit, clearly
showing $|\Re\{g^+_\nu\}|\rightarrow\infty$ in the thin film limit,
where $\alpha'_\nu=-\alpha_ \nu$ \cite{J99}.

\subsubsection{Hole arrays in thick films}
\label{holearraysin}

Periodic arrays of sufficiently narrow and spaced holes can also be
described by perpendicular electric dipoles $p$ and $p'$ and
parallel magnetic dipoles $m$ and $m'$, where primed (unprimed)
quantities are defined on the entry (exit) side of the film, as
determined by the incoming light [see Fig.\ \ref{thick-hole}(a)]. We
consider first a unit-electric-field p-polarized plane wave incident
on a hole array with parallel momentum $\kb_\parallel$ along $\xx$,
so that the external field (incident plus reflected) in the absence
of the apertures has parallel magnetic field $H_y^{\rm ext}=2$ along
the $y$ direction and perpendicular electric field $E_z^{\rm
ext}=-2k_\parallel/k$ along $z$. Then, one can generalize Eq.\
(\ref{pp}) and write a set of multiple-scattering equations for the
self-consistent dipoles \cite{CE1961,EC1962}. Symmetry
considerations demand that our magnetic and electric dipoles be
oriented as $\mb=m\yy$ and $\pb=p\zz$, respectively. Following the
notation of Sec.\ \ref{basicrelations}, we can write
\begin{eqnarray}
p &=& \aE (E_z^{\rm ext} + G_{zz} p - H m) + \apE (G_{zz} p' - H m'), \nonumber \\
p'&=& \apE (E_z^{\rm ext} + G_{zz} p - H m) + \aE (G_{zz} p' - H m'), \nonumber \\
m &=& \aM (H_y^{\rm ext} + G_{yy} m - H p) + \apM (G_{yy} m' - H p'), \nonumber \\
m'&=& \apM (H_y^{\rm ext} + G_{yy} m - H p) + \aM (G_{yy} m' + H p'), \nonumber \nonumber
\end{eqnarray}
with a new lattice sum defined as
\begin{eqnarray}
H&=&-\ii k\sum_{n\neq 0} \ee^{-\ii k_\parallel
x_n}\partial_{x_n}\frac{{\rm e}^{{\rm i} kR_n}}{R_n}. \nonumber
\end{eqnarray}
This sum stands for the interaction between mixed electric and
magnetic dipoles. We can understand the above equations in a very
intuitive way; for instance, the first one of them states that the
electric dipole on the entry side ($p$) results from the response to
the $z$-component of the self-consistent field on that side
($E_z^{\rm ext} + G_{zz} p - H m$) via the polarizability $\aE$ plus
the response to the self-consistent field on the opposite film side
($G_{zz} p' - H m'$) via $\apE$. The solution to these equations can
be readily written as
\begin{eqnarray}
p\pm p' &=& -2 [(g_{\rm M}^\pm-G_{yy}) k_\parallel/k+H]/\Delta_\pm, \label{pp1} \\
m\pm m' &=&  2 [(g_{\rm E}^\pm-G_{zz}) +H k_\parallel/k]/\Delta_\pm,
\label{mm1}
\end{eqnarray}
with
\begin{eqnarray}
\Delta_\pm=(g_{\rm E}^\pm-G_{zz})(g_{\rm M}^\pm-G_{yy})-H^2.
\nonumber
\end{eqnarray}
The zero-order transmittance of the holey film is then obtained from
the far field set up by the infinite 2D array of induced dipoles,
$T_p=|(2\pi k^2/Ak_z)(m'-p'k_\parallel/k)|^2$, where
$k_z=\sqrt{k^2-k_\parallel^2}$.

Similar considerations for s-polarized light show that its
transmittance reduces to $T=|2\pi k m'/A|^2$, with magnetic dipoles
parallel to $\kb_\parallel$ and no electric dipoles whatsoever
($E^{\rm ext}_z=0$). More precisely, $m\pm m'=(2k_z/k)/(g_{\rm
M}^\pm-G_{xx})$, from which one obtains
\begin{eqnarray}
&T_s&=\left(\frac{2\pi k_z}{A}\right)^2 \left|\frac{1}{g^+_{\rm
M}-G_{xx}} -
\frac{1}{g^-_{\rm M}-G_{xx}}\right|^2 \label{Ts} \\
&=& \left|\frac{1}{1+\frac{\ii A}{2\pi k_z} \Re\{g^+_{\rm
M}-G_{xx}\}} - \frac{1}{1+\frac{\ii A}{2\pi k_z} \Re\{g^-_{\rm
M}-G_{xx}\}}\right|^2 \nonumber
\end{eqnarray}
for the transmittance. The last identity in Eq.\ (\ref{Ts}) comes
from Eqs.\ (\ref{OT2}) and (\ref{OT}) for diffractionless arrays.

Interestingly, Eq.\ (\ref{Ts}) predicts 100\% transmission if
\begin{eqnarray}
1+\left(\frac{A}{2\pi k_z}\right)^2 \, \Re\{g^+_{\rm M}-G_{xx}\} \,
\Re\{g^-_{\rm M}-G_{xx}\}=0. \label{maxcond}
\end{eqnarray}
This is a second-order algebraic equation in $\Re\{G_{xx}\}$ that
admits positive real solutions provided
\begin{eqnarray}
  \frac{A}{4\pi k_z}\left|g^+_M-g^-_M\right| \ge 1.
\label{T1cond2}
\end{eqnarray}
Actually, $\Re\{G_{xx}\}$ can match those roots near the $l\neq 0$
singularities of Eq.\ (\ref{div}), where it can be chosen
arbitrarily large within a narrow range of wavelengths [see Eq.\
(\ref{Gana})]. It should be noted that the difference $g^+_M-g^-_M$
falls off rapidly to zero when the film thickness $h$ is made much
larger than the hole radius $b$ [see Fig.\ \ref{thick-hole}(b)].
However, if we fix both the $h/b$ ratio and the angle of incidence,
the left hand side of (\ref{T1cond2}) reduces to a positive real
constant times $\lambda A/b^3$, leading to the conclusion that 100\%
transmission is attainable at a wavelength close to the Rayleigh
condition (e.g., $\lambda\gtrsim a$ for normal incidence on a square
lattice of spacing $a$) regardless how narrow the holes are as
compared to the film thickness. Surprisingly, this requires that the
ratio of the lattice constant to the hole radius be increased for
deeper holes in order to compensate the fall in $g^+_M-g^-_M$ for
larger $h/b$.

\begin{figure}
\includegraphics[width=80mm,angle=0,clip]{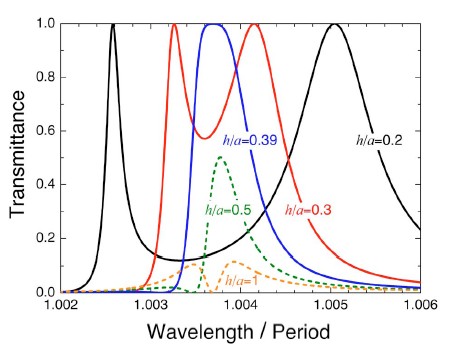}
\caption{\label{T-thick} (Color in online edition) Thickness
dependence of the normal-incidence transmittance spectra of square
arrays of circular holes drilled in perfect-conductor films,
according to Eq.\ (\ref{Ts}). The hole radius $b=0.2a$, the
wavelength $\lambda$, and the film thickness $h$ are given relative
to the period $a$ (see text insets).}
\end{figure}

The transmittance shows an interesting dependence on film thickness
$h$ \cite{MGL01}, as illustrated in Fig.\ \ref{T-thick}. The maximum
of Fig.\ \ref{maximum} is initially blue-shifted closer to
$\lambda=a$ for small $h$, accompanied by a second narrower peak at
even shorter wavelengths \footnote{In fact, there are two lattice
resonances for $h=0$, which in the language of Fano arise from
coupling to different light continua on either side of the film, but
one of these resonances has vanishing width and is placed at
$\lambda=a$ due to strong inter-side interaction.} [these are the
two solutions of Eq.\ (\ref{maxcond}) under the condition
(\ref{T1cond2})]. As $h$ increases, inter-side interaction weakens
and the two 100\% maxima approach each other. At some point only one
transmission maximum is observed when the left hand side of
(\ref{T1cond2}) is exactly 1. For even thicker films, the condition
(\ref{T1cond2}) cannot be met any longer and the transmission
maximum departs from 100\%. The Fano character of these lattice
resonances is again visible through vanishing transmission at a
wavelength immediately below the maximum.

Incidentally, perfect conductors are perfectly non-lossy, so that
light dissipation must take place only at the openings if they are
infiltrated with some dissipative material. For deep enough holes,
the transmission is negligible and the absorbance becomes $1-|r|^2$,
which can reach 100\% values under suitable resonant conditions, for
instance in the IR by combining holes drilled in noble metals
(behaving nearly as perfect conductors) infiltrated with
phonon-polariton materials. In fact, a similar effect has been
observed in the visible using Au gratings \cite{HM1976} and in the
infrared using SiC gratings \cite{GCJ02}.

\subsection{Lattice surface modes in structured metals}
\label{surfacemodes}

\begin{figure}
\includegraphics[width=80mm,angle=0,clip]{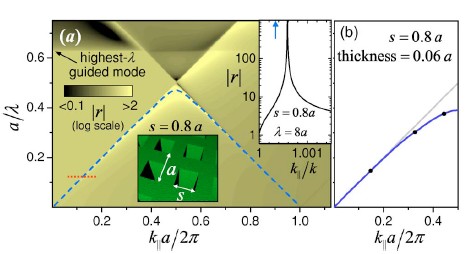}
\caption{\label{surface-mode} (Color in online edition) (a) Lattice
surface modes in a perforated semi-infinite perfect-conductor. The
contour plot shows the modulus of the specular reflection
coefficient for incident p-polarized light as a function of
wavelength $\lambda$ and parallel momentum $k_\parallel$ (see text
insets for parameters). The upper-right inset shows a detail of the
reflectivity as compared to the mode position predicted by Eqs.\
(\ref{eqPendry}) and (\ref{gamma}) (see arrow). A reflection
coefficient larger than 1 is only possible for evanescent waves
outside the light cone. (b) Lattice modes in a perforated thin film,
as measured by \textcite{UT1972} (symbols).}
\end{figure}

The flourishing area of plasmonics is demonstrating how confining
electromagnetic fields to a surface can find many potential
applications on the nanoscale \cite{O06}. Zenneck waves at radio
frequencies \cite{Z1907,B1958}, phonon-polaritons in the infrared
\cite{GCJ02,HTK02}, and plasmons in the visible are in fact
different manifestations of the same phenomenon: confinement of
electromagnetic fields to curved or planar surfaces. Even
perfect-conductor screens, which are unable to trap light when they
are flat, were experimentally shown by \textcite{UT1972} to host
confined surface modes of p polarization when molded into films
pierced by periodic arrays of holes spaced a distance much smaller
than the wavelength [see Fig.\ \ref{surface-mode}(b)].

In a recent independent development, \textcite{PMG04} have studied
surface modes in drilled semi-infinite metal, suggesting the
possibility to extend plasmon-like behavior to lower-frequency
domains via the flattening of the mode dispersion relation driven by
propagating modes of the holes, and stimulating new microwaves
observations \cite{HES05}. The analysis of \textcite{PMG04} relied
on a description of the holes based upon their lowest-order guided
modes (i.e., TE$_{1,0}$ modes), which allowed extracting local
permittivity and permeability functions in a metamaterial approach
to holey metals. However, \textcite{paper105} showed later that
higher-order modes (and in particular TM modes) are important,
giving rise to large quantitative modifications to the dispersion
relation and revealing finer details in the holey metal response
that go beyond a simple local metamaterial description (e.g., the
angular dependence of the reflection coefficient does not follow the
Fresnel equations with local optical constants).

At variance with planar perfect conductors and their lack of surface
modes, corrugated metallic surfaces can support bound states even in
the long-wavelength limit. In an intuitive picture, surface
confinement in a drilled semi-infinite perfect conductor can be
related to the evanescent penetration of the electromagnetic field
inside the holes, in much the same way as surface plasmons enter a
distance of the order of the skin depth inside a metal in the
visible and NIR regimes \cite{BS04}. Actually, these modes share
with plasmons their character of p-polarized evanescent waves.

Next, we elaborate a tutorial, analytical formulation of this
phenomenon that becomes exact in the limit of small holes of size
$s\ll a\ll\lambda$, arranged in a lattice of period $a$
\cite{paper105}. Although we focus our analysis on periodic hole
arrays drilled in a semi-infinite perfect-conductor, it must be
emphasized that periodicity is not really needed and that similar
modes should exist for patterns other than holes (e.g., small
protuberances or particles deposited on an otherwise flat surface).

Using the formalism of Sec.\ \ref{holearrays}, we find that Eqs.\
(\ref{pp1}) and (\ref{mm1}) offer a simple description of lattice
surface-bound modes in metallic films. For infinitely-deep square
holes as sketched in an inset of Fig.\ \ref{surface-mode}(a), the
surface modes must correspond to non-vanishing values of the induced
dipoles $p$ and $m$ in the absence of external fields. This can only
be accomplished if the denominator $\Delta_\pm$ is zero in those
equations, leading to
\begin{eqnarray}
(1/\aE-G_{zz})(1/\aM-G_{yy})=H^2, \label{dispersion}
\end{eqnarray}
where we have set $\alpha'_\nu=0$ for infinitely deep holes (see
Fig.\ \ref{thick-hole}). The interaction sums $G_{yy}$, $G_{zz}$,
and $H$ are generally small for $s\ll a$, except near the lattice
singularities discussed in Sec.\ \ref{latticeresonances}. In
particular, near the light line for $k_\parallel\gtrsim k$, one has
\begin{eqnarray}
\Re\{G_{zz}\}\approx \Re\{G_{yy}\}\approx \Re\{H\}\approx \frac{2\pi
k^2}{k_z a^2}, \nonumber
\end{eqnarray}
which corresponds to Eq.\ (\ref{div}) with $n=l=0$. Furthermore,
upon inspection of an expansion for $H$ similar to (\ref{GG}), we
find $\Im\{H\}=0$ outside the light cone, $k_\parallel>k$, and the
remaining imaginary parts of all quantities in Eq.\
(\ref{dispersion}) cancel out exactly because
$\Im\{G_{jj}\}=\Im\{\alpha_\nu^{-1}\}=-2k^3/3$ in that region.
Combining these results, we obtain an approximate long-wavelength
dispersion relation from Eq.\ (\ref{dispersion}):
\begin{eqnarray}
k_\parallel^2=k^2 + \Gamma \, \frac{S^3 k^4}{a^4} \label{eqPendry}
\end{eqnarray}
with
\begin{eqnarray}
\Gamma=\frac{4\pi^2}{S^3}
\left(\frac{1}{\Re\{1/\aE\}}+\frac{1}{\Re\{1/\aM\}}\right)^2.
\label{gamma}
\end{eqnarray}
Eq.\ (\ref{gamma}) is exact in the $s\ll a\ll\lambda$ limit, and it
predicts the existence of lattice surface-bound modes under the condition
$1/\Re\{1/\aE\}+1/\Re\{1/\aM\}>0$. Here, we have used the area of
the holes $S$ to make $\Gamma$ dimensionless.

\begin{figure}
\includegraphics[width=80mm,angle=0,clip]{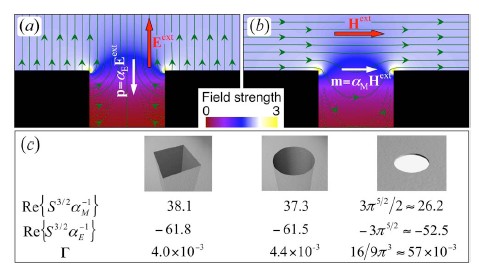}
\caption{\label{deep-hole} (Color in online edition) (a)
Electrostatic electric-field flow lines for a circular hole drilled
in a semi-infinite perfect-conductor subject to an external field
$\Eb^{\rm ext}$ perpendicular to the surface, giving rise to an
electric dipole $p=\aE\,E^{\rm ext}$ as seen from afar. (b)
Magnetostatic magnetic-field flow lines for the same hole subject to
an external parallel field $\Hb^{\rm ext}$ and leading to a magnetic
dipole $m=\aM\,H^{\rm ext}$. (c) Summary of polarizabilities for
square and circular holes in perfect-conductor surfaces, normalized
using the aperture area $S$. The values for the circular hole are
taken from the $h\gg b$ limit of Fig.\ \ref{thick-hole}. The
circular opening in a thin screen is analytical \cite{B1944,J99},
but we must correct the right-hand side of Eq.\ (\ref{gamma}) by a
factor of 4 in this case because of cooperative interaction between
both sides of the film.}
\end{figure}

Calculated values of $\Gamma$ are offered in Fig.\
\ref{deep-hole}(c) for various hole geometries. The polarizability
$\aE$ ($\aM$) is obtained from the electrostatic (magnetostatic)
far-field induced by an external electric (magnetic) field, as shown
in Fig.\ \ref{deep-hole}(a) [Fig.\ \ref{deep-hole}(b)].
Interestingly, circular and square openings of the same area give
rise to similar values of $\Gamma$. This parameter increases by an
order of magnitude when the holes are made on thin screens instead
of semi-infinite metals, producing lattice surface modes that are
further apart from the light line (see \onlinecite{UT1972}), and
therefore, more confined to the metal, as a result of cooperative
interaction between both sides of the film [see analytical solutions
for circular apertures \cite{J99} in last column of Fig.\
\ref{deep-hole}(c)]. Another suggestive possibility is offered by
split annular holes, which present resonant electric polarizability
\cite{FLL04}, and by holes filled with high-permittivity materials
(see Sec.\ \ref{interplaybetween}), for which the interaction with
single-hole modes produces large departures of the extended surface
states from the grazing light condition.

Fig.\ \ref{surface-mode}(a) shows calculated results for the
reflection coefficient of a drilled metal, obtained by rigorous
solution of Maxwell's equations in which we use a plane-wave
expansion of the field outside the metal and a guided-mode expansion
inside the holes \cite{paper105}. The lattice surface mode can be
observed as a bright region with a dashed line showing the position
at which the reflection coefficient becomes infinite. A detail of
$|r|$ for a specific wavelength (see dotted straight line) is shown
in the inset. The position of the resonance predicted by Eqs.\
(\ref{eqPendry}) and (\ref{gamma}) (see arrow in the inset) is in
reasonably close agreement with the exact calculation, considering
that the analytical model neglects neighboring-holes multipolar
interaction, which is important for openings occupying 64\% of the
surface. Finally, Fig.\ \ref{surface-mode}(b) shows experimental
results for a drilled thin film obtained by \textcite{UT1972}. These
surface modes are more bound in perforated thin films than in
semi-infinite metals, as can be seen from the values of $\Gamma$
given in Fig.\ \ref{deep-hole}(c). Actually, the measured dispersion
relation departs substantially from the light line close to the
boundary of the first Brillouin zone.

\subsection{Interplay between lattice and site resonances}
\label{interplaybetween}

\begin{figure}
\includegraphics[width=80mm,angle=0,clip]{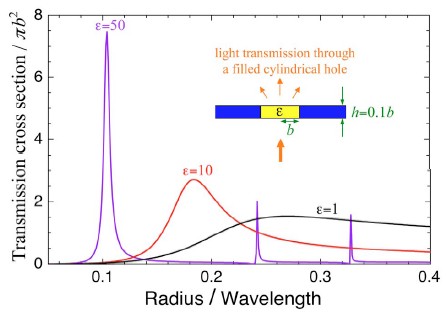}
\caption{\label{site-resonances} (Color in online edition) Enhanced
transmission driven by a localized resonance. The plot shows the
normal-incidence transmission of a circular aperture drilled in a
perfect-conductor film and filled with dielectric material for
different values of the permittivity $\epsilon$ (see labels). The
transmitted power is normalized to the incoming flux within the hole
area.}
\end{figure}

The description of extraordinary optical transmission in terms of
quasi-bound surface states driven by lattice singularities can be
extended to other types of binding. In particular, a single hole
filled with a dielectric of high permittivity can trap light in its
interior, giving rise to cavity modes even for very subwavelength
apertures, provided the permittivity is sufficiently large to shrink
the wavelength inside the dielectric to a value comparable to the
diameter of the hole. This concept is explored in Fig.\
\ref{site-resonances}, in which higher permittivities are seen to
produce larger contraction of the wavelength inside the hole, so
that the cavity mode condition is met at longer free-space
wavelengths for fixed aperture size \cite{paper069,GMP05}. This
process is accompanied by weaker coupling to external light (due in
part to higher reflectivity of the dielectric-air interface), and
therefore, narrower transmission resonances of increasingly larger
height. Original predictions of this effect \cite{paper069} have
been recently corroborated by experiment using microwaves
\cite{paper106}.

\begin{figure}
\includegraphics[width=80mm,angle=0,clip]{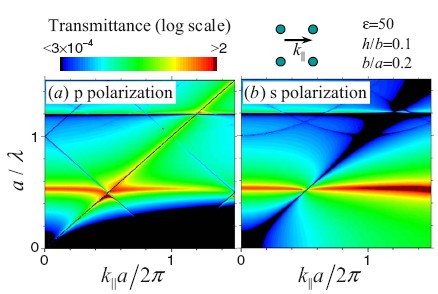}
\caption{\label{interplay} (Color in online edition) Interplay
between localized (site) and extended (lattice) resonances. The
contour plots show the zero-order beam transmittance of a square
array of circular holes drilled in a perfect-conductor film and
filled with dielectric material of permittivity $\epsilon=50$ as a
function of parallel momentum $\kb_\parallel$ and wavelength
$\lambda$. The orientation of $\kb_\parallel$ and the ratios between
the hole radius $b$, the lattice constant $a$, and the film
thickness $h$ are specified in the insets. The light is p polarized
in (a) and s polarized in (b). A transmission coefficient larger
than 1 is only possible for evanescent waves below the light cone.}
\end{figure}

An interesting situation is presented when localized modes like the
ones just described are mixed with extended lattice modes, like the
surface states underlying extraordinary optical transmission
\cite{paper106,RQ06}. The interplay between both types of modes is
illustrated in Fig.\ \ref{interplay} through the zero-order
transmittance of hole arrays filled with high-permittivity
dielectric, calculated from the formalism presented in Sec.\
\ref{holearraysin}. All incident-light polarizations interact with
the cavity modes, giving rise to omnidirectional extraordinary
transmission and invisibility behavior near the individual hole
resonance \cite{paper096,paper100}. However, only p-polarized light
couples to the $n=1$, $l=0$ lattice singularity of Fig.\
\ref{lattice-resonances}, which results in an avoided crossing of
the hybridized modes [Fig.\ \ref{interplay}(a)]. Similar avoided
crossings have been recently found in microwave experiments
\cite{HLH06}, confirming lattice surface modes and localized modes
as two distinct mechanisms leading to enhanced
transmission.\footnote{In a related context, avoided crossing of
lattice modes are well-known to occur in coinciding Wood anomalies
\cite{SG1962}.}$^,$\footnote{Incidentally, lattice modes are
observed outside the light cone for p polarization. The transmission
outside that cone is defined as the squared-amplitude ratio of
incident and transmitted evanescent waves at the exit and entrance
surfaces of the film, respectively.} Notice that s-polarized light
is immune to the $l=0$ lattice singularities of Fig.\
\ref{lattice-resonances}, and this results in a reduced number of
transmission features as compared to p polarization, in qualitative
agreement with experimental observations \cite{BMD04}.

Site resonances can occur in coaxial waveguides as well, via the
so-called TEM mode, which does not have a cutoff in wavelength
\cite{J99}. This led \textcite{RM1988} to theoretically explore the
performance of periodic annular-hole arrays as band filters. More
recently, \textcite{FZM05} have measured the increased transmission
of infrared light assisted by these modes. Similar coupling to
localized TEM modes occurs as well in slits, as we shall see in
Sec.\ \ref{slitarrays}.

The type of interplay phenomenon that we are describing has been
observed as well for localized and extended surface plasmons in the
visible regime through the absorption features of porous metals, in
which Mie modes of spherical cavities in otherwise planar surfaces
display a rich structure of hybridization and avoided crossings
\cite{KSB05,KSC06,B06_2,paper109,paper117}. The absorption can be
even complete under attainable experimental conditions
\cite{paper097}, implying black-body-like emission according to
Kirchhoff's laws of thermal radiation \cite{R1965}.

\subsection{Slit and cylinder arrays}
\label{slitarrays}

Although we have extracted conclusions for particles and holes from
his works, Wood reported his anomalies for ruled gratings rather
than 2D structures \cite{W1902,W1935}.\footnote{The reader is
referred to the papers collected by \textcite{M93} for an exciting
historical overview of XX century milestones on gratings.} In fact,
like gratings, cylinder and slit arrays exhibit lattice-resonance
phenomena. But in contrast to holes, a single arbitrarily-narrow
slit in a perfect conductor supports at least one guided wave, the
TEM mode \cite{J99}, which can couple to external p-polarized light
(magnetic field parallel to the slit) giving rise to recently
predicted \cite{T01} and observed \cite{YS02} Fabry-P\'erot
resonances in transmission. As a consequence, light passage through
slit arrays can be assisted either by coupling to the TEM mode or by
lattice resonances for p polarization \cite{PGP99}, leading to
similar interplay between localized and extended resonances as
discussed above \cite{MGC05}. Incidentally, the analogy with annular
hole arrays is clear (see Sec.\ \ref{interplaybetween}).


We shall consider first a periodic array of parallel narrow
cylinders, the axes of which define a single plane. Continuing with
our tutorial approach, and focusing for simplicity on light incident
with its electric field parallel to the cylinders, we note that
Eqs.\ (\ref{eq1})-(\ref{Gk}) are still applicable here, provided
$\aE$ and $\mathcal{G}^0$ are conveniently redefined. In particular,
the polarizability has now dimensions of area rather than volume,
and it is given for instance by $\alpha_{\rm E}^{\rm es}=\pi
b^2(\epsilon-1)$ for homogeneous cylinders of radius $b$ and
permittivity $\epsilon$ \cite{BH98}, with the optical theorem now
leading to $\Im\{1/\alpha_{{\rm E}}\}=-k^2/4$. The relevant
dipole-dipole interaction component is given by the Green function
of Helmholtz equation in two dimensions, $\mathcal{G}^0=(\ii k^2/4)
H_0^{(1)}(kR)$, where $R$ is the distance measured in a plane
perpendicular to the cylinders and $H_0^{(1)}$ is a Hankel function
\cite{AS1972}. Then, proceeding with the lattice sum
$G(\kb_\parallel)$ in a way analogous to Eq.\ (\ref{GG}), one finds
a relation similar to Eq.\ (\ref{r1}) for the reflection coefficient
of an array of lossless cylinders:
\begin{eqnarray}
r=\frac{-1}{1+\frac{2\ii a}{k}\Re\{1/\aE-G(0)\}}. \nonumber
\end{eqnarray}
Under normal incidence ($k_\parallel=0$), $G$ is found to diverge as
\begin{eqnarray}
G(0)\approx\frac{\pi}{a^2\sqrt{2}}\frac{1}{\sqrt{\lambda/a-1}}
\nonumber
\end{eqnarray}
for $\lambda\gtrsim a$, where $a$ is the lattice period. This is
similar to particle arrays [see Eq.\ (\ref{Gana})], so that the main
conclusions from our previous discussion of those arrays apply here
as well, and more precisely, the reflectivity can be made 100\% for
arbitrarily narrow or weakly-scattering ($\epsilon\gtrsim 1$)
cylinders.

A complete analysis along these lines has been recently reported for
all possible incident polarizations \cite{GLS06,LAG06}, suggesting
that similar lattice resonances, somewhat less pronounced, are
obtained for $\Eb^{\rm ext}$ perpendicular to the cylinders and with
non-vanishing projection normal to the plane of the array. However,
polarization components parallel to that plane and perpendicular to
the cylinders cannot generate lattice resonances, because the
interaction between distant dipoles aligned with their separation
vector $\Rb$ decays as $1/R^{3/2}$ in 2D, which is insufficient to
produce a divergence in $G$. \footnote{This is because
$\sum_{n=1}^\infty 1/n^{3/2}$ is finite. See also Sec.\
\ref{latticeresonances}.}

Finally, we can establish a relation between cylinder arrays and
slit arrays using arguments similar to those of Sec.\
\ref{thinscreens} for particle and hole arrays. More precisely, a
slit array cut into a thin metal screen and illuminated with
$\Eb^{\rm ext}$ perpendicular to the apertures can be analyzed using
the above results as applied to the Babinet-related stripe array
(i.e., a periodic array of stripes laying on a single imaginary
plane) for $\Eb^{\rm ext}$ parallel to the stripes. Under normal
incidence, the required component of the polarizability reads
$\aE\approx-2\pi/k^2[\ln(kb/8)+\gamma+\ii\pi/2]$, where
$\gamma=0.57721$ is the Euler constant and $b\ll\lambda$ is the
stripe width \cite{V1981}. Interestingly, $\aE$ diverges in the
electrostatic limit, so that even a single narrowing slit will
exhibit a divergent transmission cross section. This scenario can be
traced back to the abovementioned site resonances produced by the
TEM mode of slits in thick screens. As a consequence, the
interaction between slits can be very large, resulting in strong red
shifts of the transmission peaks relative to the Rayleigh condition.

\section{Real Metals versus Perfect Conductors}
\label{realmetals}

Metals of finite conductivity show significant differences with
respect to the perfect conductors considered so far, the most
remarkable of which is the existence of intrinsic surface-plasmon
excitations. The basic understanding of these differences were laid
out by \textcite{M1972} in the context of diffraction gratings (see
also \onlinecite{MM1974}, and \onlinecite{M1984}). Next, we shall
examine (in a tutorial fashion) the consequences for the interaction
between particles and holes decorating metal surfaces.

\subsection{Surface plasmons}
\label{surfaceplasmons}

Conduction electrons in metals behave like a plasma that is capable
of sustaining collective oscillations known as plasmons (e.g.,
longitudinal bulk modes, signalled by the vanishing of the
dielectric function). The existence of genuine surface plasmon
oscillations was predicted by \onlinecite{R1957}, and soon after
confirmed by electron energy-loss experiments \cite{PS1959}. Since
then, surface plasmons have developed into the rapidly growing field
of plasmonics \cite{BDE03,O06,ZSC06} owing to their potential
applicability to areas as diverse as biosensing \cite{SSA93}, signal
processing through plasmonic circuits \cite{BVD06}, or laser
technology \cite{CST03}.

\begin{figure*}
\includegraphics[width=140mm,angle=0,clip]{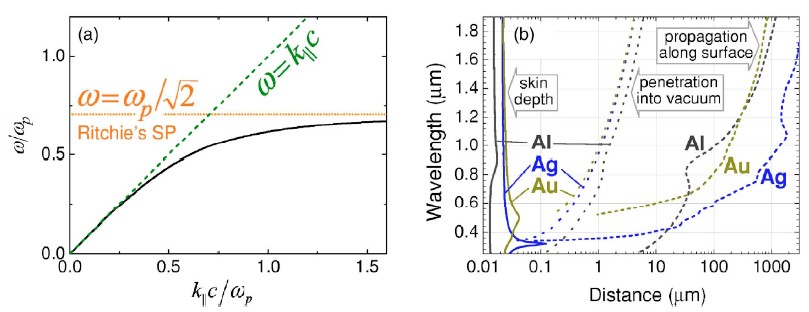}
\caption{\label{SP} (Color in online edition) (a) Surface plasmon
dispersion relation for a Drude metal of bulk plasmon frequency
$\omega_p$. (b) Extension of the plasmon field into the metal (skin
depth), into the vacuum, and along the surface (propagation
distance) for several metals, as obtained from measured optical
constants \cite{JC1972,P1985}.}
\end{figure*}

Planar surfaces possess translational invariance that provide
plasmons with well-defined parallel momentum $k_\parallel$
exceeding that of light outside the metal and thus becoming truly
surface-bound modes. Their dispersion relation can be readily
derived from the divergence of the Fresnel coefficients for p
polarization (surface-bound fields without external sources),
leading to \cite{R1988}
\begin{eqnarray}
\ksp=k\sqrt{\frac{\epsilon}{\epsilon+1}} \label{ksp}
\end{eqnarray}
for a metal-air interface. This surface plasmon dispersion relation
is represented in Fig.\ \ref{SP}(a) for a Drude metal described by
Eq.\ (\ref{Drude}). In the long $k_\parallel$ limit, the surface
plasmon frequency saturates to Ritchie's non-retarded plasmon
\cite{R1957}.

Surface plasmons are characterized by three different length scales,
as depicted in Fig.\ \ref{SP}(b): their propagation distance along
the surface ($\sim 1/2\Im\{\ksp\}$), their penetration into the
surrounding medium ($\sim 1/2\Im\{k_\perp\}$, where
$k_\perp=-k/\sqrt{\epsilon+1}$ is the normal momentum), and their
penetration into the metal (the skin depth $\sim 1/2\Im\{-\epsilon
k_\perp\}$). Interestingly, the interaction between plasmons in
either sides of a thin film gives rise to two plasmon branches, as
measured by electron microscopy \cite{VS1973,PSV1975}, one of which
has been found to propagate along very long distances thanks to
exclusion of the electric field from the metal \cite{S1981}. Well
defined plasmons require to have
$\Im\{\epsilon\}\ll\Re\{-\epsilon\}$, but similar long-range
surface-exciton polaritons exist in thin films for
$\Im\{\epsilon\}\gg|\Re\{\epsilon\}|$ \cite{YSB1990}.

Features in metal surfaces produce scattering of plasmons in a
similar way as light is dispersed by particles. This is actually a
way to couple externally incident light to plasmons, for instance
using gratings \cite{RAC1968,LMM1984}. We find a neat demonstration
of these ideas in the observation of surface-plasmon bands for
periodic surface decoration \cite{SG1962,RAC1968,KBS96} and in the
reflection of surface plasmons at point scatterers arranged as
parabolic mirrors \cite{NOY05}. Similarly, holes perforating films
have strong influence on surface plasmons, which play an important
role in their optical transmission \cite{GTG98}. However, in the
perfect-conductor limit, with $|\epsilon|\rightarrow\infty$, Eq.\
(\ref{ksp}) yields $\ksp=k$, with zero skin depth and infinite
penetration into the vacuum, that is, there are no longer
surface-bound modes. In the following we shall explore the
transition between plasmonic and perfect-conductor regimes, in an
attempt to clarify seemingly contradictory statements regarding the
role of surface plasmons to enhance \cite{SH98_2} or to suppress
\cite{CL02} extraordinary optical transmission in striped thin
films, or the heated debate opened by the explanation of recent
outstanding experiments dealing with the interaction between a slit
and a groove \cite{GAD06,GRM06,LH06}.

\subsection{Polarization schemes}
\label{polarizationschemes}

The condition that parallel electric dipoles and perpendicular
magnetic dipoles are excluded from perfect-conductor surfaces (see
Fig.\ \ref{deep-hole}) is relaxed in metals of finite conductivity.
Polarization charges in a hole for instance can lead to a net
parallel electric dipole in a thin metallic film \cite{paper118}.

\begin{figure}
\includegraphics[width=80mm,angle=0,clip]{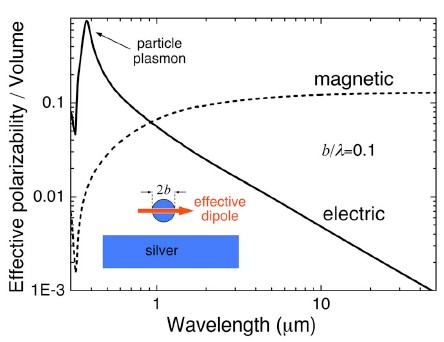}
\caption{\label{polarization} (Color in online edition) Effective
polarization strength of a silver sphere near a silver planar
surface. The sphere radius is a tenth of the wavelength. The
polarization is normalized to the sphere volume. The dielectric
function of silver is taken from \onlinecite{JC1972}.}
\end{figure}

In order to illustrate this concept, we have considered in Fig.\
\ref{polarization} the effective polarizability of a silver
spherical particle in front of a silver surface for a constant ratio
of the radius to the wavelength, $b/\lambda=0.1$. We can observe an
electric Mie mode \cite{M1908} in the visible, accompanied by
negligible magnetic response. However, the metal behaves
increasingly closer to a perfect conductor at longer wavelengths, so
that currents compete eventually with polarization, thus displaying
magnetic polarizability that becomes $\aM=-b^3/2$ for an isolated
perfect-conductor sphere in the long-wavelength limit \cite{J99}, to
be compared with the electric polarizability $\aE=b^3$.
Nevertheless, the latter is quenched by proximity of the metal flat
surface under normal-incidence illumination conditions. The onset of
magnetic response occurs when the particle becomes large compared to
the skin depth $\sim 20$ nm [see Fig.\ \ref{SP}(b)]. These results
follow from dipolar Mie scattering, conveniently corrected by
surface reflection coefficients, which qualitatively describe the
polarizability strength of the coupled particle-surface system.

This has important consequences for understanding patterned surfaces
and hole arrays. Electric dipoles dominate the response of features
smaller than the skin depth, whereas magnetic dipoles can be
significant for larger sizes, and only parallel electric dipoles and
perpendicular magnetic dipoles survive in the limit of negligible
skin depth. We are of course restricting our discussion to particles
or apertures that are small compared to the wavelength, but these
conclusions can be generalized to higher-order multipoles for bigger
features.

\subsection{Dipole-dipole interaction}
\label{dipoledipole}

\begin{figure*}
\includegraphics[width=120mm,angle=0,clip]{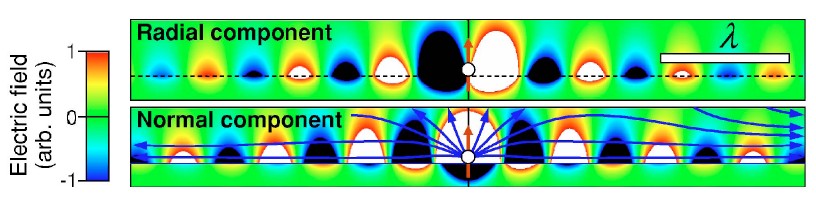}
\caption{\label{SP-maps} (Color in online edition) Instantaneous
electric field set up by a perpendicular electric dipole (see
vertical arrows) sitting at distance $\lambda/20$ from the surface
of a metal described by Eq.\ (\ref{Drude}) with $\omega_p=15$ eV and
damping $\eta=0.6$ eV (typical of Al) at frequency
$\omega=\omega_p/2$. The electric-field component parallel to the
surface (this is radial with respect to the position of the dipole)
and the component along the surface normal are represented
separately. Poynting vector flow lines are superimposed on the plot
of the normal component.}
\end{figure*}

New dipole orientations and the presence of surface plasmons in real
metals demand that we revisit the interaction between features in
tailored surfaces. In particular, the dipolar field in free space,
which decays away from the source as
\begin{eqnarray}
\mathcal{G}^0 \sim \frac{\ee^{\ii k R}}{R} \label{asym}
\end{eqnarray}
and governs the interaction between small features in
perfect-conductor surfaces (see Sec.\ \ref{basicrelations}), must be
supplemented by reflected fields near real metals, leading to an
interaction tensor of the form
\begin{eqnarray}
\mathcal{G}=\mathcal{G}^0+\mathcal{G}^{r}. \nonumber
\end{eqnarray}
As a result, light impinging on a hole can couple to circular
surface-plasmon waves \cite{W01,YVR04,CGS05,PBN05}, whose field
strength shows a rather different decay dependence with distance as
\begin{eqnarray}
\mathcal{G} \sim \frac{\ee^{\ii \ksp R}}{\sqrt{R}}. \label{asymsp}
\end{eqnarray}
This expression is consistent with energy flux conservation for any
surface-bound mode,\footnote{The Poynting vector produced by a
dipole when the fields are propagated by means of Eq.\
(\ref{asymsp}) dies off as $1/R$, if we neglect the attenuation
produced by $\Im\{\ksp\}$. Then, the integral of the radial Poynting
vector over a circle of radius $R$ centered around the dipole and
lying on the surface is independent of $R$, so that the photon flux
is conserved, indicating that we are dealing with surface-bound
propagation.} with dissipation described through the imaginary part
of $\ksp$. The slow drop of Eq.\ (\ref{asymsp}) with distance
compared to Eq.\ (\ref{asym}) can explain the observed enhancement
of the interaction between small particles in plasmonic metals
\cite{SH98}, and it is illustrated in Fig.\ \ref{SP-maps}, showing
the field produced by a dipole near a metallic surface as calculated
from a trivial extension of our tutorial approach formalism
presented below.

\begin{figure}
\includegraphics[width=80mm,angle=0,clip]{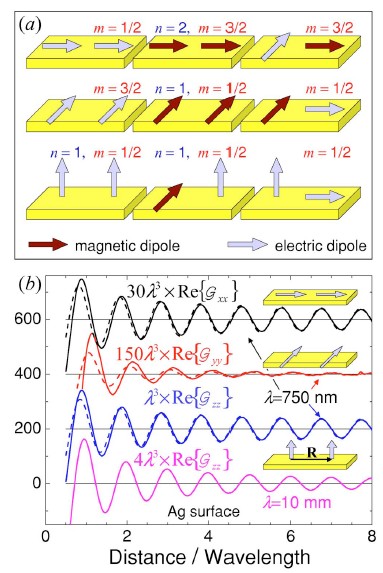}
\caption{\label{dipole-dipole} (Color in online edition) (a)
Schematic representation of the scaling of dipole-dipole
interactions for electric and magnetic dipoles with respect to their
separation $R$ near a metallic surface. The interaction decays as
$\exp(\ii kR)/R^n$ near a perfect conductor or as $\exp(\ii\ksp
R)/R^m$ near a metal with a dominant surface plasmon (see text
insets for values of the exponents $n$ and $m$). (b) Dipole-dipole
interaction near a silver surface at a wavelength of 750 nm (three
upper solid curves) as compared with the plasmon-pole approximation
(three upper dashed curves, see text). We also show the interaction
at a wavelength of 10 mm (lower curve, perfect-conductor limit). The
dipole-dipole separation vector $\Rb$ is taken along $\xx$.}
\end{figure}

The interaction between pairs of electric and magnetic dipoles near
a metal surface is analyzed in detail in Fig.\
\ref{dipole-dipole}(a) for all possible orientations except
perpendicular magnetic dipoles, which are forbidden in perfect
conductors and should take small values in real metals. Moreover,
symmetry forbids the interaction of all other pairs that are not
shown in the figure. For surface features inducing electric dipoles
under normal incidence in a plasmonic metal (see Fig.\
\ref{polarization}), the dominant interactions originate in
electric-dipole pairs aligned with their separation vector $\Rb$
(see Fig.\ \ref{dipole-dipole}), quite different from perfect
conductors, which are governed by magnetic dipoles perpendicular to
$\Rb$. However, the latter can contribute in plasmonic materials as
well for large features compared to the skin depth, as we discussed
in Sec.\ \ref{polarizationschemes}. As a thumb rule, the mutual
dipole orientations that lead to the long-range interaction
dependence given by (\ref{asymsp}) are compatible with non-vanishing
surface-plasmon field components emanating from those dipoles [i.e.,
plasmons with $m=0$ azimuthal symmetry for normal electric dipoles,
like in Fig.\ \ref{SP-maps}, or $m=\pm 1$ for parallel dipoles].

The interaction between dipoles in front of a planar surface admits
a representation in parallel momentum space similar to Eq.\
(\ref{Gxxmomentum}), but involving now the Fresnel reflection
coefficients for s and p polarization \cite{W1919,paper085},
$r_s=(k_z-k'_z)/(k_z+k'_z)$ and $r_p=(\epsilon k_z-k'_z)/(\epsilon
k_z+k'_z)$, respectively \cite{J99}, where $k_z=\sqrt{k^2-Q^2}$ and
$k'_z=\sqrt{k^2\epsilon-Q^2}$. In particular, for electric dipoles
parallel to the surface $x$ direction, one finds \cite{W1919,FW1984}
\begin{eqnarray}
\mathcal{G}^r_{xx}=\frac{\ii}{2\pi}\int \frac{d^2\Qb}{k_zQ^2}
\ee^{\ii(\Qb\cdot\Rb+k_z|z|)}[k^2Q_y^2r_s-k_z^2Q_x^2r_p],
\label{Gsp}
\end{eqnarray}
where $z$ is the sum of distances from the dipoles to the surface,
and we are interested in the $z\rightarrow 0$ limit. This expression
is general and leads to $\mathcal{G}_{xx}=0$ in perfect conductors,
for which $r_p=-r_s=1$.

The strong surface-plasmon-mediated interaction described by Eq.\
(\ref{asymsp}) arises from the pole of the Fresnel coefficient $r_p$
at $Q=\ksp$, which admits the Laurent expansion \cite{FW1984}
\begin{eqnarray}
r_p\approx \frac{2Bk}{Q-\ksp}, \label{rpdiv}
\end{eqnarray}
with
\begin{eqnarray}
B=[\epsilon/(1+\epsilon)]^{3/2}/(1-\epsilon). \nonumber
\end{eqnarray}
Performing asymptotic analysis for large $R$ and retaining only the
contribution from this pole in the integral of Eq.\ (\ref{Gsp})
(plasmon-pole approximation; see \onlinecite{FW1984}), we
obtain\footnote{It should be noted that the asymptotic behavior of
$\mathcal{G}^0$ [see Eq.\ (\ref{asym})] comes from the $Q=k$ region
of the integral in Eq.\ (\ref{Gxxmomentum}) and responds to the pole
$1/k_z$. This pole is canceled exactly by Eq.\ (\ref{Gsp}), in which
$r_p=r_s=-1$ at grazing incidence (i.e., for $Q=k$). Therefore, the
only relevant contribution to $\mathcal{G}$ for large $R$ originates
in the plasmon pole of $\mathcal{G}^r$.}
\begin{eqnarray}
\mathcal{G}_{xx}&\approx& \frac{\pi k^3B\sqrt{\epsilon}}{\epsilon+1}
\left[H_0^{(1)}(\ksp R)+H_2^{(1)}(\ksp
R)\frac{(y^2-x^2)}{R^2}\right] \nonumber \\
&\approx& \frac{-2\pi k^3B}{\epsilon+1}
\sqrt{\frac{2\epsilon}{\ii\pi\ksp}}\,\,\,\frac{\ee^{\ii \ksp
R}}{\sqrt{R}}, \label{GRsp}
\end{eqnarray}
where the second approximation comes from the asymptotic behavior of
Hankel functions for large arguments \cite{AS1972}, so that one
obtains the result anticipated in Eq.\ (\ref{asymsp}). The above
approximate expression in terms of Hankel functions is compared with
the direct numerical evaluation of Eq.\ (\ref{Gsp}), and similar
expressions for other dipole orientations, in Fig.\
\ref{dipole-dipole}(b). The agreement at $\lambda=750$ nm is
excellent for $R\gtrsim\lambda$, indicating that lattice resonances
in an array will be really dominated by surface plasmons at that
wavelength. Fig.\ \ref{dipole-dipole} illustrates as well a much
faster decay of $\mathcal{G}_{yy}$ as $1/R^{3/2}$ for electric
dipoles oriented orthogonal to $\Rb$ and parallel to the surface,
and as $1/R$ for normal electric dipoles in the perfect-conductor
limit.

\subsection{Discrepancies in lattice resonances and enhanced transmission}
\label{differencesinthelattice}

The dissimilar behavior of plasmonic metals and perfect conductors
discussed in the previous sections leads to qualitative differences
in extraordinary optical transmission, arising in part from the
$1/(Q-\ksp)$ dominant pole of the inter-hole interaction in
momentum space [see Eqs.\ (\ref{Gsp})-(\ref{GRsp})].

Considering for simplicity a square array under normal incidence, we
can analyze the lattice sum in a real metal [i.e., Eq.\ (\ref{Gk})
with $\mathcal{G}$ substituted for $\mathcal{G}^0$] following the
procedure that led to Eq. (\ref{GG}), but starting now from Eqs.
(\ref{Fourier}) and (\ref{Gsp}). In a diffrationless array, there
are just two identical singular terms in the corresponding sum over
reciprocal lattice vectors, leading to
\begin{eqnarray} G_{xx}^{\rm EE}\approx C\left(\frac{4\pi}{a\lambda}\right)^2
\frac{\lambda_{\rm SP}}{\lambda_{\rm SP}/a-1} \label{Gspana}
\end{eqnarray}
for $\Re\{\lambda_{\rm SP}\}\gtrsim a$, where $\lambda_{\rm
SP}=2\pi/\ksp$ is the surface-plasmon wavelength and $C=\ii
B/\sqrt{\epsilon+1}$. We have explicitly indicated with superscripts
that $G_{xx}^{\rm EE}$ describes the interaction between electric
dipoles (E), which can coexist with parallel magnetic dipoles (M)
(see Fig.\ \ref{dipole-dipole}). The remaining relevant lattice sums
are $G_{yy}^{\rm MM}\approx-(\epsilon+1)\,G_{xx}^{\rm EE}$ and
$G_{xy}^{\rm EM}=-G_{yx}^{\rm
ME}\approx\sqrt{\epsilon+1}\,G_{xx}^{\rm EE}$. Now, the formalism
presented in Sec.\ \ref{basicrelations} can be easily extended to
patterned surfaces and hole arrays in real metals using these
expressions of the lattice sums rather than those for perfect
conductors. \footnote{Our analysis can be applied to metals embedded
in a dielectric of refraction index $n$ simply by using the reduced
wavelength $\lambda/n$ everywhere instead of $\lambda$ and by
interpreting $\epsilon$ as the ratio of permittivities in the metal
and in the dielectric.}

In the polaritonic regime of surface plasmons, in which their
dispersion relation approaches the light line (see Fig.\ \ref{SP}),
$|\epsilon|$ is large and the dominant lattice sum scales as
$G_{yy}^{\rm MM}\sim 1/\sqrt{-\epsilon}$ in the plasmon-pole
approximation, so that for sufficiently high $|\epsilon|$ the
perfect-conductor limit of Eq.\ (\ref{Gana}) dominates over the
plasmon.

\begin{figure}
\includegraphics[width=80mm,angle=0,clip]{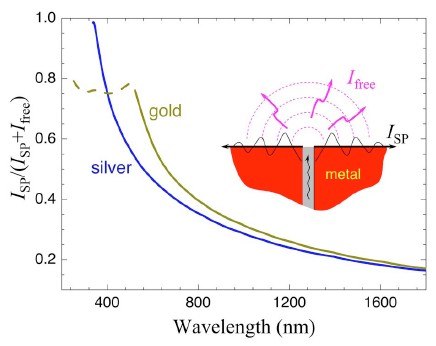}
\caption{\label{plasmon-launching} (Color in online edition)
Relation between the power radiated after transmission through a
deep subwavelength hole ($I_{\rm free}$) and the power emanating as
surface plasmons ($I_{\rm SP}$) for gold and silver, derived in the
small-hole limit. The metal dielectric function has been taken from
\textcite{JC1972}.}
\end{figure}

A descriptive example of the transition from plasmonic to
perfect-conductor behavior is offered in Fig.\
\ref{plasmon-launching}, in which the energy released by a dipole
sitting near a surface is divided into plasmon launching ($I_{\rm
SP}$) and emitted light ($I_{\rm free}$). This relates to the
question, which of the two mechanisms (plasmons or propagating
radiation) produces stronger interaction with a nearby surface
feature. Plasmon launching dominates near the electrostatic plasmon,
reaching an efficiency close to 100\% in silver. As the wavelength
advances towards to infrared, the plasmon is less bound to the
surface and has weaker coupling to our dipole. As an example of
application, when light pops out of a narrow hole after being guided
through a TE mode (e.g., in a circular hole infiltrated with a
dielectric of refraction index $n\gg 1$ and for $\lambda/n\lesssim
3.4b$), the equivalent dipole describing the hole lies parallel to
the surface. That is the situation depicted in the inset of Fig.\
\ref{plasmon-launching}.

\begin{figure}
\includegraphics[width=80mm,angle=0,clip]{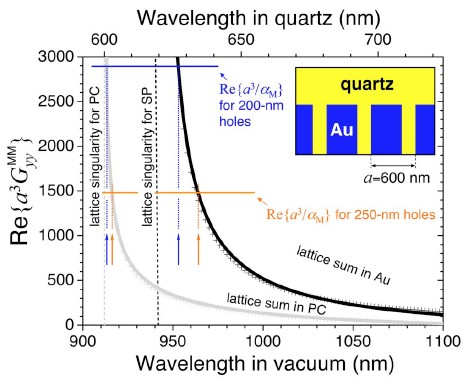}
\caption{\label{maximum-SP} (Color in online edition) Lattice sums
and lattice resonances in a square array of holes drilled in gold vs
a perfect conductor. The real part of the exact lattice sum for
interaction of parallel magnetic (M) dipoles is shown for gold
(black curve) and for a perfect conductor (PC, grey curve), as
compared to analytical approximate expressions (symbols). The
Rayleigh condition for a period $a=600$ nm is indicated by black and
grey vertical dashed lines for light in the dielectric
($\lambda/n=a$) and for surface plasmons ($\lambda_{\rm SP}=a$),
respectively. Changes in the inverse magnetic polarizability of
circular holes of different size [horizontal lines, as obtained from
Fig.\ \ref{thick-hole}(b)], lead to different wavelengths of the
lattice surface modes, as indicated by vertical arrows for the
condition that the real part of the denominator of Eq.\ (\ref{pp})
be zero.}
\end{figure}

A more explicit comparison of discrepancies between both metallic
regimes for holes is offered in Fig.\ \ref{maximum-SP}, which shows
the lattice sum for parallel magnetic dipoles [obtained by summing
Eq.\ (\ref{Gsp}) for gold, with the expression in square brackets
replaced by $k^2Q_x^2r_p-k_z^2Q_y^2r_s$], together with a
geometrical construction like in Fig.\ \ref{maximum}, applied now to
two different aperture sizes. It should be noted that the exact
calculation (solid curves) compares extremely well with analytical
expressions [symbols, obtained from Eq.\ (\ref{Gana}) for the
perfect conductor and from Eq.\ (\ref{Gspana}) for the plasmonic
metal, which needs to be multiplied by $-(\epsilon+1)$ in order to
apply it to magnetic rather than electric dipoles]. The lattice sum
singularity in perforated gold takes place to the red as compared to
the perfect-conductor case, because the surface-plasmon wavelength
is shorter than the light wavelength in the surrounding dielectric.
Moreover, the lattice sum diverges as $1/\sqrt{\lambda/n-a}$ and
$1/(\lambda_{\rm SP}-a)$ in perfect conductors and plasmonic metals,
respectively, according to Eqs. (\ref{Gana}) and (\ref{Gspana}),
thus leading to different dependence of the position of the lattice
surface resonance on hole size (see points of intersection with
horizontal lines in Fig.\ \ref{maximum-SP}); the lattice resonance
is further away from the interaction sum singularity (and a given
change in hole diameter produces larger peak shift) in the plasmonic
case considered in the figure.

\begin{figure}
\includegraphics[width=80mm,angle=0,clip]{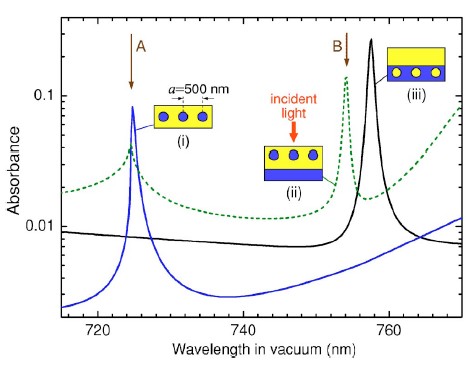}
\caption{\label{Wood} (Color in online edition) Normal-incidence
absorbance of (i) a silver particle array embedded in silica
(refraction index $n=1.45$), (ii) the same array near a planar
silver-silica interface, and (iii) an array of silica inclusions
buried in silver below a silver-silica interface. All particles are
spheres of 200 nm in diameter. The arrays have square symmetry with
lattice constant $a=500$ nm. The distance from the sphere surfaces
to the planar interface is 10 nm in the buried silica particles and
900 nm for the silver particles. The Rayleigh conditions for the
reduced wavelength of light in the silica ($\lambda/n=a$) and for
the wavelength of the silver-silica interface plasmon ($\lambda_{\rm
SP}=a$) are indicated by arrows A and B, respectively.}
\end{figure}

The crossover between both types of behavior is explored in Fig.\
\ref{Wood} through the absorbance of (i) a silver-particle array in
silica, (ii) the same array near a silver-silica interface, and
(iii) an array of silica inclusions right underneath the
metal-dielectric interface. We have done these calculations using a
layer KKR method to solve Maxwell's equations \cite{SYM98_1,SYM00}.
In the case (i) a maximum in absorption occurs near the Rayleigh
condition for light propagating in silica (i.e., $\lambda/n=a$),
whereas case (iii) shows a single maximum shifted to the right of
the Rayleigh condition for the planar interface plasmon
($\lambda_{\rm SP}=a$) \cite{GTG98}. The conclusion is that plasmons
are mediating the interaction among the dielectric inclusions, with
no signature of any anomaly near $\lambda/n=a$ whatsoever. An
intermediate situation is encountered in case (ii), showing features
near the two types of Rayleigh conditions.

It should be noted that $\lambda_{\rm SP}$ has an imaginary part
arising from absorption, and although it is small for noble metals,
in which plasmons can travel long distances along the surface, as
shown in Fig.\ \ref{SP}(b), we find that Eq.\ (\ref{Gspana}) does
not describe a divergence, but rather a Lorentzian of finite width.
This affects the height of the transmission maxima, below 100\% in
lossy metals. Furthermore, apertures perforated in metals of finite
conductivity will appear to be wider by the skin depth effect, and
their effective polarizability must be lossy.

Without entering into further considerations regarding how finite
conductivity affects the hole polarizability, let us just point out
that the wavelength at which the noted intersection takes place in
Fig.\ \ref{maximum-SP} (i.e., the wavelength of the lattice
surface-bound mode)
is in excellent agreement with the transmission peaks measured by
\textcite{KTK01} and reproduced in Fig.\ \ref{experiment}(a). The
vertical arrows in that figure indicate the predicted positions of
the transmission maxima, obtained by increasing the hole size by the
skin depth to an effective diameter of 250 nm. This agreement is
remarkable, given our neglect of higher-order multipolar terms in
the hole polarization. The shift with respect to the Rayleigh
condition for surface plasmons (vertical solid lines in Fig.\
\ref{experiment}) is significant, triggered by large,
plasmon-mediated interaction between apertures, as explained above.
Similar conclusions can be drawn for the silver film of Fig.\
\ref{experiment}(b), in which the results from the above analytical
model are shown as dashed curves (divided by a factor of 5). Only
magnetic dipoles are taken into account, with the hole
polarizability calculated for a perfect conductor. The transmittance
is obtained from Eq.\ (\ref{Ts}) with $G_{xx}$ replaced by its
plasmonic counterpart, $G_{yy}^{\rm MM}$. Although the Rayleigh
condition for plasmons (solid vertical lines in Fig.\
\ref{experiment}) agrees only with the transmission minima in silver
(presumably because gold is more dissipative in this spectral
region, so that the polarizability of the holes requires a more
realistic description including absorption), the comparison with
experiment is excellent, given the simplicity of the analytical
model, which should become exact in the limit of small scattering
features (e.g., for nanoparticle arrays on a metal substrate).

\section{Conclusion}
\label{conclusion}

Light scattering in planar periodic systems gives rise to resonant
phenomena that have common origins in particle and hole arrays, both
for reflection and for transmission. Namely, (i) the interaction
between lattice sites shows a divergent behavior when a diffracted
beam becomes grazing \cite{R1907}, producing a minimum in both the
reflectivity of particle arrays and the transmission of hole arrays;
(ii) a lattice resonance can be established at a wavelength to the
red of that condition, leading to maxima in both the reflectivity of
particle arrays and the transmission of hole arrays; (iii) these
effects have the same origin as Wood's anomalies \cite{W1935} and
they can be described in the language of Fano lineshapes
\cite{F1961}; (iv) the noted lattice resonance persists for incident
evanescent light, with the reflectivity's becoming infinite in
non-dissipative systems (e.g., patterned perfect-conductors, but
also patterned dielectrics), thus defining truly surface-bound
states \cite{UT1972,PMG04,HES05,paper105}; (v) these extended
lattice resonances mix strongly with other modes localized at
specific sites, like those created by nanoparticle and nanovoid
plasmons \cite{KSC06,paper109}; (vi) for metals with well-defined
surface plasmons, the interaction between holes or particles in the
vicinity of the surface is mediated by these excitations, so that we
have to reformulate the condition of a diffracted beam's becoming
grazing using the surface plasmon wavelength rather than the
incoming or transmitted light wavelength.

We have shown that particle arrays and hole patterns in perfect
conductors share in common the asymptotic form of their interaction,
summarized by Eq.\ (\ref{asym}), which produces singularities at the
Rayleigh condition when summed over the lattice, for instance for
$\lambda=a$ under normal incidence on square arrays, and gives rise
to surface states at slightly larger wavelengths. However, the
plasmon-mediated interaction in noble metals is more intense, as
shown in Eq.\ (\ref{asymsp}), thus producing sharper divergences and
stronger collective interaction. In this case, the singularities
occur at the band-folded plasmon lines (e.g., when $\lambda_{\rm
SP}=a$ under normal incidence on square arrays), and the lattice
surface-bound states (i.e., the plasmons of the patterned metal)
exist again to the red with respect to those lines.

All of these effects have been described here within a common
tutorial approach based upon interacting dipoles that is not only
able to explain the observed effects; its simplicity has allowed us
to extract some surprising conclusions. One of them is that
arbitrarily-weak scatterers forming a periodic structure and made of
non-dissipative materials can also produce intense lattice
resonances: given an array of arbitrarily-small particles of
positive polarizability, it is always possible to find a wavelength
(close to the period for square symmetry and normal incidence) at
which light is totally reflected; accordingly, it is possible to
obtain full transmission through holes however narrow, drilled in
arbitrarily-thick perfect-conductor films.

Interestingly, the lattice periodicity alone determines the
magnitude of the induced dipoles needed to produce complete
reflection by small particles or total transmission through narrow
holes. Moreover, the polarizability scales with the cube of the
hole/particle diameter. Combining these two statements, we find that
the self-consistent electric field acting on particles or apertures
under such resonant conditions increases when they shrink and can
reach extremely high values only limited by absorption and lattice
imperfections, thus opening new possibilities for applications in
nonlinear all-optical switching and biosensing.

The simplicity and power of the model that has been presented here
will surely find application to explain many other effects related
to light scattering in planar periodic systems and can be inspiring
for devising new phenomena.

\section*{Acknowledgments}

The author wants to thank J. J. Baumberg, A. G. Borisov, G.
G\'omez-Santos, C. L\'opez, F. Meseguer, J. B. Pendry, V. V. Popov,
J. J. S\'aenz, S. V. Shabanov, T. V. Teperik, and N. I. Zheludev for
many enjoyable and stimulating discussions. This work was supported
in part by the Spanish MEC (contract No. NAN2004-08843-C05-05) and
by the EU ({\it SPANS} STREP STRP-016881-SPANS and {\it
Metamorphose} NoE NMP3-CT-2004-500252).


\begin{thebibliography}{173}
\expandafter\ifx\csname
natexlab\endcsname\relax\def\natexlab#1{#1}\fi
\expandafter\ifx\csname bibnamefont\endcsname\relax
  \def\bibnamefont#1{#1}\fi
\expandafter\ifx\csname bibfnamefont\endcsname\relax
  \def\bibfnamefont#1{#1}\fi
\expandafter\ifx\csname citenamefont\endcsname\relax
  \def\citenamefont#1{#1}\fi
\expandafter\ifx\csname url\endcsname\relax
  \def\url#1{\texttt{#1}}\fi
\expandafter\ifx\csname urlprefix\endcsname\relax\def\urlprefix{URL
}\fi \providecommand{\bibinfo}[2]{#2}
\providecommand{\eprint}[2][]{\url{#2}}

\bibitem[{\citenamefont{Abramowitz and Stegun}(1972)}]{AS1972}
\bibinfo{author}{\bibnamefont{Abramowitz}, \bibfnamefont{M.}}, and
  \bibinfo{author}{\bibfnamefont{I.~A.} \bibnamefont{Stegun}},
  \bibinfo{year}{1972}, \emph{\bibinfo{title}{Handbook of Mathematical
  Functions}} (\bibinfo{publisher}{Dover}, \bibinfo{address}{New York}).

\bibitem[{\citenamefont{Akahane} \emph{et~al.}(2003)\citenamefont{Akahane,
  Asano, Song, and Noda}}]{AAS03}
\bibinfo{author}{\bibnamefont{Akahane}, \bibfnamefont{Y.}},
  \bibinfo{author}{\bibfnamefont{T.}~\bibnamefont{Asano}},
  \bibinfo{author}{\bibfnamefont{B.-S.} \bibnamefont{Song}}, and
  \bibinfo{author}{\bibfnamefont{S.}~\bibnamefont{Noda}}, \bibinfo{year}{2003},
  {``}\bibinfo{title}{High-Q photonic nanocavity in a two-dimensional photonic
  crystal},{''} \bibinfo{journal}{Nature} \textbf{\bibinfo{volume}{425}},
  \bibinfo{pages}{944--947}.

\bibitem[{\citenamefont{Altewischer}
  \emph{et~al.}(2002)\citenamefont{Altewischer, {van Exter}, and
  Woerdman}}]{AVW02}
\bibinfo{author}{\bibnamefont{Altewischer}, \bibfnamefont{E.}},
  \bibinfo{author}{\bibfnamefont{M.~P.} \bibnamefont{{van Exter}}}, and
  \bibinfo{author}{\bibfnamefont{J.~P.} \bibnamefont{Woerdman}},
  \bibinfo{year}{2002}, {``}\bibinfo{title}{Plasmon-assisted transmission of
  entangled photons},{''} \bibinfo{journal}{Nature}
  \textbf{\bibinfo{volume}{418}},  \bibinfo{pages}{304--306}.

\bibitem[{\citenamefont{Ashcroft and Mermin}(1976)}]{AM1976}
\bibinfo{author}{\bibnamefont{Ashcroft}, \bibfnamefont{N.~W.}}, and
  \bibinfo{author}{\bibfnamefont{N.~D.} \bibnamefont{Mermin}},
  \bibinfo{year}{1976}, \emph{\bibinfo{title}{Solid State Physics}}
  (\bibinfo{publisher}{Harcourt College Publishers}, \bibinfo{address}{New
  York}).

\bibitem[{\citenamefont{Atay} \emph{et~al.}(2004)\citenamefont{Atay, Song, and
  Nurmikko}}]{ASN04}
\bibinfo{author}{\bibnamefont{Atay}, \bibfnamefont{T.}},
  \bibinfo{author}{\bibfnamefont{J.-H.} \bibnamefont{Song}}, and
  \bibinfo{author}{\bibfnamefont{A.~V.} \bibnamefont{Nurmikko}},
  \bibinfo{year}{2004}, {``}\bibinfo{title}{Strongly interacting plasmon
  nanoparticle pairs: from dipole-dipole interaction to conductively coupled
  regime},{''} \bibinfo{journal}{Nano\ Lett.} \textbf{\bibinfo{volume}{4}},
  \bibinfo{pages}{1627--1631}.

\bibitem[{\citenamefont{Baida and {Van Labeke}}(2002)}]{BV02}
\bibinfo{author}{\bibnamefont{Baida}, \bibfnamefont{F.~I.}}, and
  \bibinfo{author}{\bibfnamefont{D.}~\bibnamefont{{Van Labeke}}},
  \bibinfo{year}{2002}, {``}\bibinfo{title}{Light transmission by subwavelength
  annular aperture arrays in metallic films},{''} \bibinfo{journal}{Opt.\
  Commun.} \textbf{\bibinfo{volume}{209}},  \bibinfo{pages}{17--22}.

\bibitem[{\citenamefont{Barlow}(1958)}]{B1958}
\bibinfo{author}{\bibnamefont{Barlow}, \bibfnamefont{H.~M.}},
  \bibinfo{year}{1958}, {``}\bibinfo{title}{Surface waves},{''}
  \bibinfo{journal}{Proc.\ IRE} \textbf{\bibinfo{volume}{46}},
  \bibinfo{pages}{1413--1417}.

\bibitem[{\citenamefont{Barnes and Sambles}(2004)}]{BS04}
\bibinfo{author}{\bibnamefont{Barnes}, \bibfnamefont{W.}}, and
  \bibinfo{author}{\bibfnamefont{R.}~\bibnamefont{Sambles}},
  \bibinfo{year}{2004}, {``}\bibinfo{title}{Only skin deep},{''}
  \bibinfo{journal}{Science} \textbf{\bibinfo{volume}{305}},
  \bibinfo{pages}{785--786}.

\bibitem[{\citenamefont{Barnes} \emph{et~al.}(2003)\citenamefont{Barnes,
  Dereux, and Ebbesen}}]{BDE03}
\bibinfo{author}{\bibnamefont{Barnes}, \bibfnamefont{W.~L.}},
  \bibinfo{author}{\bibfnamefont{A.}~\bibnamefont{Dereux}}, and
  \bibinfo{author}{\bibfnamefont{T.~W.} \bibnamefont{Ebbesen}},
  \bibinfo{year}{2003}, {``}\bibinfo{title}{Surface plasmon subwavelength
  optics},{''} \bibinfo{journal}{Nature} \textbf{\bibinfo{volume}{424}},
  \bibinfo{pages}{824--830}.

\bibitem[{\citenamefont{Barnes} \emph{et~al.}(2004)\citenamefont{Barnes,
  Murray, Dintinger, Devaux, and Ebbesen}}]{BMD04}
\bibinfo{author}{\bibnamefont{Barnes}, \bibfnamefont{W.~L.}},
  \bibinfo{author}{\bibfnamefont{W.~A.} \bibnamefont{Murray}},
  \bibinfo{author}{\bibfnamefont{J.}~\bibnamefont{Dintinger}},
  \bibinfo{author}{\bibfnamefont{E.}~\bibnamefont{Devaux}}, and
  \bibinfo{author}{\bibfnamefont{T.~W.} \bibnamefont{Ebbesen}},
  \bibinfo{year}{2004}, {``}\bibinfo{title}{Surface plasmon polaritons and
  their role in the enhanced transmission of light through periodic arrays of
  subwavelength holes in a metal film},{''} \bibinfo{journal}{Phys.\ Rev.\
  Lett.} \textbf{\bibinfo{volume}{92}},  \bibinfo{pages}{107401}.

\bibitem[{\citenamefont{Baumberg}(2006)}]{B06_2}
\bibinfo{author}{\bibnamefont{Baumberg}, \bibfnamefont{J.~J.}},
  \bibinfo{year}{2006}, {``}\bibinfo{title}{Breaking the mould: casting on the
  nanometre scale},{''} \bibinfo{journal}{Nat.\ Mater.}
  \textbf{\bibinfo{volume}{5}},  \bibinfo{pages}{2--5}.

\bibitem[{\citenamefont{Bethe}(1944)}]{B1944}
\bibinfo{author}{\bibnamefont{Bethe}, \bibfnamefont{H.~A.}},
  \bibinfo{year}{1944}, {``}\bibinfo{title}{Theory of diffraction by small
  holes},{''} \bibinfo{journal}{Phys.\ Rev.} \textbf{\bibinfo{volume}{66}},
  \bibinfo{pages}{163--182}.

\bibitem[{\citenamefont{Blanco and {Garc\'{\i}a de Abajo}}(2004)}]{paper085}
\bibinfo{author}{\bibnamefont{Blanco}, \bibfnamefont{L.~A.}}, and
  \bibinfo{author}{\bibfnamefont{F.~J.} \bibnamefont{{Garc\'{\i}a de Abajo}}},
  \bibinfo{year}{2004}, {``}\bibinfo{title}{Spontaneous light emission in
  complex nanostructures},{''} \bibinfo{journal}{Phys.\ Rev.\ B}
  \textbf{\bibinfo{volume}{69}},  \bibinfo{pages}{205414}.

\bibitem[{\citenamefont{Bohren and Huffman}(1983)}]{BH98}
\bibinfo{author}{\bibnamefont{Bohren}, \bibfnamefont{C.~F.}}, and
  \bibinfo{author}{\bibfnamefont{D.~R.} \bibnamefont{Huffman}},
  \bibinfo{year}{1983}, \emph{\bibinfo{title}{Absorption and Scattering of
  Light by Small Particles}} (\bibinfo{publisher}{Wiley-Interscience},
  \bibinfo{address}{New York}).

\bibitem[{\citenamefont{Borisov} \emph{et~al.}(2005)\citenamefont{Borisov,
  {Garc\'{\i}a de Abajo}, and Shabanov}}]{paper096}
\bibinfo{author}{\bibnamefont{Borisov}, \bibfnamefont{A.~G.}},
  \bibinfo{author}{\bibfnamefont{F.~J.} \bibnamefont{{Garc\'{\i}a de Abajo}}},
  and \bibinfo{author}{\bibfnamefont{S.~V.} \bibnamefont{Shabanov}},
  \bibinfo{year}{2005}, {``}\bibinfo{title}{Role of electromagnetic trapped
  modes in extraordinary transmission in nanostructured materials},{''}
  \bibinfo{journal}{Phys.\ Rev.\ B} \textbf{\bibinfo{volume}{71}},
  \bibinfo{pages}{075408}.

\bibitem[{\citenamefont{Born and Wolf}(1999)}]{BW99}
\bibinfo{author}{\bibnamefont{Born}, \bibfnamefont{M.}}, and
  \bibinfo{author}{\bibfnamefont{E.}~\bibnamefont{Wolf}}, \bibinfo{year}{1999},
  \emph{\bibinfo{title}{Principles of Optics: Electromagnetic Theory of
  Propagation, Interference and Diffraction of Light}}
  (\bibinfo{publisher}{Cambridge University Press},
  \bibinfo{address}{Cambridge}).

\bibitem[{\citenamefont{Bouwkamp}(1954)}]{B1954}
\bibinfo{author}{\bibnamefont{Bouwkamp}, \bibfnamefont{C.~J.}},
  \bibinfo{year}{1954}, {``}\bibinfo{title}{Diffraction theory},{''}
  \bibinfo{journal}{Rep.\ Prog.\ Phys.} \textbf{\bibinfo{volume}{17}},
  \bibinfo{pages}{35--100}.

\bibitem[{\citenamefont{Bozhevolnyi}
  \emph{et~al.}(2006)\citenamefont{Bozhevolnyi, Volkov, Devaux, Laluet, and
  Ebbesen}}]{BVD06}
\bibinfo{author}{\bibnamefont{Bozhevolnyi}, \bibfnamefont{S.~I.}},
  \bibinfo{author}{\bibfnamefont{V.~S.} \bibnamefont{Volkov}},
  \bibinfo{author}{\bibfnamefont{E.}~\bibnamefont{Devaux}},
  \bibinfo{author}{\bibfnamefont{J.-Y.} \bibnamefont{Laluet}}, and
  \bibinfo{author}{\bibfnamefont{T.~W.} \bibnamefont{Ebbesen}},
  \bibinfo{year}{2006}, {``}\bibinfo{title}{Channel plasmon subwavelength
  waveguide components including interferometers and ring resonators},{''}
  \bibinfo{journal}{Nature} \textbf{\bibinfo{volume}{440}},
  \bibinfo{pages}{508--511}.

\bibitem[{\citenamefont{Bravo-Abad}
  \emph{et~al.}(2004)\citenamefont{Bravo-Abad, {Garc\'{\i}a-Vidal}, and
  {{Mart\'{\i}n-Moreno}}}}]{BGM04}
\bibinfo{author}{\bibnamefont{Bravo-Abad}, \bibfnamefont{J.}},
  \bibinfo{author}{\bibfnamefont{F.~J.} \bibnamefont{{Garc\'{\i}a-Vidal}}}, and
  \bibinfo{author}{\bibfnamefont{L.}~\bibnamefont{{{Mart\'{\i}n-Moreno}}}},
  \bibinfo{year}{2004}, {``}\bibinfo{title}{Resonant transmission of light
  through finite chains of subwavelength holes in a metallic film},{''}
  \bibinfo{journal}{Phys.\ Rev.\ Lett.} \textbf{\bibinfo{volume}{93}},
  \bibinfo{pages}{227401}.

\bibitem[{\citenamefont{Cao and Nahata}(2004)}]{CN04}
\bibinfo{author}{\bibnamefont{Cao}, \bibfnamefont{H.}}, and
  \bibinfo{author}{\bibfnamefont{A.}~\bibnamefont{Nahata}},
  \bibinfo{year}{2004}, {``}\bibinfo{title}{Resonantly enhanced transmission of
  terahertz radiation through a periodic array of subwavelength apertures},{''}
  \bibinfo{journal}{Opt.\ Express} \textbf{\bibinfo{volume}{12}},
  \bibinfo{pages}{1004--1010}.

\bibitem[{\citenamefont{Cao and Lalanne}(2002)}]{CL02}
\bibinfo{author}{\bibnamefont{Cao}, \bibfnamefont{Q.}}, and
  \bibinfo{author}{\bibfnamefont{P.}~\bibnamefont{Lalanne}},
  \bibinfo{year}{2002}, {``}\bibinfo{title}{Negative role of surface plasmons
  in the transmission of metallic gratings with very narrow slits},{''}
  \bibinfo{journal}{Phys.\ Rev.\ Lett.} \textbf{\bibinfo{volume}{88}},
  \bibinfo{pages}{057403}.

\bibitem[{\citenamefont{Chang} \emph{et~al.}(2006)\citenamefont{Chang,
  Sarychev, and Shalaev}}]{CSS05}
\bibinfo{author}{\bibnamefont{Chang}, \bibfnamefont{C.-W.}},
  \bibinfo{author}{\bibfnamefont{A.~K.} \bibnamefont{Sarychev}}, and
  \bibinfo{author}{\bibfnamefont{V.~M.} \bibnamefont{Shalaev}},
  \bibinfo{year}{2006}, {``}\bibinfo{title}{Light diffraction by a
  subwavelength circular aperture},{''} \bibinfo{journal}{Laser\ Phys.\ Lett.}
  \textbf{\bibinfo{volume}{2}},  \bibinfo{pages}{351–--355}.

\bibitem[{\citenamefont{Chang} \emph{et~al.}(2005)\citenamefont{Chang, Gray,
  and Schatz}}]{CGS05}
\bibinfo{author}{\bibnamefont{Chang}, \bibfnamefont{S.-H.}},
  \bibinfo{author}{\bibfnamefont{S.~K.} \bibnamefont{Gray}}, and
  \bibinfo{author}{\bibfnamefont{G.~C.} \bibnamefont{Schatz}},
  \bibinfo{year}{2005}, {``}\bibinfo{title}{Surface plasmon generation and
  light transmission by isolated nanoholes and arrays of nanoholes in thin
  metal films},{''} \bibinfo{journal}{Opt.\ Express}
  \textbf{\bibinfo{volume}{13}},  \bibinfo{pages}{3150--3165}.

\bibitem[{\citenamefont{Chen}(1971)}]{C1971}
\bibinfo{author}{\bibnamefont{Chen}, \bibfnamefont{C.~C.}},
  \bibinfo{year}{1971}, {``}\bibinfo{title}{Diffraction of electromagnetic
  waves by a conducting screen perforated periodically with circular
  holes},{''} \bibinfo{journal}{IEEE\ Trans.\ Microw.\ Theory\ Tech.}
  \textbf{\bibinfo{volume}{19}},  \bibinfo{pages}{475--481}.

\bibitem[{\citenamefont{Collin and Eggimann}(1961)}]{CE1961}
\bibinfo{author}{\bibnamefont{Collin}, \bibfnamefont{R.~E.}}, and
  \bibinfo{author}{\bibfnamefont{W.~H.} \bibnamefont{Eggimann}},
  \bibinfo{year}{1961}, {``}\bibinfo{title}{Dynamic interaction fields in a
  two-dimensional lattice},{''} \bibinfo{journal}{IRE\ Trans.\ Microw.\ Theory\
  Tech.} \textbf{\bibinfo{volume}{9}},  \bibinfo{pages}{110--115}.

\bibitem[{\citenamefont{Colombelli}
  \emph{et~al.}(2003)\citenamefont{Colombelli, Srinivasan, Troccoli, Painter,
  Gmachl, Tennant, Sergent, Sivco, Cho, and Capasso}}]{CST03}
\bibinfo{author}{\bibnamefont{Colombelli}, \bibfnamefont{R.}},
  \bibinfo{author}{\bibfnamefont{K.}~\bibnamefont{Srinivasan}},
  \bibinfo{author}{\bibfnamefont{M.}~\bibnamefont{Troccoli}},
  \bibinfo{author}{\bibfnamefont{O.}~\bibnamefont{Painter}},
  \bibinfo{author}{\bibfnamefont{C.~F.} \bibnamefont{Gmachl}},
  \bibinfo{author}{\bibfnamefont{D.~M.} \bibnamefont{Tennant}},
  \bibinfo{author}{\bibfnamefont{A.~M.} \bibnamefont{Sergent}},
  \bibinfo{author}{\bibfnamefont{D.~L.} \bibnamefont{Sivco}},
  \bibinfo{author}{\bibfnamefont{A.~Y.} \bibnamefont{Cho}}, and
  \bibinfo{author}{\bibfnamefont{F.}~\bibnamefont{Capasso}},
  \bibinfo{year}{2003}, {``}\bibinfo{title}{Quantum-cascade surface-emitting
  photonic crystal laser},{''} \bibinfo{journal}{Science}
  \textbf{\bibinfo{volume}{302}},  \bibinfo{pages}{1374--1377}.

\bibitem[{\citenamefont{Cwik} \emph{et~al.}(1987)\citenamefont{Cwik, Mittra,
  Lang, and Wu}}]{CLM1987}
\bibinfo{author}{\bibnamefont{Cwik}, \bibfnamefont{T.}},
  \bibinfo{author}{\bibfnamefont{R.}~\bibnamefont{Mittra}},
  \bibinfo{author}{\bibfnamefont{K.~C.} \bibnamefont{Lang}}, and
  \bibinfo{author}{\bibfnamefont{T.~K.} \bibnamefont{Wu}},
  \bibinfo{year}{1987}, {``}\bibinfo{title}{Frequency selective screens},{''}
  \bibinfo{journal}{IEEE\ Antennas\ Propag.\ Soc.\ Newslett.}
  \textbf{\bibinfo{volume}{29}},  \bibinfo{pages}{5--10}.

\bibitem[{\citenamefont{Dawes} \emph{et~al.}(1989)\citenamefont{Dawes,
  McPhedran, and Whitbourn}}]{DMW1989}
\bibinfo{author}{\bibnamefont{Dawes}, \bibfnamefont{D.~H.}},
  \bibinfo{author}{\bibfnamefont{R.~C.} \bibnamefont{McPhedran}}, and
  \bibinfo{author}{\bibfnamefont{L.~B.} \bibnamefont{Whitbourn}},
  \bibinfo{year}{1989}, {``}\bibinfo{title}{Thin capacitive meshes on a
  dielectric boundary - theory and experiment},{''} \bibinfo{journal}{Appl.\
  Opt.} \textbf{\bibinfo{volume}{28}},  \bibinfo{pages}{3498--3510}.

\bibitem[{\citenamefont{Degiron} \emph{et~al.}(2002)\citenamefont{Degiron,
  Lezec, Barnes, and Ebbesen}}]{DLB02}
\bibinfo{author}{\bibnamefont{Degiron}, \bibfnamefont{A.}},
  \bibinfo{author}{\bibfnamefont{H.~J.} \bibnamefont{Lezec}},
  \bibinfo{author}{\bibfnamefont{W.~L.} \bibnamefont{Barnes}}, and
  \bibinfo{author}{\bibfnamefont{T.~W.} \bibnamefont{Ebbesen}},
  \bibinfo{year}{2002}, {``}\bibinfo{title}{Effects of hole depth on enhanced
  light transmission through subwavelength hole arrays},{''}
  \bibinfo{journal}{Appl.\ Phys.\ Lett.} \textbf{\bibinfo{volume}{81}},
  \bibinfo{pages}{4327--4329}.

\bibitem[{\citenamefont{Degiron} \emph{et~al.}(2004)\citenamefont{Degiron,
  Lezec, Yamamoto, and Ebbesen}}]{DLY04}
\bibinfo{author}{\bibnamefont{Degiron}, \bibfnamefont{A.}},
  \bibinfo{author}{\bibfnamefont{H.~J.} \bibnamefont{Lezec}},
  \bibinfo{author}{\bibfnamefont{N.}~\bibnamefont{Yamamoto}}, and
  \bibinfo{author}{\bibfnamefont{T.~W.} \bibnamefont{Ebbesen}},
  \bibinfo{year}{2004}, {``}\bibinfo{title}{Optical transmission properties of
  a single subwavelength aperture in a real metal},{''} \bibinfo{journal}{Opt.\
  Commun.} \textbf{\bibinfo{volume}{239}},  \bibinfo{pages}{61–--66}.

\bibitem[{\citenamefont{Dintinger}
  \emph{et~al.}(2006{\natexlab{a}})\citenamefont{Dintinger, Klein, and
  Ebbesen}}]{DKE06}
\bibinfo{author}{\bibnamefont{Dintinger}, \bibfnamefont{J.}},
  \bibinfo{author}{\bibfnamefont{S.}~\bibnamefont{Klein}}, and
  \bibinfo{author}{\bibfnamefont{T.~W.} \bibnamefont{Ebbesen}},
  \bibinfo{year}{2006}{\natexlab{a}}, {``}\bibinfo{title}{Molecule-surface
  plasmon interactions in hole arrays: enhanced absorption, refractive index
  changes, and all-optical switching},{''} \bibinfo{journal}{Adv.\ Mater.}
  \textbf{\bibinfo{volume}{18}},  \bibinfo{pages}{1267--1270}.

\bibitem[{\citenamefont{Dintinger}
  \emph{et~al.}(2006{\natexlab{b}})\citenamefont{Dintinger, Robel, Kamat,
  Genet, and Ebbesen}}]{DRK06}
\bibinfo{author}{\bibnamefont{Dintinger}, \bibfnamefont{J.}},
  \bibinfo{author}{\bibfnamefont{I.}~\bibnamefont{Robel}},
  \bibinfo{author}{\bibfnamefont{P.~V.} \bibnamefont{Kamat}},
  \bibinfo{author}{\bibfnamefont{C.}~\bibnamefont{Genet}}, and
  \bibinfo{author}{\bibfnamefont{T.~W.} \bibnamefont{Ebbesen}},
  \bibinfo{year}{2006}{\natexlab{b}}, {``}\bibinfo{title}{Terahertz all-optical
  molecule-plasmon modulation},{''} \bibinfo{journal}{Adv.\ Mater.}
  \textbf{\bibinfo{volume}{18}},  \bibinfo{pages}{1645--1648}.

\bibitem[{\citenamefont{Draine and Flatau}(1994)}]{DF94}
\bibinfo{author}{\bibnamefont{Draine}, \bibfnamefont{B.~T.}}, and
  \bibinfo{author}{\bibfnamefont{P.~J.} \bibnamefont{Flatau}},
  \bibinfo{year}{1994}, {``}\bibinfo{title}{Discrete-dipole approximation for
  scattering calculations},{''} \bibinfo{journal}{J.\ Opt.\ Soc.\ Am.\ A}
  \textbf{\bibinfo{volume}{11}},  \bibinfo{pages}{1491--1499}.

\bibitem[{\citenamefont{Ebbesen} \emph{et~al.}(1998)\citenamefont{Ebbesen,
  Lezec, Ghaemi, Thio, and Wolff}}]{ELG98}
\bibinfo{author}{\bibnamefont{Ebbesen}, \bibfnamefont{T.~W.}},
  \bibinfo{author}{\bibfnamefont{H.~J.} \bibnamefont{Lezec}},
  \bibinfo{author}{\bibfnamefont{H.~F.} \bibnamefont{Ghaemi}},
  \bibinfo{author}{\bibfnamefont{T.}~\bibnamefont{Thio}}, and
  \bibinfo{author}{\bibfnamefont{P.~A.} \bibnamefont{Wolff}},
  \bibinfo{year}{1998}, {``}\bibinfo{title}{Extraordinary optical transmission
  through sub-wavelength hole arrays},{''} \bibinfo{journal}{Nature}
  \textbf{\bibinfo{volume}{391}},  \bibinfo{pages}{667--669}.

\bibitem[{\citenamefont{Eggimann and Collin}(1962)}]{EC1962}
\bibinfo{author}{\bibnamefont{Eggimann}, \bibfnamefont{W.~H.}}, and
  \bibinfo{author}{\bibfnamefont{R.~E.} \bibnamefont{Collin}},
  \bibinfo{year}{1962}, {``}\bibinfo{title}{Electromagnetic diffraction by a
  planar array of circular disks},{''} \bibinfo{journal}{IRE\ Trans.\ Microw.\
  Theory\ Tech.} \textbf{\bibinfo{volume}{10}},  \bibinfo{pages}{528--535}.

\bibitem[{\citenamefont{Ekinci} \emph{et~al.}(2007)\citenamefont{Ekinci, Solak,
  and David}}]{ESD07}
\bibinfo{author}{\bibnamefont{Ekinci}, \bibfnamefont{Y.}},
  \bibinfo{author}{\bibfnamefont{H.~H.} \bibnamefont{Solak}}, and
  \bibinfo{author}{\bibfnamefont{C.}~\bibnamefont{David}},
  \bibinfo{year}{2007}, {``}\bibinfo{title}{Extraordinary optical transmission
  in the ultraviolet region through aluminum hole arrays},{''}
  \bibinfo{journal}{Opt.\ Lett.} \textbf{\bibinfo{volume}{32}},
  \bibinfo{pages}{172--174}.

\bibitem[{\citenamefont{Elliott} \emph{et~al.}(2004)\citenamefont{Elliott,
  Smolyaninov, Zheludev, and Zayats}}]{ESZ04}
\bibinfo{author}{\bibnamefont{Elliott}, \bibfnamefont{J.}},
  \bibinfo{author}{\bibfnamefont{I.~I.} \bibnamefont{Smolyaninov}},
  \bibinfo{author}{\bibfnamefont{N.~I.} \bibnamefont{Zheludev}}, and
  \bibinfo{author}{\bibfnamefont{A.~V.} \bibnamefont{Zayats}},
  \bibinfo{year}{2004}, {``}\bibinfo{title}{Polarization control of optical
  transmission of a periodic array of elliptical nanoholes in a metal
  film},{''} \bibinfo{journal}{Opt.\ Lett.} \textbf{\bibinfo{volume}{29}},
  \bibinfo{pages}{1414--1416}.

\bibitem[{\citenamefont{Falcone} \emph{et~al.}(2004)\citenamefont{Falcone,
  Lopetegi, Laso, Baena, Bonache, Beruete, {Marqu\'{e}s}, {Mart\'{\i}n}, and
  Sorolla}}]{FLL04}
\bibinfo{author}{\bibnamefont{Falcone}, \bibfnamefont{F.}},
  \bibinfo{author}{\bibfnamefont{T.}~\bibnamefont{Lopetegi}},
  \bibinfo{author}{\bibfnamefont{M.~A.~G.} \bibnamefont{Laso}},
  \bibinfo{author}{\bibfnamefont{J.~D.} \bibnamefont{Baena}},
  \bibinfo{author}{\bibfnamefont{J.}~\bibnamefont{Bonache}},
  \bibinfo{author}{\bibfnamefont{M.}~\bibnamefont{Beruete}},
  \bibinfo{author}{\bibfnamefont{R.}~\bibnamefont{{Marqu\'{e}s}}},
  \bibinfo{author}{\bibfnamefont{F.}~\bibnamefont{{Mart\'{\i}n}}}, and
  \bibinfo{author}{\bibfnamefont{M.}~\bibnamefont{Sorolla}},
  \bibinfo{year}{2004}, {``}\bibinfo{title}{Babinet principle applied to the
  design of metasurfaces and metamaterials},{''} \bibinfo{journal}{Phys.\ Rev.\
  Lett.} \textbf{\bibinfo{volume}{93}},  \bibinfo{pages}{197401}.

\bibitem[{\citenamefont{Fan} \emph{et~al.}(2005)\citenamefont{Fan, Zhang,
  Minhas, Malloy, and Brueck}}]{FZM05}
\bibinfo{author}{\bibnamefont{Fan}, \bibfnamefont{W.}},
  \bibinfo{author}{\bibfnamefont{S.}~\bibnamefont{Zhang}},
  \bibinfo{author}{\bibfnamefont{B.}~\bibnamefont{Minhas}},
  \bibinfo{author}{\bibfnamefont{K.~J.} \bibnamefont{Malloy}}, and
  \bibinfo{author}{\bibfnamefont{S.~R.~J.} \bibnamefont{Brueck}},
  \bibinfo{year}{2005}, {``}\bibinfo{title}{Enhanced infrared transmission
  through subwavelength coaxial metallic arrays},{''} \bibinfo{journal}{Phys.\
  Rev.\ Lett.} \textbf{\bibinfo{volume}{94}},  \bibinfo{pages}{033902}.

\bibitem[{\citenamefont{Fano}(1936)}]{F1936}
\bibinfo{author}{\bibnamefont{Fano}, \bibfnamefont{U.}}, \bibinfo{year}{1936},
  {``}\bibinfo{title}{Some theoretical considerations on anomalous diffraction
  gratings},{''} \bibinfo{journal}{Phys.\ Rev.} \textbf{\bibinfo{volume}{50}},
  \bibinfo{pages}{573--573}.

\bibitem[{\citenamefont{Fano}(1941)}]{F1941}
\bibinfo{author}{\bibnamefont{Fano}, \bibfnamefont{U.}}, \bibinfo{year}{1941},
  {``}\bibinfo{title}{The theory of anomalous diffraction gratings and of
  quasi-stationary waves on metallic surfaces (Sommerfeld's waves)},{''}
  \bibinfo{journal}{J.\ Opt.\ Soc.\ Am.} \textbf{\bibinfo{volume}{31}},
  \bibinfo{pages}{213--222}.

\bibitem[{\citenamefont{Fano}(1961)}]{F1961}
\bibinfo{author}{\bibnamefont{Fano}, \bibfnamefont{U.}}, \bibinfo{year}{1961},
  {``}\bibinfo{title}{Effects of configuration interaction on intensities and
  phase shifts},{''} \bibinfo{journal}{Phys.\ Rev.}
  \textbf{\bibinfo{volume}{124}},  \bibinfo{pages}{1866--1878}.

\bibitem[{\citenamefont{Far\'{\i}as and Rieder}(1998)}]{FR98}
\bibinfo{author}{\bibnamefont{Far\'{\i}as}, \bibfnamefont{D.}}, and
  \bibinfo{author}{\bibfnamefont{K.-H.} \bibnamefont{Rieder}},
  \bibinfo{year}{1998}, {``}\bibinfo{title}{Atomic beam diffraction from solid
  surfaces},{''} \bibinfo{journal}{Rep.\ Prog.\ Phys.}
  \textbf{\bibinfo{volume}{61}},  \bibinfo{pages}{1575--1664}.

\bibitem[{\citenamefont{Ford and Weber}(1984)}]{FW1984}
\bibinfo{author}{\bibnamefont{Ford}, \bibfnamefont{G.~W.}}, and
  \bibinfo{author}{\bibfnamefont{W.~H.} \bibnamefont{Weber}},
  \bibinfo{year}{1984}, {``}\bibinfo{title}{Electromagnetic interactions of
  molecules with metal surfaces},{''} \bibinfo{journal}{Phys.\ Rep.}
  \textbf{\bibinfo{volume}{113}},  \bibinfo{pages}{195--287}.

\bibitem[{\citenamefont{{Garc\'{\i}a de Abajo}}(1999)}]{paper040}
\bibinfo{author}{\bibnamefont{{Garc\'{\i}a de Abajo}}, \bibfnamefont{F.~J.}},
  \bibinfo{year}{1999}, {``}\bibinfo{title}{Interaction of radiation and fast
  electrons with clusters of dielectrics: a multiple scattering approach},{''}
  \bibinfo{journal}{Phys.\ Rev.\ Lett.} \textbf{\bibinfo{volume}{82}},
  \bibinfo{pages}{2776--2779}.

\bibitem[{\citenamefont{{Garc\'{\i}a de Abajo}}(2002)}]{paper069}
\bibinfo{author}{\bibnamefont{{Garc\'{\i}a de Abajo}}, \bibfnamefont{F.~J.}},
  \bibinfo{year}{2002}, {``}\bibinfo{title}{Light transmission through a single
  cylindrical hole in a metallic film},{''} \bibinfo{journal}{Opt.\ Express}
  \textbf{\bibinfo{volume}{10}},  \bibinfo{pages}{1475--1484}.

\bibitem[{\citenamefont{{Garc\'{\i}a de Abajo}}
  \emph{et~al.}(2005{\natexlab{a}})\citenamefont{{Garc\'{\i}a de Abajo},
  {G\'{o}mez-Medina}, and S\'{a}enz}}]{paper102}
\bibinfo{author}{\bibnamefont{{Garc\'{\i}a de Abajo}}, \bibfnamefont{F.~J.}},
  \bibinfo{author}{\bibfnamefont{R.}~\bibnamefont{{G\'{o}mez-Medina}}}, and
  \bibinfo{author}{\bibfnamefont{J.~J.} \bibnamefont{S\'{a}enz}},
  \bibinfo{year}{2005}{\natexlab{a}}, {``}\bibinfo{title}{Full transmission
  through perfect-conductor subwavelength hole arrays},{''}
  \bibinfo{journal}{Phys.\ Rev.\ E} \textbf{\bibinfo{volume}{72}},
  \bibinfo{pages}{016608}.

\bibitem[{\citenamefont{{Garc\'{\i}a de Abajo}}
  \emph{et~al.}(2005{\natexlab{b}})\citenamefont{{Garc\'{\i}a de Abajo},
  {G\'{o}mez-Santos}, Blanco, Borisov, and Shabanov}}]{paper100}
\bibinfo{author}{\bibnamefont{{Garc\'{\i}a de Abajo}}, \bibfnamefont{F.~J.}},
  \bibinfo{author}{\bibfnamefont{G.}~\bibnamefont{{G\'{o}mez-Santos}}},
  \bibinfo{author}{\bibfnamefont{L.~A.} \bibnamefont{Blanco}},
  \bibinfo{author}{\bibfnamefont{A.~G.} \bibnamefont{Borisov}}, and
  \bibinfo{author}{\bibfnamefont{S.~V.} \bibnamefont{Shabanov}},
  \bibinfo{year}{2005}{\natexlab{b}}, {``}\bibinfo{title}{Tunneling mechanism
  of light transmission through metallic films},{''} \bibinfo{journal}{Phys.\
  Rev.\ Lett.} \textbf{\bibinfo{volume}{95}},  \bibinfo{pages}{067403}.

\bibitem[{\citenamefont{{Garc\'{\i}a de Abajo} and S\'{a}enz}(2005)}]{paper105}
\bibinfo{author}{\bibnamefont{{Garc\'{\i}a de Abajo}}, \bibfnamefont{F.~J.}},
  and \bibinfo{author}{\bibfnamefont{J.~J.} \bibnamefont{S\'{a}enz}},
  \bibinfo{year}{2005}, {``}\bibinfo{title}{Electromagnetic surface modes in
  structured perfect-conductor surfaces},{''} \bibinfo{journal}{Phys.\ Rev.\
  Lett.} \textbf{\bibinfo{volume}{95}},  \bibinfo{pages}{233901}.

\bibitem[{\citenamefont{{Garc\'{\i}a de Abajo}}
  \emph{et~al.}(2006)\citenamefont{{Garc\'{\i}a de Abajo}, S\'{a}enz, Campillo,
  and Dolado}}]{paper106}
\bibinfo{author}{\bibnamefont{{Garc\'{\i}a de Abajo}}, \bibfnamefont{F.~J.}},
  \bibinfo{author}{\bibfnamefont{J.~J.} \bibnamefont{S\'{a}enz}},
  \bibinfo{author}{\bibfnamefont{I.}~\bibnamefont{Campillo}}, and
  \bibinfo{author}{\bibfnamefont{J.~S.} \bibnamefont{Dolado}},
  \bibinfo{year}{2006}, {``}\bibinfo{title}{Site and lattice resonances in
  metallic hole arrays},{''} \bibinfo{journal}{Opt.\ Express}
  \textbf{\bibinfo{volume}{14}},  \bibinfo{pages}{7--18}.

\bibitem[{\citenamefont{{Garc\'{\i}a-Vidal}}
  \emph{et~al.}(2005)\citenamefont{{Garc\'{\i}a-Vidal}, Moreno, Porto, and
  {{Mart\'{\i}n-Moreno}}}}]{GMP05}
\bibinfo{author}{\bibnamefont{{Garc\'{\i}a-Vidal}}, \bibfnamefont{F.~J.}},
  \bibinfo{author}{\bibfnamefont{E.}~\bibnamefont{Moreno}},
  \bibinfo{author}{\bibfnamefont{J.~A.} \bibnamefont{Porto}}, and
  \bibinfo{author}{\bibfnamefont{L.}~\bibnamefont{{{Mart\'{\i}n-Moreno}}}},
  \bibinfo{year}{2005}, {``}\bibinfo{title}{Transmission of light through a
  single rectangular hole},{''} \bibinfo{journal}{Phys.\ Rev.\ Lett.}
  \textbf{\bibinfo{volume}{95}},  \bibinfo{pages}{103901}.

\bibitem[{\citenamefont{{Garc\'{\i}a-Vidal}}
  \emph{et~al.}(2006)\citenamefont{{Garc\'{\i}a-Vidal}, Rodrigo, and
  {Mart\'{\i}n-Moreno}}}]{GRM06}
\bibinfo{author}{\bibnamefont{{Garc\'{\i}a-Vidal}}, \bibfnamefont{F.~J.}},
  \bibinfo{author}{\bibfnamefont{S.~G.} \bibnamefont{Rodrigo}}, and
  \bibinfo{author}{\bibfnamefont{L.}~\bibnamefont{{Mart\'{\i}n-Moreno}}},
  \bibinfo{year}{2006}, {``}\bibinfo{title}{Foundations of the composite
  diffracted evanescent wave model},{''} \bibinfo{journal}{Nat.\ Phys.}
  \textbf{\bibinfo{volume}{2}},  \bibinfo{pages}{790--790}.

\bibitem[{\citenamefont{Gay} \emph{et~al.}(2006)\citenamefont{Gay, Alloschery,
  {De Lesegno}, {O'Dwyer}, Weiner, and Lezec}}]{GAD06}
\bibinfo{author}{\bibnamefont{Gay}, \bibfnamefont{G.}},
  \bibinfo{author}{\bibfnamefont{O.}~\bibnamefont{Alloschery}},
  \bibinfo{author}{\bibfnamefont{B.~V.} \bibnamefont{{De Lesegno}}},
  \bibinfo{author}{\bibfnamefont{C.}~\bibnamefont{{O'Dwyer}}},
  \bibinfo{author}{\bibfnamefont{J.}~\bibnamefont{Weiner}}, and
  \bibinfo{author}{\bibfnamefont{H.~J.} \bibnamefont{Lezec}},
  \bibinfo{year}{2006}, {``}\bibinfo{title}{The optical response of
  nanostructured surfaces and the composite diffracted evanescent wave
  model},{''} \bibinfo{journal}{Nat.\ Phys.} \textbf{\bibinfo{volume}{2}},
  \bibinfo{pages}{262--267}.

\bibitem[{\citenamefont{Genet and Ebbesen}(2007)}]{GE07}
\bibinfo{author}{\bibnamefont{Genet}, \bibfnamefont{C.}}, and
  \bibinfo{author}{\bibfnamefont{T.~W.} \bibnamefont{Ebbesen}},
  \bibinfo{year}{2007}, {``}\bibinfo{title}{Light in tiny holes},{''}
  \bibinfo{journal}{Nature} \textbf{\bibinfo{volume}{445}},
  \bibinfo{pages}{39--46}.

\bibitem[{\citenamefont{Genet} \emph{et~al.}(2003)\citenamefont{Genet, {van
  Exter}, and Woerdman}}]{GEW03}
\bibinfo{author}{\bibnamefont{Genet}, \bibfnamefont{C.}},
  \bibinfo{author}{\bibfnamefont{M.~P.} \bibnamefont{{van Exter}}}, and
  \bibinfo{author}{\bibfnamefont{J.~P.} \bibnamefont{Woerdman}},
  \bibinfo{year}{2003}, {``}\bibinfo{title}{Fano-type interpretation of red
  shifts and red tails in hole array transmission spectra},{''}
  \bibinfo{journal}{Opt.\ Commun.} \textbf{\bibinfo{volume}{225}},
  \bibinfo{pages}{331–--336}.

\bibitem[{\citenamefont{Ghaemi} \emph{et~al.}(1998)\citenamefont{Ghaemi, Thio,
  Grupp, Ebbesen, and Lezec}}]{GTG98}
\bibinfo{author}{\bibnamefont{Ghaemi}, \bibfnamefont{H.~F.}},
  \bibinfo{author}{\bibfnamefont{T.}~\bibnamefont{Thio}},
  \bibinfo{author}{\bibfnamefont{D.~E.} \bibnamefont{Grupp}},
  \bibinfo{author}{\bibfnamefont{T.~W.} \bibnamefont{Ebbesen}}, and
  \bibinfo{author}{\bibfnamefont{H.~J.} \bibnamefont{Lezec}},
  \bibinfo{year}{1998}, {``}\bibinfo{title}{Surface plasmons enhance optical
  transmission through subwavelength holes},{''} \bibinfo{journal}{Phys.\ Rev.\
  B} \textbf{\bibinfo{volume}{58}},  \bibinfo{pages}{6779--6782}.

\bibitem[{\citenamefont{Glasser and Zucker}(1980)}]{GZ1980}
\bibinfo{author}{\bibnamefont{Glasser}, \bibfnamefont{M.~L.}}, and
  \bibinfo{author}{\bibfnamefont{I.~J.} \bibnamefont{Zucker}},
  \bibinfo{year}{1980}, in \emph{\bibinfo{booktitle}{Theoretical Chemistry:
  Advances and Perspectives}}, edited by
  \bibinfo{editor}{\bibfnamefont{H.}~\bibnamefont{Eyring}} and
  \bibinfo{editor}{\bibfnamefont{D.}~\bibnamefont{Henderson}}
  (\bibinfo{publisher}{Academic Press}, \bibinfo{address}{New York}),
  volume~\bibinfo{volume}{5},  \bibinfo{pages}{67--139}.

\bibitem[{\citenamefont{{G\'omez-Medina}}
  \emph{et~al.}(2006)\citenamefont{{G\'omez-Medina}, Laroche, and
  {S\'aenz}}}]{GLS06}
\bibinfo{author}{\bibnamefont{{G\'omez-Medina}}, \bibfnamefont{R.}},
  \bibinfo{author}{\bibfnamefont{M.}~\bibnamefont{Laroche}}, and
  \bibinfo{author}{\bibfnamefont{J.~J.} \bibnamefont{{S\'aenz}}},
  \bibinfo{year}{2006}, {``}\bibinfo{title}{Extraordinary optical reflection
  from sub-wavelength cylinder arrays},{''} \bibinfo{journal}{Opt.\ Express}
  \textbf{\bibinfo{volume}{14}},  \bibinfo{pages}{3730--3737}.

\bibitem[{\citenamefont{{G\'{o}mez-Rivas}}
  \emph{et~al.}(2003)\citenamefont{{G\'{o}mez-Rivas}, Schotsch, {Haring
  Bolivar}, and Kurz}}]{GSH03}
\bibinfo{author}{\bibnamefont{{G\'{o}mez-Rivas}}, \bibfnamefont{J.}},
  \bibinfo{author}{\bibfnamefont{C.}~\bibnamefont{Schotsch}},
  \bibinfo{author}{\bibfnamefont{P.}~\bibnamefont{{Haring Bolivar}}}, and
  \bibinfo{author}{\bibfnamefont{H.}~\bibnamefont{Kurz}}, \bibinfo{year}{2003},
  {``}\bibinfo{title}{Enhanced transmission of THz radiation through
  subwavelength holes},{''} \bibinfo{journal}{Phys.\ Rev.\ B}
  \textbf{\bibinfo{volume}{68}},  \bibinfo{pages}{201306(R)}.

\bibitem[{\citenamefont{Gordon} \emph{et~al.}(2004)\citenamefont{Gordon, Brolo,
  McKinnon, Rajora, Leathem, and Kavanagh}}]{GBM04}
\bibinfo{author}{\bibnamefont{Gordon}, \bibfnamefont{R.}},
  \bibinfo{author}{\bibfnamefont{A.~G.} \bibnamefont{Brolo}},
  \bibinfo{author}{\bibfnamefont{A.}~\bibnamefont{McKinnon}},
  \bibinfo{author}{\bibfnamefont{A.}~\bibnamefont{Rajora}},
  \bibinfo{author}{\bibfnamefont{B.}~\bibnamefont{Leathem}}, and
  \bibinfo{author}{\bibfnamefont{K.~L.} \bibnamefont{Kavanagh}},
  \bibinfo{year}{2004}, {``}\bibinfo{title}{Strong polarization in the optical
  transmission through elliptical nanohole arrays},{''}
  \bibinfo{journal}{Phys.\ Rev.\ Lett.} \textbf{\bibinfo{volume}{92}},
  \bibinfo{pages}{037401}.

\bibitem[{\citenamefont{Gradshteyn and Ryzhik}(1980)}]{GR1980}
\bibinfo{author}{\bibnamefont{Gradshteyn}, \bibfnamefont{I.~S.}}, and
  \bibinfo{author}{\bibfnamefont{I.~M.} \bibnamefont{Ryzhik}},
  \bibinfo{year}{1980}, \emph{\bibinfo{title}{Table of Integrals, Series, and
  Products}} (\bibinfo{publisher}{Academic Press}, \bibinfo{address}{London}).

\bibitem[{\citenamefont{Greffet} \emph{et~al.}(2002)\citenamefont{Greffet,
  Carminati, Joulain, Mulet, Mainguy, and Chen}}]{GCJ02}
\bibinfo{author}{\bibnamefont{Greffet}, \bibfnamefont{J.-J.}},
  \bibinfo{author}{\bibfnamefont{R.}~\bibnamefont{Carminati}},
  \bibinfo{author}{\bibfnamefont{K.}~\bibnamefont{Joulain}},
  \bibinfo{author}{\bibfnamefont{J.-P.} \bibnamefont{Mulet}},
  \bibinfo{author}{\bibfnamefont{S.}~\bibnamefont{Mainguy}}, and
  \bibinfo{author}{\bibfnamefont{Y.}~\bibnamefont{Chen}}, \bibinfo{year}{2002},
  {``}\bibinfo{title}{Coherent emission of light by thermal sources},{''}
  \bibinfo{journal}{Nature} \textbf{\bibinfo{volume}{416}},
  \bibinfo{pages}{61--64}.

\bibitem[{\citenamefont{Grigorenko}
  \emph{et~al.}(2005)\citenamefont{Grigorenko, Geim, Gleeson, Zhang, Firsov,
  Khrushchev, and Petrovic}}]{GGG05}
\bibinfo{author}{\bibnamefont{Grigorenko}, \bibfnamefont{A.~N.}},
  \bibinfo{author}{\bibfnamefont{A.~K.} \bibnamefont{Geim}},
  \bibinfo{author}{\bibfnamefont{H.~F.} \bibnamefont{Gleeson}},
  \bibinfo{author}{\bibfnamefont{Y.}~\bibnamefont{Zhang}},
  \bibinfo{author}{\bibfnamefont{A.~A.} \bibnamefont{Firsov}},
  \bibinfo{author}{\bibfnamefont{I.~Y.} \bibnamefont{Khrushchev}}, and
  \bibinfo{author}{\bibfnamefont{J.}~\bibnamefont{Petrovic}},
  \bibinfo{year}{2005}, {``}\bibinfo{title}{Nanofabricated media with negative
  permeability at visible frequencies},{''} \bibinfo{journal}{Nature}
  \textbf{\bibinfo{volume}{438}},  \bibinfo{pages}{335--338}.

\bibitem[{\citenamefont{Henke} \emph{et~al.}(1993)\citenamefont{Henke,
  Gullikson, and Davis}}]{HGD93}
\bibinfo{author}{\bibnamefont{Henke}, \bibfnamefont{B.~L.}},
  \bibinfo{author}{\bibfnamefont{E.~M.} \bibnamefont{Gullikson}}, and
  \bibinfo{author}{\bibfnamefont{J.~C.} \bibnamefont{Davis}},
  \bibinfo{year}{1993}, {``}\bibinfo{title}{X-ray interactions:
  photoabsorption, scattering, transmission, and reflection at $E=50-30,000$
  eV, $Z=1-92$},{''} \bibinfo{journal}{Atom.\ Data\ Nucl.\ Data\ Tables}
  \textbf{\bibinfo{volume}{54}},  \bibinfo{pages}{181--342}.

\bibitem[{\citenamefont{Hibbins} \emph{et~al.}(2005)\citenamefont{Hibbins,
  Evans, and Sambles}}]{HES05}
\bibinfo{author}{\bibnamefont{Hibbins}, \bibfnamefont{A.~P.}},
  \bibinfo{author}{\bibfnamefont{B.~R.} \bibnamefont{Evans}}, and
  \bibinfo{author}{\bibfnamefont{J.~R.} \bibnamefont{Sambles}},
  \bibinfo{year}{2005}, {``}\bibinfo{title}{Experimental verification of
  designer surface plasmons},{''} \bibinfo{journal}{Science}
  \textbf{\bibinfo{volume}{308}},  \bibinfo{pages}{670--672}.

\bibitem[{\citenamefont{Hibbins} \emph{et~al.}(2006)\citenamefont{Hibbins,
  Lockyear, Hooper, and Sambles}}]{HLH06}
\bibinfo{author}{\bibnamefont{Hibbins}, \bibfnamefont{A.~P.}},
  \bibinfo{author}{\bibfnamefont{M.~J.} \bibnamefont{Lockyear}},
  \bibinfo{author}{\bibfnamefont{I.~R.} \bibnamefont{Hooper}}, and
  \bibinfo{author}{\bibfnamefont{J.~R.} \bibnamefont{Sambles}},
  \bibinfo{year}{2006}, {``}\bibinfo{title}{Waveguide arrays as plasmonic
  metamaterials: transmission below cutoff},{''} \bibinfo{journal}{Phys.\ Rev.\
  Lett.} \textbf{\bibinfo{volume}{96}},  \bibinfo{pages}{073904}.

\bibitem[{\citenamefont{Hicks} \emph{et~al.}(2005)\citenamefont{Hicks, Zou,
  Schatz, Spears, {Van Duyne}, Gunnarsson, Rindzevicius, Kasemo, and
  K\"all}}]{HZS05}
\bibinfo{author}{\bibnamefont{Hicks}, \bibfnamefont{E.~M.}},
  \bibinfo{author}{\bibfnamefont{S.}~\bibnamefont{Zou}},
  \bibinfo{author}{\bibfnamefont{G.~C.} \bibnamefont{Schatz}},
  \bibinfo{author}{\bibfnamefont{K.~G.} \bibnamefont{Spears}},
  \bibinfo{author}{\bibfnamefont{R.~P.} \bibnamefont{{Van Duyne}}},
  \bibinfo{author}{\bibfnamefont{L.}~\bibnamefont{Gunnarsson}},
  \bibinfo{author}{\bibfnamefont{T.}~\bibnamefont{Rindzevicius}},
  \bibinfo{author}{\bibfnamefont{B.}~\bibnamefont{Kasemo}}, and
  \bibinfo{author}{\bibfnamefont{M.}~\bibnamefont{K\"all}},
  \bibinfo{year}{2005}, {``}\bibinfo{title}{Controlling plasmon line shapes
  through diffractive coupling in linear arrays of cylindrical nanoparticles
  fabricated by electron beam lithography},{''} \bibinfo{journal}{Nano\ Lett.}
  \textbf{\bibinfo{volume}{5}},  \bibinfo{pages}{1065--1070}.

\bibitem[{\citenamefont{Hillenbrand}
  \emph{et~al.}(2002)\citenamefont{Hillenbrand, Taubner, and Keilmann}}]{HTK02}
\bibinfo{author}{\bibnamefont{Hillenbrand}, \bibfnamefont{R.}},
  \bibinfo{author}{\bibfnamefont{T.}~\bibnamefont{Taubner}}, and
  \bibinfo{author}{\bibfnamefont{F.}~\bibnamefont{Keilmann}},
  \bibinfo{year}{2002}, {``}\bibinfo{title}{Phonon-enhanced light-matter
  interaction at the nanometer scale},{''} \bibinfo{journal}{Nature}
  \textbf{\bibinfo{volume}{418}},  \bibinfo{pages}{159--162}.

\bibitem[{\citenamefont{Huang} \emph{et~al.}(2007)\citenamefont{Huang,
  Zheludev, Chen, and {Garc\'{\i}a de Abajo}}}]{paper119}
\bibinfo{author}{\bibnamefont{Huang}, \bibfnamefont{F.~M.}},
  \bibinfo{author}{\bibfnamefont{N.}~\bibnamefont{Zheludev}},
  \bibinfo{author}{\bibfnamefont{Y.}~\bibnamefont{Chen}}, and
  \bibinfo{author}{\bibfnamefont{F.~J.} \bibnamefont{{Garc\'{\i}a de Abajo}}},
  \bibinfo{year}{2007}, {``}\bibinfo{title}{Focusing of light by a nano-hole
  array},{''} \bibinfo{journal}{Appl.\ Phys.\ Lett.}
  \textbf{\bibinfo{volume}{90}},  \bibinfo{pages}{091119}.

\bibitem[{\citenamefont{Hutley and Maystre}(1976)}]{HM1976}
\bibinfo{author}{\bibnamefont{Hutley}, \bibfnamefont{M.~C.}}, and
  \bibinfo{author}{\bibfnamefont{D.}~\bibnamefont{Maystre}},
  \bibinfo{year}{1976}, {``}\bibinfo{title}{The total absorption of light by a
  diffraction grating},{''} \bibinfo{journal}{Opt.\ Commun.}
  \textbf{\bibinfo{volume}{19}},  \bibinfo{pages}{431--436}.

\bibitem[{\citenamefont{Jackson}(1999)}]{J99}
\bibinfo{author}{\bibnamefont{Jackson}, \bibfnamefont{J.~D.}},
  \bibinfo{year}{1999}, \emph{\bibinfo{title}{Classical Electrodynamics}}
  (\bibinfo{publisher}{Wiley}, \bibinfo{address}{New York}).

\bibitem[{\citenamefont{James}(1977)}]{J1977}
\bibinfo{author}{\bibnamefont{James}, \bibfnamefont{G.~L.}},
  \bibinfo{year}{1977}, {``}\bibinfo{title}{Radiation properties of 90 degrees
  conical horns},{''} \bibinfo{journal}{Electron.\ Lett.}
  \textbf{\bibinfo{volume}{13}},  \bibinfo{pages}{293--294}.

\bibitem[{\citenamefont{Janke} \emph{et~al.}(2005)\citenamefont{Janke, {G\'omez
  Rivas}, {Haring Bolivar}, and Kurz}}]{JGH05}
\bibinfo{author}{\bibnamefont{Janke}, \bibfnamefont{C.}},
  \bibinfo{author}{\bibfnamefont{J.}~\bibnamefont{{G\'omez Rivas}}},
  \bibinfo{author}{\bibfnamefont{P.}~\bibnamefont{{Haring Bolivar}}}, and
  \bibinfo{author}{\bibfnamefont{H.}~\bibnamefont{Kurz}}, \bibinfo{year}{2005},
  {``}\bibinfo{title}{All-optical switching of the transmission of
  electromagnetic radiation through subwavelength apertures},{''}
  \bibinfo{journal}{Opt.\ Lett.} \textbf{\bibinfo{volume}{30}},
  \bibinfo{pages}{2357--2359}.

\bibitem[{\citenamefont{Joannopoulos}
  \emph{et~al.}(1997)\citenamefont{Joannopoulos, Villeneuve, and Fan}}]{JVF97}
\bibinfo{author}{\bibnamefont{Joannopoulos}, \bibfnamefont{J.~D.}},
  \bibinfo{author}{\bibfnamefont{P.~R.} \bibnamefont{Villeneuve}}, and
  \bibinfo{author}{\bibfnamefont{S.~H.} \bibnamefont{Fan}},
  \bibinfo{year}{1997}, {``}\bibinfo{title}{Photonic crystals: putting a new
  twist on light},{''} \bibinfo{journal}{Nature}
  \textbf{\bibinfo{volume}{386}},  \bibinfo{pages}{143--149}.

\bibitem[{\citenamefont{Johnson and Christy}(1972)}]{JC1972}
\bibinfo{author}{\bibnamefont{Johnson}, \bibfnamefont{P.~B.}}, and
  \bibinfo{author}{\bibfnamefont{R.~W.} \bibnamefont{Christy}},
  \bibinfo{year}{1972}, {``}\bibinfo{title}{Optical constants of the noble
  metals},{''} \bibinfo{journal}{Phys.\ Rev.\ B} \textbf{\bibinfo{volume}{6}},
  \bibinfo{pages}{4370--4379}.

\bibitem[{\citenamefont{Jones}(1945)}]{J1945}
\bibinfo{author}{\bibnamefont{Jones}, \bibfnamefont{R.~C.}},
  \bibinfo{year}{1945}, {``}\bibinfo{title}{A generalization of the dielectric
  ellipsoid problem},{''} \bibinfo{journal}{Phys.\ Rev.}
  \textbf{\bibinfo{volume}{68}},  \bibinfo{pages}{93--96}.

\bibitem[{\citenamefont{Kambe}(1968)}]{K1968}
\bibinfo{author}{\bibnamefont{Kambe}, \bibfnamefont{K.}}, \bibinfo{year}{1968},
  {``}\bibinfo{title}{Theory of low-energy electron diffraction 2. Cellular
  method for complex monolayers and multilayers},{''} \bibinfo{journal}{Z.\
  Naturfors.\ A} \textbf{\bibinfo{volume}{23}},  \bibinfo{pages}{1280--1294}.

\bibitem[{\citenamefont{Kelf} \emph{et~al.}(2005)\citenamefont{Kelf, Sugawara,
  Baumberg, Abdelsalam, and Bartlett}}]{KSB05}
\bibinfo{author}{\bibnamefont{Kelf}, \bibfnamefont{T.~A.}},
  \bibinfo{author}{\bibfnamefont{Y.}~\bibnamefont{Sugawara}},
  \bibinfo{author}{\bibfnamefont{J.~J.} \bibnamefont{Baumberg}},
  \bibinfo{author}{\bibfnamefont{M.}~\bibnamefont{Abdelsalam}}, and
  \bibinfo{author}{\bibfnamefont{P.~N.} \bibnamefont{Bartlett}},
  \bibinfo{year}{2005}, {``}\bibinfo{title}{Plasmonic band gaps and trapped
  plasmons on nanostructured metal surfaces},{''} \bibinfo{journal}{Phys.\
  Rev.\ Lett.} \textbf{\bibinfo{volume}{95}},  \bibinfo{pages}{116802}.

\bibitem[{\citenamefont{Kelf} \emph{et~al.}(2006)\citenamefont{Kelf, Sugawara,
  Cole, Baumberg, Abdelsalam, Cintra, Mahajan, Russell, and Bartlett}}]{KSC06}
\bibinfo{author}{\bibnamefont{Kelf}, \bibfnamefont{T.~A.}},
  \bibinfo{author}{\bibfnamefont{Y.}~\bibnamefont{Sugawara}},
  \bibinfo{author}{\bibfnamefont{R.~M.} \bibnamefont{Cole}},
  \bibinfo{author}{\bibfnamefont{J.~J.} \bibnamefont{Baumberg}},
  \bibinfo{author}{\bibfnamefont{M.~E.} \bibnamefont{Abdelsalam}},
  \bibinfo{author}{\bibfnamefont{S.}~\bibnamefont{Cintra}},
  \bibinfo{author}{\bibfnamefont{S.}~\bibnamefont{Mahajan}},
  \bibinfo{author}{\bibfnamefont{A.~E.} \bibnamefont{Russell}}, and
  \bibinfo{author}{\bibfnamefont{P.~N.} \bibnamefont{Bartlett}},
  \bibinfo{year}{2006}, {``}\bibinfo{title}{Localized and delocalized plasmons
  in metallic nanovoids},{''} \bibinfo{journal}{Phys.\ Rev.\ B}
  \textbf{\bibinfo{volume}{74}},  \bibinfo{pages}{245415}.

\bibitem[{\citenamefont{Kitson} \emph{et~al.}(1996)\citenamefont{Kitson,
  Barnes, and Sambles}}]{KBS96}
\bibinfo{author}{\bibnamefont{Kitson}, \bibfnamefont{S.~C.}},
  \bibinfo{author}{\bibfnamefont{W.~L.} \bibnamefont{Barnes}}, and
  \bibinfo{author}{\bibfnamefont{J.~R.} \bibnamefont{Sambles}},
  \bibinfo{year}{1996}, {``}\bibinfo{title}{Full photonic band gap for surface
  modes in the visible},{''} \bibinfo{journal}{Phys.\ Rev.\ Lett.}
  \textbf{\bibinfo{volume}{77}},  \bibinfo{pages}{2670--2673}.

\bibitem[{\citenamefont{{Klein Koerkamp}}
  \emph{et~al.}(2004)\citenamefont{{Klein Koerkamp}, Enoch, Segerink, {van
  Hulst}, and Kuipers}}]{KES04}
\bibinfo{author}{\bibnamefont{{Klein Koerkamp}}, \bibfnamefont{K.~J.}},
  \bibinfo{author}{\bibfnamefont{S.}~\bibnamefont{Enoch}},
  \bibinfo{author}{\bibfnamefont{F.~B.} \bibnamefont{Segerink}},
  \bibinfo{author}{\bibfnamefont{N.~F.} \bibnamefont{{van Hulst}}}, and
  \bibinfo{author}{\bibfnamefont{L.}~\bibnamefont{Kuipers}},
  \bibinfo{year}{2004}, {``}\bibinfo{title}{Strong influence of hole shape on
  extraordinary transmission through periodic arrays of subwavelength
  holes},{''} \bibinfo{journal}{Phys.\ Rev.\ Lett.}
  \textbf{\bibinfo{volume}{92}},  \bibinfo{pages}{183901}.

\bibitem[{\citenamefont{Krasavin} \emph{et~al.}(2005)\citenamefont{Krasavin,
  Schwanecke, Zheludev, Reichelt, Stroucken, Koch, and Wright}}]{KSZ05}
\bibinfo{author}{\bibnamefont{Krasavin}, \bibfnamefont{A.~V.}},
  \bibinfo{author}{\bibfnamefont{A.~S.} \bibnamefont{Schwanecke}},
  \bibinfo{author}{\bibfnamefont{N.~I.} \bibnamefont{Zheludev}},
  \bibinfo{author}{\bibfnamefont{M.}~\bibnamefont{Reichelt}},
  \bibinfo{author}{\bibfnamefont{T.}~\bibnamefont{Stroucken}},
  \bibinfo{author}{\bibfnamefont{S.~W.} \bibnamefont{Koch}}, and
  \bibinfo{author}{\bibfnamefont{E.~M.} \bibnamefont{Wright}},
  \bibinfo{year}{2005}, {``}\bibinfo{title}{Polarization conversion and
  "focusing" of light propagating through a small chiral hole in a metallic
  screen},{''} \bibinfo{journal}{Appl.\ Phys.\ Lett.}
  \textbf{\bibinfo{volume}{86}},  \bibinfo{pages}{201105}.

\bibitem[{\citenamefont{Krenn} \emph{et~al.}(1999)\citenamefont{Krenn, Dereux,
  Weeber, Bourillot, Lacroute, Goudonnet, Schider, Gotschy, Leitner, Aussenegg,
  and Girard}}]{KDW99}
\bibinfo{author}{\bibnamefont{Krenn}, \bibfnamefont{J.~R.}},
  \bibinfo{author}{\bibfnamefont{A.}~\bibnamefont{Dereux}},
  \bibinfo{author}{\bibfnamefont{J.~C.} \bibnamefont{Weeber}},
  \bibinfo{author}{\bibfnamefont{E.}~\bibnamefont{Bourillot}},
  \bibinfo{author}{\bibfnamefont{Y.}~\bibnamefont{Lacroute}},
  \bibinfo{author}{\bibfnamefont{J.~P.} \bibnamefont{Goudonnet}},
  \bibinfo{author}{\bibfnamefont{G.}~\bibnamefont{Schider}},
  \bibinfo{author}{\bibfnamefont{W.}~\bibnamefont{Gotschy}},
  \bibinfo{author}{\bibfnamefont{A.}~\bibnamefont{Leitner}},
  \bibinfo{author}{\bibfnamefont{F.~R.} \bibnamefont{Aussenegg}}, and
  \bibinfo{author}{\bibfnamefont{C.}~\bibnamefont{Girard}},
  \bibinfo{year}{1999}, {``}\bibinfo{title}{Squeezing the optical near-field
  zone by plasmon coupling of metallic nanoparticles},{''}
  \bibinfo{journal}{Phys.\ Rev.\ Lett.} \textbf{\bibinfo{volume}{82}},
  \bibinfo{pages}{2590--2593}.

\bibitem[{\citenamefont{Krishnan} \emph{et~al.}(2001)\citenamefont{Krishnan,
  Thio, Kim, Lezec, Ebbesen, Wolff, Pendry, {Mart\'{\i}n-Moreno}, and
  {Garc\'{\i}a-Vidal}}}]{KTK01}
\bibinfo{author}{\bibnamefont{Krishnan}, \bibfnamefont{A.}},
  \bibinfo{author}{\bibfnamefont{T.}~\bibnamefont{Thio}},
  \bibinfo{author}{\bibfnamefont{T.~J.} \bibnamefont{Kim}},
  \bibinfo{author}{\bibfnamefont{H.~J.} \bibnamefont{Lezec}},
  \bibinfo{author}{\bibfnamefont{T.~W.} \bibnamefont{Ebbesen}},
  \bibinfo{author}{\bibfnamefont{P.~A.} \bibnamefont{Wolff}},
  \bibinfo{author}{\bibfnamefont{J.}~\bibnamefont{Pendry}},
  \bibinfo{author}{\bibfnamefont{L.}~\bibnamefont{{Mart\'{\i}n-Moreno}}}, and
  \bibinfo{author}{\bibfnamefont{F.~J.} \bibnamefont{{Garc\'{\i}a-Vidal}}},
  \bibinfo{year}{2001}, {``}\bibinfo{title}{Evanescently coupled resonance in
  surface plasmon enhanced transmission},{''} \bibinfo{journal}{Opt.\ Commun.}
  \textbf{\bibinfo{volume}{200}},  \bibinfo{pages}{1--7}.

\bibitem[{\citenamefont{Lalanne and Hugonin}(2006)}]{LH06}
\bibinfo{author}{\bibnamefont{Lalanne}, \bibfnamefont{P.}}, and
  \bibinfo{author}{\bibfnamefont{J.~P.} \bibnamefont{Hugonin}},
  \bibinfo{year}{2006}, {``}\bibinfo{title}{Interaction between optical
  nano-objects at metallo-dielectric interfaces},{''} \bibinfo{journal}{Nat.\
  Phys.} \textbf{\bibinfo{volume}{2}},  \bibinfo{pages}{551--556}.

\bibitem[{\citenamefont{Laroche} \emph{et~al.}(2006)\citenamefont{Laroche,
  Albaladejo, {G\'omez-Medina}, and {S\'aenz}}}]{LAG06}
\bibinfo{author}{\bibnamefont{Laroche}, \bibfnamefont{M.}},
  \bibinfo{author}{\bibfnamefont{S.}~\bibnamefont{Albaladejo}},
  \bibinfo{author}{\bibfnamefont{R.}~\bibnamefont{{G\'omez-Medina}}}, and
  \bibinfo{author}{\bibfnamefont{J.~J.} \bibnamefont{{S\'aenz}}},
  \bibinfo{year}{2006}, {``}\bibinfo{title}{Tuning the optical response of
  nanocylinder arrays: an analytical study},{''} \bibinfo{journal}{Phys.\ Rev.\
  B} \textbf{\bibinfo{volume}{74}},  \bibinfo{pages}{245422}.

\bibitem[{\citenamefont{Lezec} \emph{et~al.}(2002)\citenamefont{Lezec, Degiron,
  Devaux, Linke, Mart\'{\i}n-Moreno, {Garc\'{\i}a-Vidal}, and Ebbesen}}]{LDD02}
\bibinfo{author}{\bibnamefont{Lezec}, \bibfnamefont{H.~J.}},
  \bibinfo{author}{\bibfnamefont{A.}~\bibnamefont{Degiron}},
  \bibinfo{author}{\bibfnamefont{E.}~\bibnamefont{Devaux}},
  \bibinfo{author}{\bibfnamefont{R.~A.} \bibnamefont{Linke}},
  \bibinfo{author}{\bibfnamefont{L.}~\bibnamefont{Mart\'{\i}n-Moreno}},
  \bibinfo{author}{\bibfnamefont{F.~J.} \bibnamefont{{Garc\'{\i}a-Vidal}}}, and
  \bibinfo{author}{\bibfnamefont{T.~W.} \bibnamefont{Ebbesen}},
  \bibinfo{year}{2002}, {``}\bibinfo{title}{Beaming light from a subwavelength
  aperture},{''} \bibinfo{journal}{Science} \textbf{\bibinfo{volume}{297}},
  \bibinfo{pages}{820--822}.

\bibitem[{\citenamefont{Lezec and Thio}(2004)}]{LT04}
\bibinfo{author}{\bibnamefont{Lezec}, \bibfnamefont{H.~J.}}, and
  \bibinfo{author}{\bibfnamefont{T.}~\bibnamefont{Thio}}, \bibinfo{year}{2004},
  {``}\bibinfo{title}{Diffracted evanescent wave model for enhanced and
  suppressed optical transmission through subwavelength hole arrays},{''}
  \bibinfo{journal}{Opt.\ Express} \textbf{\bibinfo{volume}{12}},
  \bibinfo{pages}{3629--3651}.

\bibitem[{\citenamefont{Liz-Marz\'an}(2006)}]{L06}
\bibinfo{author}{\bibnamefont{Liz-Marz\'an}, \bibfnamefont{L.~M.}},
  \bibinfo{year}{2006}, {``}\bibinfo{title}{Tailoring surface plasmon through
  the morphology and assembly of metal nanoparticles},{''}
  \bibinfo{journal}{Langmuir} \textbf{\bibinfo{volume}{22}},
  \bibinfo{pages}{32--41}.

\bibitem[{\citenamefont{Loewen} \emph{et~al.}(1984)\citenamefont{Loewen,
  McKinney, and McPhedran}}]{LMM1984}
\bibinfo{author}{\bibnamefont{Loewen}, \bibfnamefont{E.~G.}},
  \bibinfo{author}{\bibfnamefont{W.~R.} \bibnamefont{McKinney}}, and
  \bibinfo{author}{\bibfnamefont{R.}~\bibnamefont{McPhedran}},
  \bibinfo{year}{1984}, in \emph{\bibinfo{booktitle}{Application, Theory, and
  Fabrication of Periodic Structures}}, edited by
  \bibinfo{editor}{\bibfnamefont{J.~M.} \bibnamefont{Lerner}}
  (\bibinfo{publisher}{SPIE}), volume \bibinfo{volume}{503},
  \bibinfo{pages}{187--197}.

\bibitem[{\citenamefont{{L\'opez}}(2003)}]{C03}
\bibinfo{author}{\bibnamefont{{L\'opez}}, \bibfnamefont{C.}},
  \bibinfo{year}{2003}, {``}\bibinfo{title}{Materials aspects of photonic
  crystals},{''} \bibinfo{journal}{Adv.\ Mater.} \textbf{\bibinfo{volume}{15}},
   \bibinfo{pages}{1679--1704}.

\bibitem[{\citenamefont{{Lord Rayleigh}}(1907)}]{R1907}
\bibinfo{author}{\bibnamefont{{Lord Rayleigh}}}, \bibinfo{year}{1907},
  {``}\bibinfo{title}{Note on the remarkable case of diffraction spectra
  described by Prof. Wood},{''} \bibinfo{journal}{Philos.\ Mag.}
  \textbf{\bibinfo{volume}{14}},  \bibinfo{pages}{60--65}.

\bibitem[{\citenamefont{Maier} \emph{et~al.}(2001)\citenamefont{Maier,
  Brongersma, Kik, Meltzer, Requicha, and Atwater}}]{MBA01}
\bibinfo{author}{\bibnamefont{Maier}, \bibfnamefont{S.~A.}},
  \bibinfo{author}{\bibfnamefont{M.~L.} \bibnamefont{Brongersma}},
  \bibinfo{author}{\bibfnamefont{P.~G.} \bibnamefont{Kik}},
  \bibinfo{author}{\bibfnamefont{S.}~\bibnamefont{Meltzer}},
  \bibinfo{author}{\bibfnamefont{A.~A.~G.} \bibnamefont{Requicha}}, and
  \bibinfo{author}{\bibfnamefont{H.~A.} \bibnamefont{Atwater}},
  \bibinfo{year}{2001}, {``}\bibinfo{title}{Plasmonics - a route to nanoscale
  optical devices},{''} \bibinfo{journal}{Adv.\ Mater.}
  \textbf{\bibinfo{volume}{13}},  \bibinfo{pages}{1501--1505}.

\bibitem[{\citenamefont{Maier} \emph{et~al.}(2003)\citenamefont{Maier, Kik,
  Atwater, Meltzer, Harel, Koel, and Requicha}}]{MKA03}
\bibinfo{author}{\bibnamefont{Maier}, \bibfnamefont{S.~A.}},
  \bibinfo{author}{\bibfnamefont{P.~G.} \bibnamefont{Kik}},
  \bibinfo{author}{\bibfnamefont{H.~A.} \bibnamefont{Atwater}},
  \bibinfo{author}{\bibfnamefont{S.}~\bibnamefont{Meltzer}},
  \bibinfo{author}{\bibfnamefont{E.}~\bibnamefont{Harel}},
  \bibinfo{author}{\bibfnamefont{B.~E.} \bibnamefont{Koel}}, and
  \bibinfo{author}{\bibfnamefont{A.~A.~G.} \bibnamefont{Requicha}},
  \bibinfo{year}{2003}, {``}\bibinfo{title}{Local detection of electromagnetic
  energy transport below the diffraction limit in metal nanoparticle plasmon
  waveguides},{''} \bibinfo{journal}{Nat.\ Mater.}
  \textbf{\bibinfo{volume}{2}},  \bibinfo{pages}{229--232}.

\bibitem[{\citenamefont{Marquier} \emph{et~al.}(2005)\citenamefont{Marquier,
  Greffet, Collin, Pardo, and Pelouard}}]{MGC05}
\bibinfo{author}{\bibnamefont{Marquier}, \bibfnamefont{F.}},
  \bibinfo{author}{\bibfnamefont{J.-J.} \bibnamefont{Greffet}},
  \bibinfo{author}{\bibfnamefont{S.}~\bibnamefont{Collin}},
  \bibinfo{author}{\bibfnamefont{F.}~\bibnamefont{Pardo}}, and
  \bibinfo{author}{\bibfnamefont{J.~L.} \bibnamefont{Pelouard}},
  \bibinfo{year}{2005}, {``}\bibinfo{title}{Resonant transmission through a
  metallic film due to coupled modes},{''} \bibinfo{journal}{Opt.\ Express}
  \textbf{\bibinfo{volume}{13}},  \bibinfo{pages}{70--76}.

\bibitem[{\citenamefont{{Mart\'{\i}n-Moreno}}
  \emph{et~al.}(2001)\citenamefont{{Mart\'{\i}n-Moreno}, {Garc\'{\i}a-Vidal},
  Lezec, Pellerin, Thio, Pendry, and Ebbesen}}]{MGL01}
\bibinfo{author}{\bibnamefont{{Mart\'{\i}n-Moreno}}, \bibfnamefont{L.}},
  \bibinfo{author}{\bibfnamefont{F.~J.} \bibnamefont{{Garc\'{\i}a-Vidal}}},
  \bibinfo{author}{\bibfnamefont{H.~J.} \bibnamefont{Lezec}},
  \bibinfo{author}{\bibfnamefont{K.~M.} \bibnamefont{Pellerin}},
  \bibinfo{author}{\bibfnamefont{T.}~\bibnamefont{Thio}},
  \bibinfo{author}{\bibfnamefont{J.~B.} \bibnamefont{Pendry}}, and
  \bibinfo{author}{\bibfnamefont{T.~W.} \bibnamefont{Ebbesen}},
  \bibinfo{year}{2001}, {``}\bibinfo{title}{Theory of extraordinary optical
  transmission through subwavelength hole arrays},{''} \bibinfo{journal}{Phys.\
  Rev.\ Lett.} \textbf{\bibinfo{volume}{86}},  \bibinfo{pages}{1114--1117}.

\bibitem[{\citenamefont{{Mart\'{\i}nez-Sala}}
  \emph{et~al.}(1995)\citenamefont{{Mart\'{\i}nez-Sala}, Sancho, S\'anchez,
  G\'omez, Llinares, and Meseguer}}]{MSS95}
\bibinfo{author}{\bibnamefont{{Mart\'{\i}nez-Sala}}, \bibfnamefont{R.}},
  \bibinfo{author}{\bibfnamefont{J.}~\bibnamefont{Sancho}},
  \bibinfo{author}{\bibfnamefont{J.~V.} \bibnamefont{S\'anchez}},
  \bibinfo{author}{\bibfnamefont{V.}~\bibnamefont{G\'omez}},
  \bibinfo{author}{\bibfnamefont{J.}~\bibnamefont{Llinares}}, and
  \bibinfo{author}{\bibfnamefont{F.}~\bibnamefont{Meseguer}},
  \bibinfo{year}{1995}, {``}\bibinfo{title}{Sound attenuation by
  sculpture},{''} \bibinfo{journal}{Nature} \textbf{\bibinfo{volume}{378}},
  \bibinfo{pages}{241--241}.

\bibitem[{\citenamefont{Matsui} \emph{et~al.}(2007)\citenamefont{Matsui,
  Agrawal, Nahata, and Vardeny}}]{MAN07}
\bibinfo{author}{\bibnamefont{Matsui}, \bibfnamefont{T.}},
  \bibinfo{author}{\bibfnamefont{A.}~\bibnamefont{Agrawal}},
  \bibinfo{author}{\bibfnamefont{A.}~\bibnamefont{Nahata}}, and
  \bibinfo{author}{\bibfnamefont{Z.~V.} \bibnamefont{Vardeny}},
  \bibinfo{year}{2007}, {``}\bibinfo{title}{Transmission resonances through
  aperiodic arrays of subwavelength apertures},{''} \bibinfo{journal}{Nature}
  \textbf{\bibinfo{volume}{446}},  \bibinfo{pages}{517–--521}.

\bibitem[{\citenamefont{Maystre}(1972)}]{M1972}
\bibinfo{author}{\bibnamefont{Maystre}, \bibfnamefont{D.}},
  \bibinfo{year}{1972}, {``}\bibinfo{title}{Sur la diffraction d'une onde plane
  par un reseau metallique de conductivite finie},{''} \bibinfo{journal}{Opt.\
  Commun.} \textbf{\bibinfo{volume}{6}},  \bibinfo{pages}{50--54}.

\bibitem[{\citenamefont{Maystre}(1980)}]{M1980_2}
\bibinfo{author}{\bibnamefont{Maystre}, \bibfnamefont{D.}},
  \bibinfo{year}{1980}, \bibinfo{note}{in {\it Electromagnetic Theory of
  Gratings}, edited by R. Petit (Springer-Verlag, Berlin), pp. 63-100.}

\bibitem[{\citenamefont{Maystre}(1984)}]{M1984}
\bibinfo{author}{\bibnamefont{Maystre}, \bibfnamefont{D.}},
  \bibinfo{year}{1984}, {``}\bibinfo{title}{Rigorous vector theories of
  diffraction gratings},{''} \bibinfo{journal}{Prog.\ Opt.}
  \textbf{\bibinfo{volume}{21}},  \bibinfo{pages}{1--67}.

\bibitem[{\citenamefont{Maystre}(1993)}]{M93}
\bibinfo{author}{\bibnamefont{Maystre}, \bibfnamefont{D.}},
  \bibinfo{year}{1993}, \bibinfo{note}{{\it Diffraction gratings}, SPIE
  Milestones Series, volume MS 83.}

\bibitem[{\citenamefont{McPhedran} \emph{et~al.}(1980)\citenamefont{McPhedran,
  Derrick, and Botten}}]{M1980}
\bibinfo{author}{\bibnamefont{McPhedran}, \bibfnamefont{R.~C.}},
  \bibinfo{author}{\bibfnamefont{G.~H.} \bibnamefont{Derrick}}, and
  \bibinfo{author}{\bibfnamefont{L.~C.} \bibnamefont{Botten}},
  \bibinfo{year}{1980}, \bibinfo{note}{in {\it Electromagnetic Theory of
  Gratings}, edited by R. Petit (Springer-Verlag, Berlin), pp. 227-276.}

\bibitem[{\citenamefont{McPhedran and Maystre}(1974)}]{MM1974}
\bibinfo{author}{\bibnamefont{McPhedran}, \bibfnamefont{R.~C.}}, and
  \bibinfo{author}{\bibfnamefont{D.}~\bibnamefont{Maystre}},
  \bibinfo{year}{1974}, {``}\bibinfo{title}{A detailed theoretical study of the
  anomalies of a sinusoidal diffraction grating},{''} \bibinfo{journal}{Opt.\
  Acta} \textbf{\bibinfo{volume}{21}},  \bibinfo{pages}{413--421}.

\bibitem[{\citenamefont{Mie}(1908)}]{M1908}
\bibinfo{author}{\bibnamefont{Mie}, \bibfnamefont{G.}}, \bibinfo{year}{1908},
  {``}\bibinfo{title}{Beitr\"{a}ge zur Optik tr\"{u}ber Medien, speziell
  kolloidaler Metall\"{o}sungen},{''} \bibinfo{journal}{Ann.\ Phys.\ (Leipzig)}
  \textbf{\bibinfo{volume}{25}},  \bibinfo{pages}{377--445}.

\bibitem[{\citenamefont{Milton}(2002)}]{M02}
\bibinfo{author}{\bibnamefont{Milton}, \bibfnamefont{G.~W.}},
  \bibinfo{year}{2002}, \emph{\bibinfo{title}{The Theory of Composites}}
  (\bibinfo{publisher}{Cambridge University Press},
  \bibinfo{address}{Cambridge}).

\bibitem[{\citenamefont{Mittra} \emph{et~al.}(1988)\citenamefont{Mittra, Chan,
  and Cwik}}]{MCC1988}
\bibinfo{author}{\bibnamefont{Mittra}, \bibfnamefont{R.}},
  \bibinfo{author}{\bibfnamefont{C.~H.} \bibnamefont{Chan}}, and
  \bibinfo{author}{\bibfnamefont{T.}~\bibnamefont{Cwik}}, \bibinfo{year}{1988},
  {``}\bibinfo{title}{Techniques for analyzing frequency selective surfaces - a
  review},{''} \bibinfo{journal}{Proc.\ IEEE} \textbf{\bibinfo{volume}{76}},
  \bibinfo{pages}{1593--1615}.

\bibitem[{\citenamefont{Miyamaru and Hangyo}(2004)}]{MH04}
\bibinfo{author}{\bibnamefont{Miyamaru}, \bibfnamefont{F.}}, and
  \bibinfo{author}{\bibfnamefont{M.}~\bibnamefont{Hangyo}},
  \bibinfo{year}{2004}, {``}\bibinfo{title}{Finite size effect of transmission
  property for metal hole arrays in subterahertz region},{''}
  \bibinfo{journal}{Appl.\ Phys.\ Lett.} \textbf{\bibinfo{volume}{84}},
  \bibinfo{pages}{2742--2744}.

\bibitem[{\citenamefont{Nomura} \emph{et~al.}(2005)\citenamefont{Nomura, Ohtsu,
  and Yatsui}}]{NOY05}
\bibinfo{author}{\bibnamefont{Nomura}, \bibfnamefont{W.}},
  \bibinfo{author}{\bibfnamefont{M.}~\bibnamefont{Ohtsu}}, and
  \bibinfo{author}{\bibfnamefont{T.}~\bibnamefont{Yatsui}},
  \bibinfo{year}{2005}, {``}\bibinfo{title}{Nanodot coupler with a surface
  plasmon polariton condenser for optical far/near-field conversion},{''}
  \bibinfo{journal}{Appl.\ Phys.\ Lett.} \textbf{\bibinfo{volume}{86}},
  \bibinfo{pages}{181108}.

\bibitem[{\citenamefont{Nordlander}
  \emph{et~al.}(2004)\citenamefont{Nordlander, Oubre, Prodan, Li, and
  Stockman}}]{NOP04}
\bibinfo{author}{\bibnamefont{Nordlander}, \bibfnamefont{P.}},
  \bibinfo{author}{\bibfnamefont{C.}~\bibnamefont{Oubre}},
  \bibinfo{author}{\bibfnamefont{E.}~\bibnamefont{Prodan}},
  \bibinfo{author}{\bibfnamefont{K.}~\bibnamefont{Li}}, and
  \bibinfo{author}{\bibfnamefont{M.~I.} \bibnamefont{Stockman}},
  \bibinfo{year}{2004}, {``}\bibinfo{title}{Plasmon hybridizaton in
  nanoparticle dimers},{''} \bibinfo{journal}{Nano\ Lett.}
  \textbf{\bibinfo{volume}{4}},  \bibinfo{pages}{899--903}.

\bibitem[{\citenamefont{Oberm\"uller and Karrai}(1995)}]{OK95}
\bibinfo{author}{\bibnamefont{Oberm\"uller}, \bibfnamefont{C.}}, and
  \bibinfo{author}{\bibfnamefont{K.}~\bibnamefont{Karrai}},
  \bibinfo{year}{1995}, {``}\bibinfo{title}{Far field characterization of
  diffracting circular apertures},{''} \bibinfo{journal}{Appl.\ Phys.\ Lett.}
  \textbf{\bibinfo{volume}{67}},  \bibinfo{pages}{3408--3410}.

\bibitem[{\citenamefont{Ozbay}(2006)}]{O06}
\bibinfo{author}{\bibnamefont{Ozbay}, \bibfnamefont{E.}}, \bibinfo{year}{2006},
  {``}\bibinfo{title}{Plasmonics: merging photonics and electronics at
  nanoscale dimensions},{''} \bibinfo{journal}{Science}
  \textbf{\bibinfo{volume}{311}},  \bibinfo{pages}{189--193}.

\bibitem[{\citenamefont{Palik}(1985)}]{P1985}
\bibinfo{author}{\bibnamefont{Palik}, \bibfnamefont{E.~D.}},
  \bibinfo{year}{1985}, \emph{\bibinfo{title}{Handbook of Optical Constants of
  Solids}} (\bibinfo{publisher}{Academic Press}, \bibinfo{address}{New York}).

\bibitem[{\citenamefont{Pendry}(1974)}]{P1974}
\bibinfo{author}{\bibnamefont{Pendry}, \bibfnamefont{J.~B.}},
  \bibinfo{year}{1974}, \emph{\bibinfo{title}{Low Energy Electron Diffraction}}
  (\bibinfo{publisher}{Academic Press}, \bibinfo{address}{London}).

\bibitem[{\citenamefont{Pendry} \emph{et~al.}(2004)\citenamefont{Pendry,
  Mart\'{\i}n-Moreno, and {Garc\'{\i}a-Vidal}}}]{PMG04}
\bibinfo{author}{\bibnamefont{Pendry}, \bibfnamefont{J.~B.}},
  \bibinfo{author}{\bibfnamefont{L.}~\bibnamefont{Mart\'{\i}n-Moreno}}, and
  \bibinfo{author}{\bibfnamefont{F.~J.} \bibnamefont{{Garc\'{\i}a-Vidal}}},
  \bibinfo{year}{2004}, {``}\bibinfo{title}{Mimicking surface plasmons with
  structured surfaces},{''} \bibinfo{journal}{Science}
  \textbf{\bibinfo{volume}{305}},  \bibinfo{pages}{847--848}.

\bibitem[{\citenamefont{Pettit} \emph{et~al.}(1975)\citenamefont{Pettit,
  Silcox, and Vincent}}]{PSV1975}
\bibinfo{author}{\bibnamefont{Pettit}, \bibfnamefont{R.~B.}},
  \bibinfo{author}{\bibfnamefont{J.}~\bibnamefont{Silcox}}, and
  \bibinfo{author}{\bibfnamefont{R.}~\bibnamefont{Vincent}},
  \bibinfo{year}{1975}, {``}\bibinfo{title}{Measurement of surface-plasmon
  dispersion in oxidized aluminum films},{''} \bibinfo{journal}{Phys.\ Rev.\ B}
  \textbf{\bibinfo{volume}{11}},  \bibinfo{pages}{3116--3123}.

\bibitem[{\citenamefont{Popov} \emph{et~al.}(2005)\citenamefont{Popov, Bonod,
  Nevi\`ere, Rigneault, Lenne, and Chaumet}}]{PBN05}
\bibinfo{author}{\bibnamefont{Popov}, \bibfnamefont{E.}},
  \bibinfo{author}{\bibfnamefont{N.}~\bibnamefont{Bonod}},
  \bibinfo{author}{\bibfnamefont{M.}~\bibnamefont{Nevi\`ere}},
  \bibinfo{author}{\bibfnamefont{H.}~\bibnamefont{Rigneault}},
  \bibinfo{author}{\bibfnamefont{P.-F.} \bibnamefont{Lenne}}, and
  \bibinfo{author}{\bibfnamefont{P.}~\bibnamefont{Chaumet}},
  \bibinfo{year}{2005}, {``}\bibinfo{title}{Surface plasmon excitation on a
  single subwavelength hole in a metallic sheet},{''} \bibinfo{journal}{Appl.\
  Opt.} \textbf{\bibinfo{volume}{44}},  \bibinfo{pages}{2332--2337}.

\bibitem[{\citenamefont{Popov} \emph{et~al.}(2000)\citenamefont{Popov,
  Nevi\`{e}re, Enoch, and Reinisch}}]{PNE00}
\bibinfo{author}{\bibnamefont{Popov}, \bibfnamefont{E.}},
  \bibinfo{author}{\bibfnamefont{M.}~\bibnamefont{Nevi\`{e}re}},
  \bibinfo{author}{\bibfnamefont{S.}~\bibnamefont{Enoch}}, and
  \bibinfo{author}{\bibfnamefont{R.}~\bibnamefont{Reinisch}},
  \bibinfo{year}{2000}, {``}\bibinfo{title}{Theory of light transmission
  through subwavelength periodic hole arrays},{''} \bibinfo{journal}{Phys.\
  Rev.\ B} \textbf{\bibinfo{volume}{62}},  \bibinfo{pages}{16100--16108}.

\bibitem[{\citenamefont{Porto} \emph{et~al.}(1999)\citenamefont{Porto,
  {Garc\'{\i}a-Vidal}, and Pendry}}]{PGP99}
\bibinfo{author}{\bibnamefont{Porto}, \bibfnamefont{J.~A.}},
  \bibinfo{author}{\bibfnamefont{F.~J.} \bibnamefont{{Garc\'{\i}a-Vidal}}}, and
  \bibinfo{author}{\bibfnamefont{J.~B.} \bibnamefont{Pendry}},
  \bibinfo{year}{1999}, {``}\bibinfo{title}{Transmission resonances on metallic
  gratings with very narrow slits},{''} \bibinfo{journal}{Phys.\ Rev.\ Lett.}
  \textbf{\bibinfo{volume}{83}},  \bibinfo{pages}{2845--2848}.

\bibitem[{\citenamefont{Powell and Swan}(1959)}]{PS1959}
\bibinfo{author}{\bibnamefont{Powell}, \bibfnamefont{C.~J.}}, and
  \bibinfo{author}{\bibfnamefont{J.~B.} \bibnamefont{Swan}},
  \bibinfo{year}{1959}, {``}\bibinfo{title}{Origin of the characteristic
  electron energy losses in aluminum},{''} \bibinfo{journal}{Phys.\ Rev.}
  \textbf{\bibinfo{volume}{115}},  \bibinfo{pages}{869--875}.

\bibitem[{\citenamefont{Przybilla}
  \emph{et~al.}(2006{\natexlab{a}})\citenamefont{Przybilla, Degiron, Laluet,
  Genet, and Ebbesen}}]{PDL06}
\bibinfo{author}{\bibnamefont{Przybilla}, \bibfnamefont{F.}},
  \bibinfo{author}{\bibfnamefont{A.}~\bibnamefont{Degiron}},
  \bibinfo{author}{\bibfnamefont{J.-Y.} \bibnamefont{Laluet}},
  \bibinfo{author}{\bibfnamefont{C.}~\bibnamefont{Genet}}, and
  \bibinfo{author}{\bibfnamefont{T.~W.} \bibnamefont{Ebbesen}},
  \bibinfo{year}{2006}{\natexlab{a}}, {``}\bibinfo{title}{Optical transmission
  in perforated noble and transition metal films},{''} \bibinfo{journal}{J.\
  Opt.\ A-Pure\ Appl.\ Opt.} \textbf{\bibinfo{volume}{8}},
  \bibinfo{pages}{458--463}.

\bibitem[{\citenamefont{Przybilla}
  \emph{et~al.}(2006{\natexlab{b}})\citenamefont{Przybilla, Genet, and
  Ebbesen}}]{PGE06}
\bibinfo{author}{\bibnamefont{Przybilla}, \bibfnamefont{F.}},
  \bibinfo{author}{\bibfnamefont{C.}~\bibnamefont{Genet}}, and
  \bibinfo{author}{\bibfnamefont{T.~W.} \bibnamefont{Ebbesen}},
  \bibinfo{year}{2006}{\natexlab{b}}, {``}\bibinfo{title}{Enhanced transmission
  through Penrose subwavelength hole arrays},{''} \bibinfo{journal}{Appl.\
  Phys.\ Lett.} \textbf{\bibinfo{volume}{89}},  \bibinfo{pages}{121115}.

\bibitem[{\citenamefont{Purcell and Pennypacker}(1973)}]{PP1973}
\bibinfo{author}{\bibnamefont{Purcell}, \bibfnamefont{E.~M.}}, and
  \bibinfo{author}{\bibfnamefont{C.~R.} \bibnamefont{Pennypacker}},
  \bibinfo{year}{1973}, {``}\bibinfo{title}{Scattering and absorption of light
  by nonspherical dielectric grains},{''} \bibinfo{journal}{Astrophysical J.}
  \textbf{\bibinfo{volume}{186}},  \bibinfo{pages}{705--714}.

\bibitem[{\citenamefont{Quinten} \emph{et~al.}(1998)\citenamefont{Quinten,
  Leitner, Krenn, and Aussenegg}}]{QLK98}
\bibinfo{author}{\bibnamefont{Quinten}, \bibfnamefont{M.}},
  \bibinfo{author}{\bibfnamefont{A.}~\bibnamefont{Leitner}},
  \bibinfo{author}{\bibfnamefont{J.~R.} \bibnamefont{Krenn}}, and
  \bibinfo{author}{\bibfnamefont{F.~R.} \bibnamefont{Aussenegg}},
  \bibinfo{year}{1998}, {``}\bibinfo{title}{Electromagnetic energy transport
  via linear chains of silver nanoparticles},{''} \bibinfo{journal}{Opt.\
  Lett.} \textbf{\bibinfo{volume}{23}},  \bibinfo{pages}{1331--1333}.

\bibitem[{\citenamefont{Raether}(1988)}]{R1988}
\bibinfo{author}{\bibnamefont{Raether}, \bibfnamefont{H.}},
  \bibinfo{year}{1988}, \emph{\bibinfo{title}{Surface Plasmons on Smooth and
  Rough Surfaces and on Gratings}}, volume \bibinfo{volume}{111} of
  \emph{\bibinfo{series}{Springer Tracks in Modern Physics}}
  (\bibinfo{publisher}{Springer-Verlag}, \bibinfo{address}{Berlin}).

\bibitem[{\citenamefont{Reif}(1965)}]{R1965}
\bibinfo{author}{\bibnamefont{Reif}, \bibfnamefont{F.}}, \bibinfo{year}{1965},
  \emph{\bibinfo{title}{Fundamentals of Statistical and Thermal Physics}}
  (\bibinfo{publisher}{McGraw-Hill}, \bibinfo{address}{New York}).

\bibitem[{\citenamefont{Rindzevicius}
  \emph{et~al.}(2007)\citenamefont{Rindzevicius, Alaverdyan, Sepulveda,
  Pakizeh, K\"{a}ll, Hillenbrand, Aizpurua, and {Garc\'{\i}a de
  Abajo}}}]{paper118}
\bibinfo{author}{\bibnamefont{Rindzevicius}, \bibfnamefont{T.}},
  \bibinfo{author}{\bibfnamefont{Y.}~\bibnamefont{Alaverdyan}},
  \bibinfo{author}{\bibfnamefont{B.}~\bibnamefont{Sepulveda}},
  \bibinfo{author}{\bibfnamefont{T.}~\bibnamefont{Pakizeh}},
  \bibinfo{author}{\bibfnamefont{M.}~\bibnamefont{K\"{a}ll}},
  \bibinfo{author}{\bibfnamefont{R.}~\bibnamefont{Hillenbrand}},
  \bibinfo{author}{\bibfnamefont{J.}~\bibnamefont{Aizpurua}}, and
  \bibinfo{author}{\bibfnamefont{F.~J.} \bibnamefont{{Garc\'{\i}a de Abajo}}},
  \bibinfo{year}{2007}, {``}\bibinfo{title}{Nanohole plasmons in optically thin
  gold films},{''} \bibinfo{journal}{J.\ Phys. Chem.\ C}
  \textbf{\bibinfo{volume}{111}},  \bibinfo{pages}{1207--1212}.

\bibitem[{\citenamefont{Ritchie}(1957)}]{R1957}
\bibinfo{author}{\bibnamefont{Ritchie}, \bibfnamefont{R.~H.}},
  \bibinfo{year}{1957}, {``}\bibinfo{title}{Plasma losses by fast electrons in
  thin films},{''} \bibinfo{journal}{Phys.\ Rev.}
  \textbf{\bibinfo{volume}{106}},  \bibinfo{pages}{874--881}.

\bibitem[{\citenamefont{Ritchie} \emph{et~al.}(1968)\citenamefont{Ritchie,
  Arakawa, Cowan, and Hamm}}]{RAC1968}
\bibinfo{author}{\bibnamefont{Ritchie}, \bibfnamefont{R.~H.}},
  \bibinfo{author}{\bibfnamefont{E.~T.} \bibnamefont{Arakawa}},
  \bibinfo{author}{\bibfnamefont{J.~J.} \bibnamefont{Cowan}}, and
  \bibinfo{author}{\bibfnamefont{R.~N.} \bibnamefont{Hamm}},
  \bibinfo{year}{1968}, {``}\bibinfo{title}{Surface-plasmon resonance effect in
  grating diffraction},{''} \bibinfo{journal}{Phys.\ Rev.\ Lett.}
  \textbf{\bibinfo{volume}{21}},  \bibinfo{pages}{1530--1533}.

\bibitem[{\citenamefont{Roberts}(1987)}]{R1987}
\bibinfo{author}{\bibnamefont{Roberts}, \bibfnamefont{A.}},
  \bibinfo{year}{1987}, {``}\bibinfo{title}{Electromagnetic theory of
  diffraction by a circular aperture in a thick, perfectly conducting
  screen},{''} \bibinfo{journal}{J.\ Opt.\ Soc.\ Am.\ A}
  \textbf{\bibinfo{volume}{4}},  \bibinfo{pages}{1970--1983}.

\bibitem[{\citenamefont{Roberts and McPhedran}(1988)}]{RM1988}
\bibinfo{author}{\bibnamefont{Roberts}, \bibfnamefont{A.}}, and
  \bibinfo{author}{\bibfnamefont{R.~C.} \bibnamefont{McPhedran}},
  \bibinfo{year}{1988}, {``}\bibinfo{title}{Bandpass grids with annular
  apertures},{''} \bibinfo{journal}{IEEE\ Trans.\ Antennas\ Propag.}
  \textbf{\bibinfo{volume}{36}},  \bibinfo{pages}{607--611}.

\bibitem[{\citenamefont{Romero} \emph{et~al.}(2006)\citenamefont{Romero,
  Aizpurua, Bryant, and {Garc\'{\i}a de Abajo}}}]{paper114}
\bibinfo{author}{\bibnamefont{Romero}, \bibfnamefont{I.}},
  \bibinfo{author}{\bibfnamefont{J.}~\bibnamefont{Aizpurua}},
  \bibinfo{author}{\bibfnamefont{G.~W.} \bibnamefont{Bryant}}, and
  \bibinfo{author}{\bibfnamefont{F.~J.} \bibnamefont{{Garc\'{\i}a de Abajo}}},
  \bibinfo{year}{2006}, {``}\bibinfo{title}{Plasmons in nearly touching
  metallic nanoparticles: singular response in the limit of touching
  dimers},{''} \bibinfo{journal}{Opt.\ Express} \textbf{\bibinfo{volume}{14}},
  \bibinfo{pages}{9988--9999}.

\bibitem[{\citenamefont{Ruan and Qiu}(2006)}]{RQ06}
\bibinfo{author}{\bibnamefont{Ruan}, \bibfnamefont{Z.}}, and
  \bibinfo{author}{\bibfnamefont{M.}~\bibnamefont{Qiu}}, \bibinfo{year}{2006},
  {``}\bibinfo{title}{Enhanced transmission through periodic arrays of
  subwavelength holes: The role of localized waveguide resonances},{''}
  \bibinfo{journal}{Phys.\ Rev.\ Lett.} \textbf{\bibinfo{volume}{96}},
  \bibinfo{pages}{233901}.

\bibitem[{\citenamefont{Salomon} \emph{et~al.}(2001)\citenamefont{Salomon,
  Grillot, Zayats, and {de Fornel}}}]{SGZ01}
\bibinfo{author}{\bibnamefont{Salomon}, \bibfnamefont{L.}},
  \bibinfo{author}{\bibfnamefont{F.}~\bibnamefont{Grillot}},
  \bibinfo{author}{\bibfnamefont{A.~V.} \bibnamefont{Zayats}}, and
  \bibinfo{author}{\bibfnamefont{F.}~\bibnamefont{{de Fornel}}},
  \bibinfo{year}{2001}, {``}\bibinfo{title}{Near-field distribution of optical
  transmission of periodic subwavelength holes in a metal film},{''}
  \bibinfo{journal}{Phys.\ Rev.\ Lett.} \textbf{\bibinfo{volume}{86}},
  \bibinfo{pages}{1110--1113}.

\bibitem[{\citenamefont{Sarid}(1981)}]{S1981}
\bibinfo{author}{\bibnamefont{Sarid}, \bibfnamefont{D.}}, \bibinfo{year}{1981},
  {``}\bibinfo{title}{Long-range surface-plasma waves on very thin metal
  films},{''} \bibinfo{journal}{Phys.\ Rev.\ Lett.}
  \textbf{\bibinfo{volume}{47}},  \bibinfo{pages}{1927--1930}.

\bibitem[{\citenamefont{Sarrazin} \emph{et~al.}(2003)\citenamefont{Sarrazin,
  Vigneron, and Vigoureux}}]{SVV03}
\bibinfo{author}{\bibnamefont{Sarrazin}, \bibfnamefont{M.}},
  \bibinfo{author}{\bibfnamefont{J.-P.} \bibnamefont{Vigneron}}, and
  \bibinfo{author}{\bibfnamefont{J.-M.} \bibnamefont{Vigoureux}},
  \bibinfo{year}{2003}, {``}\bibinfo{title}{Role of Wood anomalies in optical
  properties of thin metallic films with a bidimensional array of subwavelength
  holes},{''} \bibinfo{journal}{Phys.\ Rev.\ B} \textbf{\bibinfo{volume}{67}},
  \bibinfo{pages}{085415}.

\bibitem[{\citenamefont{{Schr\"oter} and Heitmann}(1998)}]{SH98_2}
\bibinfo{author}{\bibnamefont{{Schr\"oter}}, \bibfnamefont{U.}}, and
  \bibinfo{author}{\bibfnamefont{D.}~\bibnamefont{Heitmann}},
  \bibinfo{year}{1998}, {``}\bibinfo{title}{Surface-plasmon-enhanced
  transmission through metallic gratings},{''} \bibinfo{journal}{Phys.\ Rev.\
  B} \textbf{\bibinfo{volume}{58}},  \bibinfo{pages}{15419--15421}.

\bibitem[{\citenamefont{Schuster} \emph{et~al.}(1993)\citenamefont{Schuster,
  Swanson, Alex, Bourret, and Simon}}]{SSA93}
\bibinfo{author}{\bibnamefont{Schuster}, \bibfnamefont{S.~C.}},
  \bibinfo{author}{\bibfnamefont{R.~V.} \bibnamefont{Swanson}},
  \bibinfo{author}{\bibfnamefont{L.~A.} \bibnamefont{Alex}},
  \bibinfo{author}{\bibfnamefont{R.~B.} \bibnamefont{Bourret}}, and
  \bibinfo{author}{\bibfnamefont{M.~I.} \bibnamefont{Simon}},
  \bibinfo{year}{1993}, {``}\bibinfo{title}{Assembly and function of a
  quaternary signal-transduction complex monitored by surface-plasmon
  resonance},{''} \bibinfo{journal}{Nature} \textbf{\bibinfo{volume}{365}},
  \bibinfo{pages}{343--347}.

\bibitem[{\citenamefont{Schwanecke}
  \emph{et~al.}(2006)\citenamefont{Schwanecke, Papasimakis, Fedotov, Huang,
  Chen, {Garc\'{\i}a de Abajo}, and Zheludev}}]{SPF2006}
\bibinfo{author}{\bibnamefont{Schwanecke}, \bibfnamefont{A.~S.}},
  \bibinfo{author}{\bibfnamefont{N.}~\bibnamefont{Papasimakis}},
  \bibinfo{author}{\bibfnamefont{V.~A.} \bibnamefont{Fedotov}},
  \bibinfo{author}{\bibfnamefont{F.}~\bibnamefont{Huang}},
  \bibinfo{author}{\bibfnamefont{Y.}~\bibnamefont{Chen}},
  \bibinfo{author}{\bibfnamefont{F.~J.} \bibnamefont{{Garc\'{\i}a de Abajo}}},
  and \bibinfo{author}{\bibfnamefont{N.~I.} \bibnamefont{Zheludev}},
  \bibinfo{year}{2006}, \bibinfo{note}{nanophotonics topical meeting NANO at
  IPRA/NANO OSA Collocated Topical Meetings, Uncasville, CT}.

\bibitem[{\citenamefont{Selcuk} \emph{et~al.}(2006)\citenamefont{Selcuk, Woo,
  Tanner, Hebard, Borisov, and Shabanov}}]{SWT06}
\bibinfo{author}{\bibnamefont{Selcuk}, \bibfnamefont{S.}},
  \bibinfo{author}{\bibfnamefont{K.}~\bibnamefont{Woo}},
  \bibinfo{author}{\bibfnamefont{D.~B.} \bibnamefont{Tanner}},
  \bibinfo{author}{\bibfnamefont{A.~F.} \bibnamefont{Hebard}},
  \bibinfo{author}{\bibfnamefont{A.~G.} \bibnamefont{Borisov}}, and
  \bibinfo{author}{\bibfnamefont{S.~V.} \bibnamefont{Shabanov}},
  \bibinfo{year}{2006}, {``}\bibinfo{title}{Trapped electromagnetic modes and
  scaling in the transmittance of perforated metal films},{''}
  \bibinfo{journal}{Phys.\ Rev.\ Lett.} \textbf{\bibinfo{volume}{97}},
  \bibinfo{pages}{067403}.

\bibitem[{\citenamefont{Smith} \emph{et~al.}(2004)\citenamefont{Smith, Pendry,
  and Wiltshire}}]{YPF04}
\bibinfo{author}{\bibnamefont{Smith}, \bibfnamefont{D.~R.}},
  \bibinfo{author}{\bibfnamefont{J.~B.} \bibnamefont{Pendry}}, and
  \bibinfo{author}{\bibfnamefont{M.~C.~K.} \bibnamefont{Wiltshire}},
  \bibinfo{year}{2004}, {``}\bibinfo{title}{Metamaterials and negative
  refractive index},{''} \bibinfo{journal}{Science}
  \textbf{\bibinfo{volume}{305}},  \bibinfo{pages}{788--792}.

\bibitem[{\citenamefont{Smolyaninov}
  \emph{et~al.}(2002)\citenamefont{Smolyaninov, Zayats, Stanishevsky, and
  Davis}}]{SZS02}
\bibinfo{author}{\bibnamefont{Smolyaninov}, \bibfnamefont{I.~I.}},
  \bibinfo{author}{\bibfnamefont{A.~V.} \bibnamefont{Zayats}},
  \bibinfo{author}{\bibfnamefont{A.}~\bibnamefont{Stanishevsky}}, and
  \bibinfo{author}{\bibfnamefont{C.~C.} \bibnamefont{Davis}},
  \bibinfo{year}{2002}, {``}\bibinfo{title}{Optical control of photon tunneling
  through an array of nanometer-scale cylindrical channels},{''}
  \bibinfo{journal}{Phys.\ Rev.\ B} \textbf{\bibinfo{volume}{66}},
  \bibinfo{pages}{205414}.

\bibitem[{\citenamefont{Stefanou} \emph{et~al.}(1998)\citenamefont{Stefanou,
  Yannopapas, and Modinos}}]{SYM98_1}
\bibinfo{author}{\bibnamefont{Stefanou}, \bibfnamefont{N.}},
  \bibinfo{author}{\bibfnamefont{V.}~\bibnamefont{Yannopapas}}, and
  \bibinfo{author}{\bibfnamefont{A.}~\bibnamefont{Modinos}},
  \bibinfo{year}{1998}, {``}\bibinfo{title}{Heterostructures of photonic
  crystals: frequency bands and transmission coefficients},{''}
  \bibinfo{journal}{Comput. Phys. Commun.} \textbf{\bibinfo{volume}{113}},
  \bibinfo{pages}{49--77}.

\bibitem[{\citenamefont{Stefanou} \emph{et~al.}(2000)\citenamefont{Stefanou,
  Yannopapas, and Modinos}}]{SYM00}
\bibinfo{author}{\bibnamefont{Stefanou}, \bibfnamefont{N.}},
  \bibinfo{author}{\bibfnamefont{V.}~\bibnamefont{Yannopapas}}, and
  \bibinfo{author}{\bibfnamefont{A.}~\bibnamefont{Modinos}},
  \bibinfo{year}{2000}, {``}\bibinfo{title}{MULTEM 2: a new version of the
  program for transmission and band-structure calculations of photonic
  crystals},{''} \bibinfo{journal}{Comput. Phys. Commun.}
  \textbf{\bibinfo{volume}{132}},  \bibinfo{pages}{189--196}.

\bibitem[{\citenamefont{Stewart and Gallaway}(1962)}]{SG1962}
\bibinfo{author}{\bibnamefont{Stewart}, \bibfnamefont{J.~E.}}, and
  \bibinfo{author}{\bibfnamefont{W.~S.} \bibnamefont{Gallaway}},
  \bibinfo{year}{1962}, {``}\bibinfo{title}{Diffraction anomalies in grating
  spectrophotometers},{''} \bibinfo{journal}{Appl.\ Opt.}
  \textbf{\bibinfo{volume}{1}},  \bibinfo{pages}{421--429}.

\bibitem[{\citenamefont{Stuart and Hall}(1998)}]{SH98}
\bibinfo{author}{\bibnamefont{Stuart}, \bibfnamefont{H.~R.}}, and
  \bibinfo{author}{\bibfnamefont{D.~G.} \bibnamefont{Hall}},
  \bibinfo{year}{1998}, {``}\bibinfo{title}{Enhanced dipole-dipole interaction
  between elementary radiators near a surface},{''} \bibinfo{journal}{Phys.\
  Rev.\ Lett.} \textbf{\bibinfo{volume}{80}},  \bibinfo{pages}{5663--5666}.

\bibitem[{\citenamefont{Sun} \emph{et~al.}(2006)\citenamefont{Sun, Tian, Li,
  Cheng, Zhang, Jin, and Yang}}]{STL06}
\bibinfo{author}{\bibnamefont{Sun}, \bibfnamefont{M.}},
  \bibinfo{author}{\bibfnamefont{J.}~\bibnamefont{Tian}},
  \bibinfo{author}{\bibfnamefont{Z.-Y.} \bibnamefont{Li}},
  \bibinfo{author}{\bibfnamefont{B.-Y.} \bibnamefont{Cheng}},
  \bibinfo{author}{\bibfnamefont{D.-Z.} \bibnamefont{Zhang}},
  \bibinfo{author}{\bibfnamefont{A.-Z.} \bibnamefont{Jin}}, and
  \bibinfo{author}{\bibfnamefont{H.-F.} \bibnamefont{Yang}},
  \bibinfo{year}{2006}, {``}\bibinfo{title}{The role of periodicity in enhanced
  transmission through subwavelength hole arrays},{''} \bibinfo{journal}{Chin.\
  Phys.\ Lett.} \textbf{\bibinfo{volume}{23}},  \bibinfo{pages}{486--488}.

\bibitem[{\citenamefont{Takakura}(2001)}]{T01}
\bibinfo{author}{\bibnamefont{Takakura}, \bibfnamefont{Y.}},
  \bibinfo{year}{2001}, {``}\bibinfo{title}{Optical resonance in a narrow slit
  in a thick metallic screen},{''} \bibinfo{journal}{Phys.\ Rev.\ Lett.}
  \textbf{\bibinfo{volume}{86}},  \bibinfo{pages}{5601--5603}.

\bibitem[{\citenamefont{Talbot}(1836)}]{T1836}
\bibinfo{author}{\bibnamefont{Talbot}, \bibfnamefont{H.~F.}},
  \bibinfo{year}{1836}, {``}\bibinfo{title}{Facts relating to optical science,
  No. IV},{''} \bibinfo{journal}{Philos.\ Mag.} \textbf{\bibinfo{volume}{9}},
  \bibinfo{pages}{401--407}.

\bibitem[{\citenamefont{Teperik} \emph{et~al.}(2005)\citenamefont{Teperik,
  Popov, and {Garc\'{\i}a de Abajo}}}]{paper097}
\bibinfo{author}{\bibnamefont{Teperik}, \bibfnamefont{T.~V.}},
  \bibinfo{author}{\bibfnamefont{V.~V.} \bibnamefont{Popov}}, and
  \bibinfo{author}{\bibfnamefont{F.~J.} \bibnamefont{{Garc\'{\i}a de Abajo}}},
  \bibinfo{year}{2005}, {``}\bibinfo{title}{Void plasmons and total absorption
  of light in nanoporous metallic films},{''} \bibinfo{journal}{Phys.\ Rev.\ B}
  \textbf{\bibinfo{volume}{71}},  \bibinfo{pages}{085408}.

\bibitem[{\citenamefont{Teperik}
  \emph{et~al.}(2006{\natexlab{a}})\citenamefont{Teperik, Popov, {Garc\'{\i}a
  de Abajo}, Abdelsalam, Barlett, Kelf, Sugawara, and Baumberg}}]{paper109}
\bibinfo{author}{\bibnamefont{Teperik}, \bibfnamefont{T.~V.}},
  \bibinfo{author}{\bibfnamefont{V.~V.} \bibnamefont{Popov}},
  \bibinfo{author}{\bibfnamefont{F.~J.} \bibnamefont{{Garc\'{\i}a de Abajo}}},
  \bibinfo{author}{\bibfnamefont{M.}~\bibnamefont{Abdelsalam}},
  \bibinfo{author}{\bibfnamefont{P.~N.} \bibnamefont{Barlett}},
  \bibinfo{author}{\bibfnamefont{T.~A.} \bibnamefont{Kelf}},
  \bibinfo{author}{\bibfnamefont{Y.}~\bibnamefont{Sugawara}}, and
  \bibinfo{author}{\bibfnamefont{J.~J.} \bibnamefont{Baumberg}},
  \bibinfo{year}{2006}{\natexlab{a}}, {``}\bibinfo{title}{Strong coupling of
  light to flat metals via a buried nanovoid lattice: the interplay of
  localized and free plasmons},{''} \bibinfo{journal}{Opt.\ Express}
  \textbf{\bibinfo{volume}{14}},  \bibinfo{pages}{1965--1972}.

\bibitem[{\citenamefont{Teperik}
  \emph{et~al.}(2006{\natexlab{b}})\citenamefont{Teperik, Popov, {Garc\'{\i}a
  de Abajo}, Kelf, Sugawara, Baumberg, Abdelsalam, and Bartlett}}]{paper117}
\bibinfo{author}{\bibnamefont{Teperik}, \bibfnamefont{T.~V.}},
  \bibinfo{author}{\bibfnamefont{V.~V.} \bibnamefont{Popov}},
  \bibinfo{author}{\bibfnamefont{F.~J.} \bibnamefont{{Garc\'{\i}a de Abajo}}},
  \bibinfo{author}{\bibfnamefont{T.~A.} \bibnamefont{Kelf}},
  \bibinfo{author}{\bibfnamefont{Y.}~\bibnamefont{Sugawara}},
  \bibinfo{author}{\bibfnamefont{J.~J.} \bibnamefont{Baumberg}},
  \bibinfo{author}{\bibfnamefont{M.}~\bibnamefont{Abdelsalam}}, and
  \bibinfo{author}{\bibfnamefont{P.~N.} \bibnamefont{Bartlett}},
  \bibinfo{year}{2006}{\natexlab{b}}, {``}\bibinfo{title}{Mie plasmon enhanced
  diffraction of light from nanoporous metal surfaces},{''}
  \bibinfo{journal}{Opt.\ Express} \textbf{\bibinfo{volume}{14}},
  \bibinfo{pages}{11964--11971}.

\bibitem[{\citenamefont{Treacy}(1999)}]{T99}
\bibinfo{author}{\bibnamefont{Treacy}, \bibfnamefont{M.~M.~J.}},
  \bibinfo{year}{1999}, {``}\bibinfo{title}{Dynamical diffraction in metallic
  optical gratings},{''} \bibinfo{journal}{Appl.\ Phys.\ Lett.}
  \textbf{\bibinfo{volume}{75}},  \bibinfo{pages}{606--608}.

\bibitem[{\citenamefont{Treacy}(2002)}]{T02}
\bibinfo{author}{\bibnamefont{Treacy}, \bibfnamefont{M.~M.~J.}},
  \bibinfo{year}{2002}, {``}\bibinfo{title}{Dynamical diffraction explanation
  of the anomalous transmission of light through metallic gratings},{''}
  \bibinfo{journal}{Phys.\ Rev.\ B} \textbf{\bibinfo{volume}{66}},
  \bibinfo{pages}{195105}.

\bibitem[{\citenamefont{Ulrich}(1967)}]{U1967}
\bibinfo{author}{\bibnamefont{Ulrich}, \bibfnamefont{R.}},
  \bibinfo{year}{1967}, {``}\bibinfo{title}{Far-infrared properties of metallic
  mesh and its complementary structure},{''} \bibinfo{journal}{Infrared Phys.}
  \textbf{\bibinfo{volume}{7}},  \bibinfo{pages}{37--55}.

\bibitem[{\citenamefont{Ulrich and Tacke}(1972)}]{UT1972}
\bibinfo{author}{\bibnamefont{Ulrich}, \bibfnamefont{R.}}, and
  \bibinfo{author}{\bibfnamefont{M.}~\bibnamefont{Tacke}},
  \bibinfo{year}{1972}, {``}\bibinfo{title}{Submillimeter waveguiding on
  periodic metal structure},{''} \bibinfo{journal}{Appl.\ Phys.\ Lett.}
  \textbf{\bibinfo{volume}{22}},  \bibinfo{pages}{251--253}.

\bibitem[{\citenamefont{{van Coevorden}} \emph{et~al.}(1996)\citenamefont{{van
  Coevorden}, Sprik, Tip, and Lagendijk}}]{VST96}
\bibinfo{author}{\bibnamefont{{van Coevorden}}, \bibfnamefont{D.~V.}},
  \bibinfo{author}{\bibfnamefont{R.}~\bibnamefont{Sprik}},
  \bibinfo{author}{\bibfnamefont{A.}~\bibnamefont{Tip}}, and
  \bibinfo{author}{\bibfnamefont{A.}~\bibnamefont{Lagendijk}},
  \bibinfo{year}{1996}, {``}\bibinfo{title}{Photonic band structure of atomic
  lattices},{''} \bibinfo{journal}{Phys.\ Rev.\ Lett.}
  \textbf{\bibinfo{volume}{77}},  \bibinfo{pages}{2412--2415}.

\bibitem[{\citenamefont{{van de Hulst}}(1981)}]{V1981}
\bibinfo{author}{\bibnamefont{{van de Hulst}}, \bibfnamefont{H.~C.}},
  \bibinfo{year}{1981}, \emph{\bibinfo{title}{Light Scattering by Small
  Particles}} (\bibinfo{publisher}{Dover}, \bibinfo{address}{New York}).

\bibitem[{\citenamefont{{van der Molen}} \emph{et~al.}(2005)\citenamefont{{van
  der Molen}, {Klein Koerkamp}, Enoch, Segerink, {van Hulst}, and
  Kuipers}}]{VKE05}
\bibinfo{author}{\bibnamefont{{van der Molen}}, \bibfnamefont{K.~L.}},
  \bibinfo{author}{\bibfnamefont{K.~J.} \bibnamefont{{Klein Koerkamp}}},
  \bibinfo{author}{\bibfnamefont{S.}~\bibnamefont{Enoch}},
  \bibinfo{author}{\bibfnamefont{F.~B.} \bibnamefont{Segerink}},
  \bibinfo{author}{\bibfnamefont{N.~F.} \bibnamefont{{van Hulst}}}, and
  \bibinfo{author}{\bibfnamefont{L.}~\bibnamefont{Kuipers}},
  \bibinfo{year}{2005}, {``}\bibinfo{title}{Role of shape and localized
  resonances in extraordinary transmission through periodic arrays of
  subwavelength holes: experiment and theory},{''} \bibinfo{journal}{Phys.\
  Rev.\ B} \textbf{\bibinfo{volume}{72}},  \bibinfo{pages}{045421}.

\bibitem[{\citenamefont{Vincent and Silcox}(1973)}]{VS1973}
\bibinfo{author}{\bibnamefont{Vincent}, \bibfnamefont{R.}}, and
  \bibinfo{author}{\bibfnamefont{J.}~\bibnamefont{Silcox}},
  \bibinfo{year}{1973}, {``}\bibinfo{title}{Dispersion of radiative surface
  plasmons in aluminum films by electron scattering},{''}
  \bibinfo{journal}{Phys.\ Rev.\ Lett.} \textbf{\bibinfo{volume}{31}},
  \bibinfo{pages}{1487--1490}.

\bibitem[{\citenamefont{Wannemacher}(2001)}]{W01}
\bibinfo{author}{\bibnamefont{Wannemacher}, \bibfnamefont{R.}},
  \bibinfo{year}{2001}, {``}\bibinfo{title}{Plasmon-supported transmission of
  light through nanometric holes in metallic thin films},{''}
  \bibinfo{journal}{Opt.\ Commun.} \textbf{\bibinfo{volume}{195}},
  \bibinfo{pages}{107--118}.

\bibitem[{\citenamefont{Webb and Li}(2006)}]{WL06}
\bibinfo{author}{\bibnamefont{Webb}, \bibfnamefont{K.~J.}}, and
  \bibinfo{author}{\bibfnamefont{J.}~\bibnamefont{Li}}, \bibinfo{year}{2006},
  {``}\bibinfo{title}{Analysis of transmission through small apertures in
  conducting films},{''} \bibinfo{journal}{Phys.\ Rev.\ B}
  \textbf{\bibinfo{volume}{73}},  \bibinfo{pages}{033401}.

\bibitem[{\citenamefont{Weyl}(1919)}]{W1919}
\bibinfo{author}{\bibnamefont{Weyl}, \bibfnamefont{H.}}, \bibinfo{year}{1919},
  {``}\bibinfo{title}{Ausbreitung elektromagnetisher Wellen ueber einem. ebenen
  Leiter},{''} \bibinfo{journal}{Ann.\ Phys.\ (Leipzig)}
  \textbf{\bibinfo{volume}{60}},  \bibinfo{pages}{481--500}.

\bibitem[{\citenamefont{Wood}(1902)}]{W1902}
\bibinfo{author}{\bibnamefont{Wood}, \bibfnamefont{R.~W.}},
  \bibinfo{year}{1902}, {``}\bibinfo{title}{On a remarkable case of uneven
  distribution of light in a diffraction grating spectrum},{''}
  \bibinfo{journal}{Philos.\ Mag.} \textbf{\bibinfo{volume}{4}},
  \bibinfo{pages}{396--402}.

\bibitem[{\citenamefont{Wood}(1912)}]{W1912}
\bibinfo{author}{\bibnamefont{Wood}, \bibfnamefont{R.~W.}},
  \bibinfo{year}{1912}, {``}\bibinfo{title}{Diffraction gratings with
  controlled groove form and abnormal distribution of intensity},{''}
  \bibinfo{journal}{Philos.\ Mag.} \textbf{\bibinfo{volume}{23}},
  \bibinfo{pages}{310--317}.

\bibitem[{\citenamefont{Wood}(1935)}]{W1935}
\bibinfo{author}{\bibnamefont{Wood}, \bibfnamefont{R.~W.}},
  \bibinfo{year}{1935}, {``}\bibinfo{title}{Anomalous diffraction
  gratings},{''} \bibinfo{journal}{Phys.\ Rev.} \textbf{\bibinfo{volume}{48}},
  \bibinfo{pages}{928--936}.

\bibitem[{\citenamefont{Yang and Sambles}(2002)}]{YS02}
\bibinfo{author}{\bibnamefont{Yang}, \bibfnamefont{F.}}, and
  \bibinfo{author}{\bibfnamefont{J.~R.} \bibnamefont{Sambles}},
  \bibinfo{year}{2002}, {``}\bibinfo{title}{Resonant transmission of microwaves
  through a narrow metallic slit},{''} \bibinfo{journal}{Phys.\ Rev.\ Lett.}
  \textbf{\bibinfo{volume}{89}},  \bibinfo{pages}{063901}.

\bibitem[{\citenamefont{Yang} \emph{et~al.}(1990)\citenamefont{Yang, Sambles,
  and Bradberry}}]{YSB1990}
\bibinfo{author}{\bibnamefont{Yang}, \bibfnamefont{F.}},
  \bibinfo{author}{\bibfnamefont{J.~R.} \bibnamefont{Sambles}}, and
  \bibinfo{author}{\bibfnamefont{G.~W.} \bibnamefont{Bradberry}},
  \bibinfo{year}{1990}, {``}\bibinfo{title}{Long-range coupled surface exciton
  polaritons},{''} \bibinfo{journal}{Phys.\ Rev.\ Lett.}
  \textbf{\bibinfo{volume}{64}},  \bibinfo{pages}{559--562}.

\bibitem[{\citenamefont{Yin} \emph{et~al.}(2004)\citenamefont{Yin,
  Vlasko-Vlasov, Rydh, Pearson, Welp, Chang, Gray, Schatz, Brown, and
  Kimball}}]{YVR04}
\bibinfo{author}{\bibnamefont{Yin}, \bibfnamefont{L.}},
  \bibinfo{author}{\bibfnamefont{V.~K.} \bibnamefont{Vlasko-Vlasov}},
  \bibinfo{author}{\bibfnamefont{A.}~\bibnamefont{Rydh}},
  \bibinfo{author}{\bibfnamefont{J.}~\bibnamefont{Pearson}},
  \bibinfo{author}{\bibfnamefont{U.}~\bibnamefont{Welp}},
  \bibinfo{author}{\bibfnamefont{S.-H.} \bibnamefont{Chang}},
  \bibinfo{author}{\bibfnamefont{S.~K.} \bibnamefont{Gray}},
  \bibinfo{author}{\bibfnamefont{G.~C.} \bibnamefont{Schatz}},
  \bibinfo{author}{\bibfnamefont{D.~B.} \bibnamefont{Brown}}, and
  \bibinfo{author}{\bibfnamefont{C.~W.} \bibnamefont{Kimball}},
  \bibinfo{year}{2004}, {``}\bibinfo{title}{Surface plasmons at single
  nanoholes in Au films},{''} \bibinfo{journal}{Appl.\ Phys.\ Lett.}
  \textbf{\bibinfo{volume}{85}},  \bibinfo{pages}{467--469}.

\bibitem[{\citenamefont{Zenneck}(1907)}]{Z1907}
\bibinfo{author}{\bibnamefont{Zenneck}, \bibfnamefont{J.}},
  \bibinfo{year}{1907}, {``}\bibinfo{title}{Uber die Fortpflanzung ebener
  elektromagnetischer Wellen langs einer ebenen Leiterflache und ihre Beziehung
  zur drahtlosen Telegraphie},{''} \bibinfo{journal}{Ann.\ Phys.\ (Leipzig)}
  \textbf{\bibinfo{volume}{23}},  \bibinfo{pages}{846--866}.

\bibitem[{\citenamefont{Zia} \emph{et~al.}(2006)\citenamefont{Zia, Schuller,
  Chandran, and Brongersma}}]{ZSC06}
\bibinfo{author}{\bibnamefont{Zia}, \bibfnamefont{R.}},
  \bibinfo{author}{\bibfnamefont{J.~A.} \bibnamefont{Schuller}},
  \bibinfo{author}{\bibfnamefont{A.}~\bibnamefont{Chandran}}, and
  \bibinfo{author}{\bibfnamefont{M.~L.} \bibnamefont{Brongersma}},
  \bibinfo{year}{2006}, {``}\bibinfo{title}{Plasmonics: the next chip-scale
  technology},{''} \bibinfo{journal}{Materials\ Today}
  \textbf{\bibinfo{volume}{9}},  \bibinfo{pages}{20--27}.

\bibitem[{\citenamefont{Zou} \emph{et~al.}(2004)\citenamefont{Zou, Janel, and
  Schatz}}]{ZJS04}
\bibinfo{author}{\bibnamefont{Zou}, \bibfnamefont{S.}},
  \bibinfo{author}{\bibfnamefont{N.}~\bibnamefont{Janel}}, and
  \bibinfo{author}{\bibfnamefont{G.~C.} \bibnamefont{Schatz}},
  \bibinfo{year}{2004}, {``}\bibinfo{title}{Silver nanoparticle array
  structures that produce remarkably narrow plasmon lineshapes},{''}
  \bibinfo{journal}{J.\ Chem.\ Phys.} \textbf{\bibinfo{volume}{120}},
  \bibinfo{pages}{10871--10875}.

\bibitem[{\citenamefont{Zou and Schatz}(2004)}]{ZS04}
\bibinfo{author}{\bibnamefont{Zou}, \bibfnamefont{S.}}, and
  \bibinfo{author}{\bibfnamefont{G.~C.} \bibnamefont{Schatz}},
  \bibinfo{year}{2004}, {``}\bibinfo{title}{Narrow plasmonic/photonic
  extinction and scattering line shapes for one and two dimensional silver
  nanoparticle arrays},{''} \bibinfo{journal}{J.\ Chem.\ Phys.}
  \textbf{\bibinfo{volume}{121}},  \bibinfo{pages}{12606--12612}.

\end{thebibliography}

\end{document}